\documentclass{article}

\usepackage{arxiv}

\usepackage[utf8]{inputenc} 
\usepackage[T1]{fontenc}    
\usepackage{hyperref}       
\usepackage{url}            
\usepackage{booktabs}       
\usepackage{amsfonts}       
\usepackage{nicefrac}       
\usepackage{microtype}      
\usepackage{lipsum}
\usepackage{graphicx}

\usepackage{amsmath}
\usepackage[version=4]{mhchem}
\usepackage{siunitx}
\usepackage{longtable,tabularx}
\usepackage{subcaption}
\usepackage{chemformula}
\usepackage{threeparttable}
\usepackage{enumitem}
\graphicspath{ {./images/} }

\title{RAPRAL v1.0: RAdiation Prediction using RAy tracing and Line-by-line methods for hypersonic air flows}

\author{
 Yuzhe Zhang \\
  State Key Laboratory of High Temperature Gas Dynamics\\ Institute of Mechanics, Chinese Academy of Science, Beijing 100190, China \\
  \texttt{zhangyuzhe@imech.ac.cn} \\
   \And
 Qizhen Hong \\
  State Key Laboratory of High Temperature Gas Dynamics\\ Institute of Mechanics, Chinese Academy of Science, Beijing 100190, China \\
  \texttt{hongqizhen@imech.ac.cn} \\
  \And
 Xiaoyong Wang \\
  State Key Laboratory of High Temperature Gas Dynamics\\ Institute of Mechanics, Chinese Academy of Science, Beijing 100190, China \\
  \texttt{wangxy@imech.ac.cn} \\
  \And
 Quanhua Sun \\
  State Key Laboratory of High Temperature Gas Dynamics\\ Institute of Mechanics, Chinese Academy of Science, Beijing 100190, China \\
  \texttt{qsun@imech.ac.cn} 
}

\begin{document}
\maketitle
\begin{abstract}
A new radiation solver, RAPRAL (RAdiation Prediction based on RAy tracing and Line-by-line) implemented in C++, is developed for simulating high-temperature thermochemical nonequilibrium radiative processes. RAPRAL integrates detailed line-by-line spectral modeling with a ray-tracing solution of the radiative transfer equation, enabling accurate resolution of both spectral features and spatial radiation transport. The adopted methods and their implementation are described in detail. 
To assess the overall capability and accuracy of RAPRAL, we first focus on the computation of atomic and molecular bulk spectral coefficients. Through comparison with the established code in the literature, RAPRAL demonstrates its ability to accurately capture key spectral features across a wide range of conditions.
Moreover, RAPRAL is applied to predict afterbody radiative heating in the Fire II flight experiment, based on a two-temperature, 11-species air flowfield. The results demonstrate that the present approach provides reliable predictions of radiative heat flux and effectively captures the dominant radiation mechanisms.
Overall, the presented results demonstrate that RAPRAL is a robust tool for simulating radiative processes in hypersonic air flows, and future versions will extend its capabilities to include species relevant to planetary atmospheres.
\end{abstract}


\section{Introduction}
In hypersonic flows, such as those encountered during atmospheric reentry, the contribution of radiative heating increases significantly with both freestream velocity and vehicle characteristic length. Therefore, in addition to convective heating, which have been extensively investigated \cite{fay1958theory,hoshizaki1975critical,gkimisis2023data,teixeira2023catalytic,tang2024aerodynamic}, radiative heat transfer progressively becomes a non-negligible, and in some cases dominant, component of the total heat load \cite{johnston2015features,johnston2016refinements}.

An early estimate of radiative heat flux was proposed by Martin \cite{martin1966atmospheric} in the form of an empirical correlation:
\begin{equation}
    q_{rad} = CR_{nose}{\left(\frac{\rho}{\rho^*}\right)}^{1.6}{\left(\frac{V_{\infty}}{10^4}\right)}^{8.5}~ {\rm W/cm^2},
\end{equation}
where $R_{nose}$ is the nose radius, $\rho$ and $\rho^*$ denote the local and sea-level atmospheric densities, respectively, $V_{\infty}$ is the freestream velocity, and $C$ is a unit-dependent constant. This expression highlights the strong power-law dependence of radiative heating on both velocity and characteristic length scale.
Brandis and Johnston \cite{brandis2014characterization} showed that, for a vehicle with a 1 m nose radius, radiative heating exceeds 50\% of the total heat flux at velocities above 12 km/s, while for a 5 m nose radius this threshold decreases to approximately 10 km/s. Sutton \cite{sutton1984air}, based on inviscid and chemical equilibrium flow-radiation coupling simulations, reported that approximately 35\% of the peak heat flux in the Fire II flight experiment is attributable to radiation. Furthermore, Lee and Goodrich \cite{lee1972aerothermodynamic} measured a peak radiative heat flux of 115 $\rm W/cm^2$ at the stagnation point during the Apollo 17 mission, accounting for nearly 25\% of the total heat load. Collectively, these results demonstrate that radiative heat transfer plays a critical role in high-enthalpy entry flows.
With the increasing interest in deep-space exploration missions, such as Mars sample return \cite{brandis2013investigation} and asteroid sample return \cite{atkins1997stardust}, entry velocities commonly exceed 9 km/s, further amplifying the need for accurate radiative heat transfer predictions in thermal protection system (TPS) design.

Currently, widely used radiation prediction tools in the aerospace community include the NEQAIR code \cite{park1985nonequilibrium,whiting1996neqair96}, as well as several open-source programs such as Photaura \cite{potter2011modelling} developed at University of Queensland and SPARK \cite{lopez2016spark} developed at University of Lisbon. Table \ref{tab:radSoftware} summarizes representative radiation prediction tools.
\begin{table}[htb]
    \caption{\enspace Summary of widely used radiation prediction codes and their key features.}
    \footnotesize
    \setlength{\tabcolsep}{4pt}
    \renewcommand{\arraystretch}{1.5}
    \centering
    \begin{threeparttable}
        \begin{tabular}{lcccccc}
            \hline
            Codes & Country & State & Populations method & Molecular spectral method & Ref. \\
            \hline
            HARA & USA & In-house & Eq. \& Neq. & Band model \& LBL & Johnston \cite{johnston2006nonequilibrium} \\
            SPECAIR & USA & Commercial & Eq. \& Neq. & LBL & Laux et al. \cite{laux2003optical} \\
            SPRADIAN07 & JPN \& KOR & In-house & Eq. \& Neq. & LBL & Fujita and Abe \cite{fujita1997spradian} \\
            NEQAIR & USA & In-house & Eq. \& Neq. & LBL & Park and Whiting \cite{park1985nonequilibrium, whiting1996neqair96} \\
            RADIS & FRA & Open-source & Eq.* & LBL & Pannier and Laux \cite{pannier2019radis} \\
            PHOTAURA & AUS & Open-source & Eq. \& Neq. & LBL & Potter \cite{potter2011modelling} \\
            SPARK & PRT & Open-source & Eq.** & LBL & Lopez et al. \cite{lopez2016spark} \\
            \hline
        \end{tabular}

        \begin{tablenotes}
            \item *: Electronic ground states for all species; nonequilibrium rovibrational populations only for \ch{CO} and \ch{CO2}, with all others in equilibrium distribution.
            \item **: Nonequilibrium excited-state populations are not computed internally; they should be supplied as external number density inputs.
        \end{tablenotes}
    \end{threeparttable}
    \label{tab:radSoftware}
\end{table}

Existing radiation prediction codes have been successfully developed for a wide range of applications, with modeling strategies often tailored to specific spectral regimes and underlying physical assumptions.
For example, among open-source codes, RADIS \cite{pannier2019radis} is primarily oriented toward infrared molecular radiation, where vibrational-rotational transitions dominate, and typically assumes electronic ground states for most species; nonequilibrium rovibrational treatments are only available for selected molecules such as \ch{CO} and \ch{CO2}.
Spark \cite{lopez2016spark} provides a flexible framework for handling atomic and molecular discrete and continuum radiation processes; however, nonequilibrium population distributions are generally supplied through coupling with external solvers.
Photaura \cite{potter2011modelling} offers a high level of integration with flow solvers such as Poshax \cite{gollan2008computational} and Eilmer \cite{gollan2026eilmer}, enabling both equilibrium and nonequilibrium radiation modeling within coupled flow-radiation simulations. However, its design is primarily oriented toward tightly coupled simulations, which may introduce additional complexity when standalone spectral analysis or code portability is desired. Its line-by-line (LBL) spectral calculations are typically performed in a serial manner, which may limit computational efficiency in large-scale, multi-species applications. Furthermore, based on publicly available information, subsequent development activity appears limited, and the code has not been carried forward into more recent projects \cite{GDTK} by the original development team.
Given the status of existing open-source codes and the demands of hypersonic nonequilibrium radiation analysis in the VUV/UV/Vis spectral range, a new radiation solver, RAPRAL (RAdiation Prediction using RAy-tracing and Line-by-line methods), is developed in this work. 
The current version of the code focuses on air species and supports both equilibrium and nonequilibrium electronic state populations for atomic and diatomic species. Rovibrational populations are described using Boltzmann distributions characterized by vibrational and rotational temperatures, with extensible interfaces for future nonequilibrium treatments in these modes. A LBL spectral model is coupled with a ray-tracing framework to solve the radiative transfer equation (RTE) on structured-grid flowfields. 
To enhance computational efficiency, OpenMP-based parallelization is applied to both spectral property evaluation and radiative transfer calculations. Additionally, a unified flowfield data interface allows the code to operate independently of specific CFD solvers, improving portability and applicability across different simulation environments. The solver is implemented in C++, with Python utilities used for species and spectroscopic database generation. 

The present study focuses on the numerical implementation of RAPRAL and its validation against benchmark cases. Particular emphasis is placed on the detailed formulation of absorption and emission coefficient models, as well as the numerical treatment of the RTE within a nonequilibrium framework.
The underlying algorithms are described with attention to both mathematical consistency and computational efficiency. The solver’s reliability and accuracy are assessed through comparisons with established open-source code and experimental data across representative test cases.

\section{Methodology}
\label{sec:headings}

\subsection{Collisional module}
The determination of internal state populations for molecules is essential for evaluating radiative source terms. For atomic species, only electronic energy levels are relevant, whereas for diatomic or polyatomic molecules, electronic, vibrational, and rotational energy modes should be considered.
Depending on the degree of thermochemical nonequilibrium in the flowfield, different approaches can be employed to determine these populations. In regions where the local thermodynamic equilibrium (LTE) assumption holds, excited-state populations can be described by the Boltzmann distribution. 
In contrast, under strong nonequilibrium conditions, a collisional-radiative (CR) model \cite{potter2011modelling} is required to account for the underlying elementary processes (see below) to obtain physically consistent state populations.

\begin{figure}[htb]
    \centering
    \includegraphics[width=0.7\linewidth]{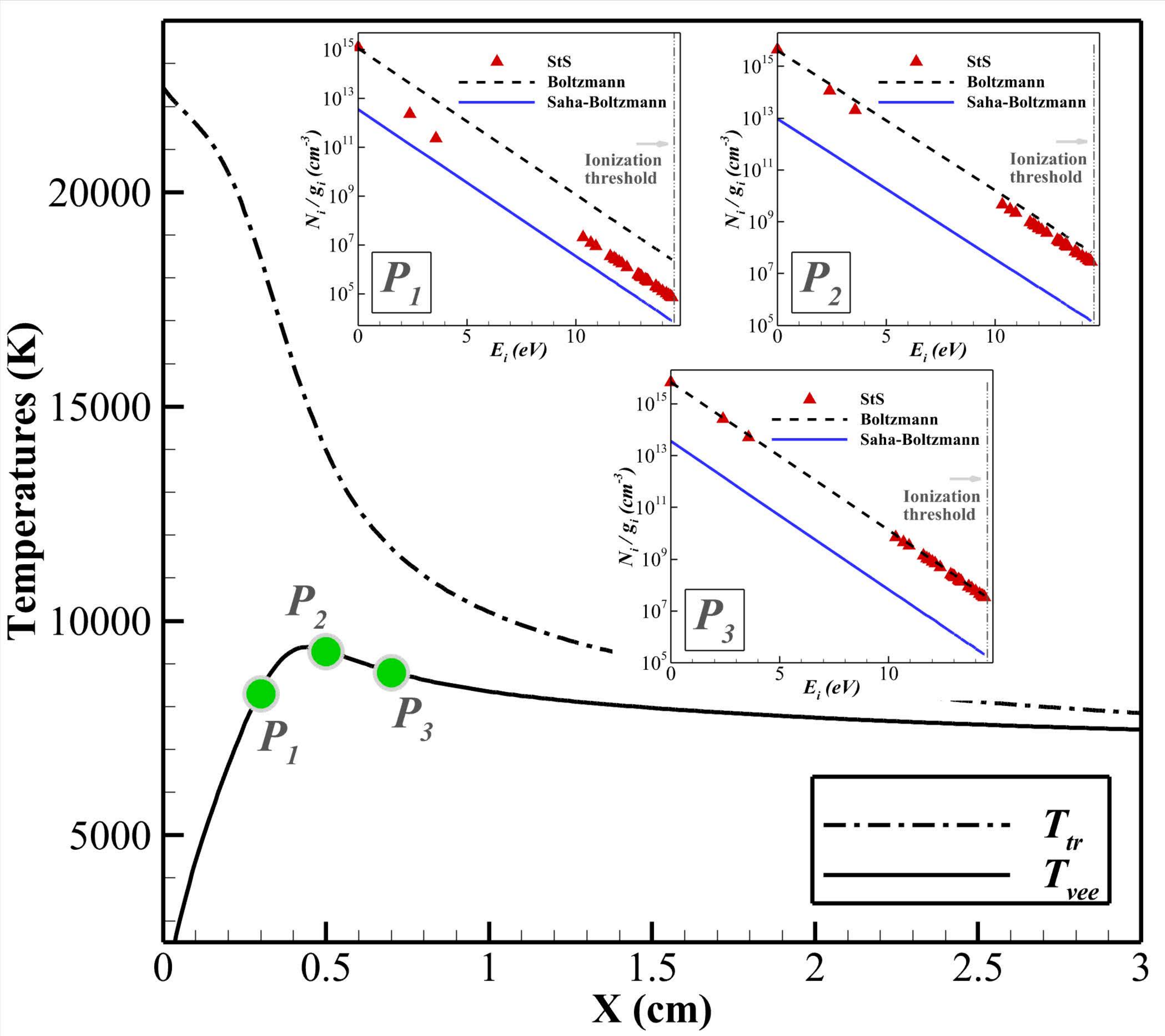}
    \caption{\enspace Populations of atomic N electronic states at three positions ($P_1, P_2, P_3$) behind a normal shock ($X = 0$) under condition of $V_{\infty} = 6.88~{\rm km/s}$ and $p = 0.2~{\rm Torr}$.}
    \label{fig:popN_nonBoltz}
\end{figure}

For atomic species, under equilibrium conditions, the electronic state populations follow the Boltzmann distribution in the absence of free electrons, or the Saha-Boltzmann distribution when ionization is considered. The Saha-Boltzmann formulation incorporates ionization equilibrium and typically considers multi-stage ionization processes. For instance, in the case of atomic oxygen, free-free transitions generally require inclusion of ionization up to the second stage, i.e., $\rm O \Leftrightarrow O^+ + e^-, O^+ \Leftrightarrow O^{2+} + e^-$.
\begin{equation}
    {\rm Boltzmann}: N_i^B = N\frac{Q_{ele, i}}{\sum{Q_{ele, j}}} = N\frac{g_i exp\left(\frac{-hcE_i}{kT_{ee}}\right)}{\sum{g_j exp\left(\frac{-hcE_j}{kT_{ee}}\right)}},
\end{equation}
\begin{equation}
    \begin{split}
        & {\rm Saha-Boltzmann}: N_i^{\rm S-B} =  1\times{10^{6}}N_+ N_{e^-}{\left(\frac{h^2}{2\pi m\times{10^{-3}}kT_{ee}}\right)^{1.5}}\frac{g_i exp\left(\frac{-hc\left(E_i - E_{\rm ionize}\right)}{kT_{ee}}\right)}{2Q_+},
    \end{split}
\end{equation}
where $N$ denotes the total number density of all electronic states of a given atomic species ($\text{cm}^{-3}$), $k = 1.380\times10^{-16}~\rm{g\cdot cm^2/s^2/K}$ is the Boltzmann constant, $c = 2.998\times10^{10}~\rm{cm/s}$ is the speed of light, and $h = 6.626\times10^{-27}~\rm{g\cdot cm^2/s}$ is the Planck constant. $N_+$ and $N_{e^-}$ represent the number densities ($\text{cm}^{-3}$) of ions (e.g. \ch{N+}) and electrons, respectively, $m$ is the electron mass (g), $E_{\rm ionize}$ and $Q_+$ are the ionization energy (cm$^{-1}$) and the partition function of the ionized species, $g_{i}$ denotes the electronic degeneracy, $E_i$ corresponds to the electronic energy (in $\rm cm^{-1}$), and $T_{ee}$ is electron-electronic temperature.
Under strongly nonequilibrium conditions, the electronic state populations should be determined using more detailed approaches, such as the CR model (also referred to as state-to-state, StS, modeling), as described in the following section.
Figure~\ref{fig:popN_nonBoltz} illustrates the temperature distributions and the population distributions of atomic nitrogen electronic states as functions of energy level at three locations ($P_1$, $P_2$, and $P_3$) downstream of a normal shock ($X = 0$) in a pure nitrogen flow ($V_{\infty} = 6.88~\mathrm{km/s},~p = 0.2~\mathrm{Torr}$). The electronic StS results correspond to populations directly obtained from the state-specific modeling, in which the rate mechanism is adopted from the sensitivity-analysis-optimized model proposed by the authors \cite{zhang2025uncertainty}. Based on these results, the corresponding Boltzmann and Saha-Boltzmann distributions are reconstructed for comparison.
At location $P_1$, all excited-state populations, except for the ground state ($E_i = 0$), exhibit significant deviations from the Boltzmann distribution. Moreover, the populations of higher excited states tend to approach the Saha-Boltzmann distribution, indicating that the flow is closer to ionization equilibrium in this regime. As the flow evolves downstream, the populations gradually relax toward the Boltzmann distribution. The pronounced nonequilibrium in excited-state populations persists over a finite region behind the shock and can have a significant impact on radiation predictions, as will be discussed in a later section.

For molecular species, the quantum-state description is more complex than for atomic species. For instance, a diatomic molecule such as $\mathrm{N}_2$, neglecting spin effects, can be a specific quantum state uniquely identified by the electronic level $i$, vibrational quantum number $V$, and rotational quantum number $J$. As with atomic species, the determination of molecular state populations depends on the degree of thermochemical nonequilibrium. Unlike atomic systems, diatomic molecules do not generally admit a universal closed-form Saha-Boltzmann distribution due to the additional vibrational and rotational energy modes. 
Under equilibrium conditions, the population of molecular electronic states can be described by a Boltzmann distribution as follows:
\begin{equation}
    \begin{split}
        & N_{i,mol}^{B} = N_{mol} \frac{Q_{ele,i}Q_{vr,i}}{\sum{Q_{ele,j}Q_{vr,j}}} = N_{mol} \frac{g_i exp\left(\frac{-hcE_i}{kT_{ee}}\right)Q_{vr,i}}{\sum{g_j exp\left(\frac{-hcE_j}{kT_{ee}}\right)Q_{vr,j}}}, \\
        & Q_{vr,i} = \sum_{V=0}^{V_{max}}\{exp\left(\frac{-hcE_V}{kT_v}\right) \frac{1}{\sigma} \sum_{J=J_{min}}^{J_{max}}{\left[\left(2J+1\right)L_{alter}exp\left(\frac{-hcE_J}{kT_r}\right)\right]}\},
    \end{split}
\end{equation}
where the subscript $mol$ denotes molecular species, and $Q_{vr,i}$ is the rovibrational partition function associated with electronic state $i$. The quantities $V_{\max}$, $J_{\min}$, and $J_{\max}$ represent the maximum vibrational quantum number for electronic state $i$, and the minimum and maximum rotational quantum numbers for vibrational state $V$, respectively. The symmetry factor $\sigma$ accounts for molecular symmetry, taking a value of 2 for homonuclear diatomic molecules and 1 for heteronuclear diatomic molecules. 
The terms $E_v$ and $E_J$ correspond to the vibrational and rotational energies (in $\rm cm^{-1}$), respectively, while $T_v$ and $T_r$ denote the vibrational and rotational temperatures. Detailed formulations of these quantities are provided in a later section.
In practice, the population of a specific rotational level for a given electronic and vibrational state is of particular interest, as it directly determines the intensity of bound–bound spectral lines. The corresponding number density of a rovibronic level can be expressed as:
\begin{equation}
    N_{\left(i,V,J\right), mol}^{B} = N_{i,{mol}}^{B} \frac{S_{multi}}{\sigma} \frac{\left(2J+1\right)L_{alter}exp\left(\frac{-hcE_V}{kT_v} + \frac{-hcE_J}{kT_r}\right)}{Q_{vr,i}},
\end{equation}
where $S_{\text{multi}}$ accounts for the effect of spin multiplicity.

\begin{figure}[!htb]
    \centering
    \includegraphics[width=0.7\linewidth]{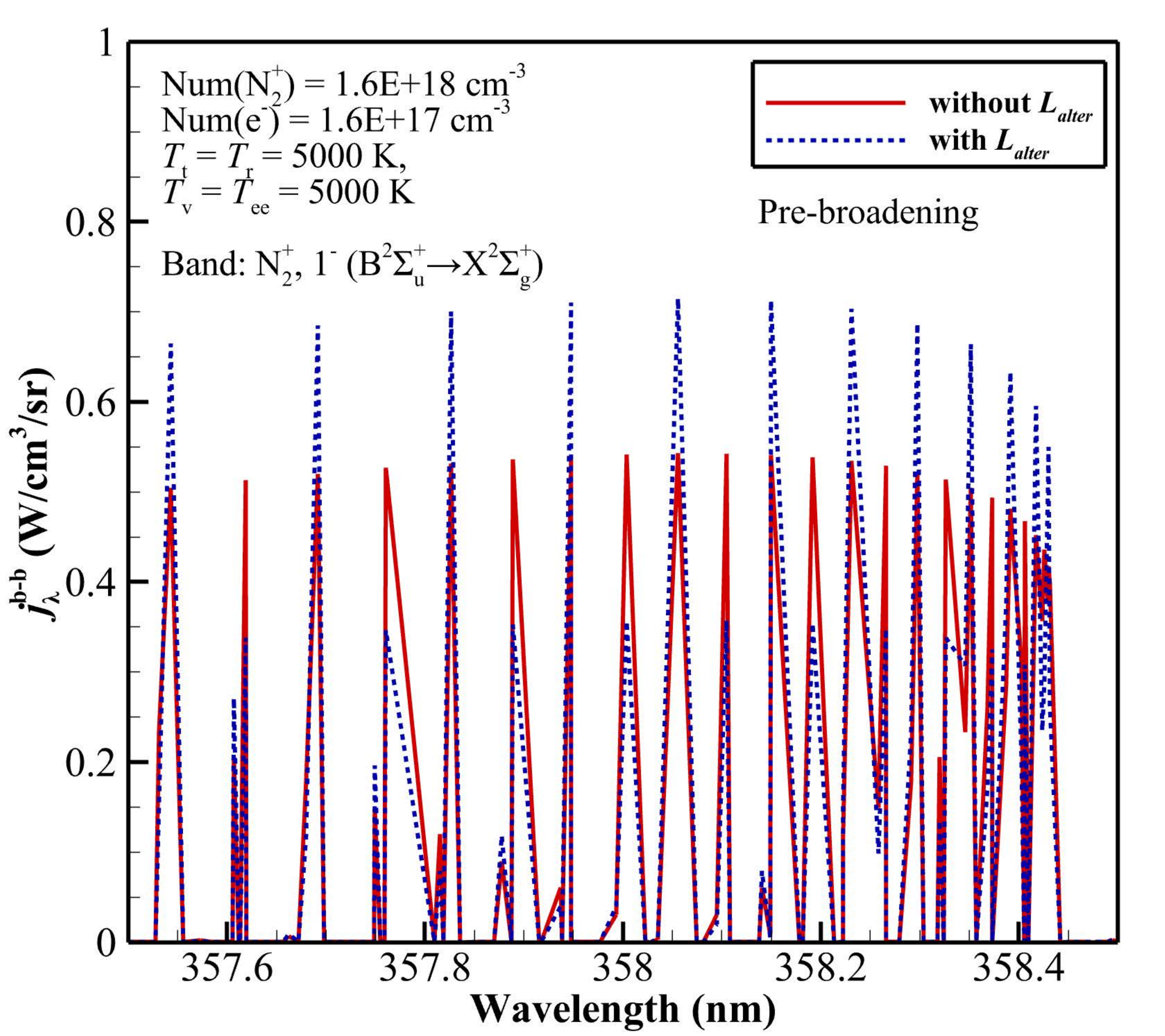}
    \caption{\enspace Effect of the line alternation factor on the emission intensity of selected bands of the $\rm N_2^+$ first negative system.}
    \label{fig:factorLineAlter}
\end{figure}

The line alternation factor $L_{\text{alter}}$ (dimensionless) is introduced to account for the alternating variation in spectral line intensities with rotational quantum number. This phenomenon originates from nuclear spin statistics associated with the identical nuclei forming the molecule. A more detailed discussion can be found in Refs. \cite{whiting1996neqair96, herzberg2013molecular}; here, only the formulation adopted in the present work is briefly summarized.
As summarized in Table~\ref{tab:factorLineAlter}, $L$ denotes the electronic orbital angular momentum quantum number ($L = 0, 1, 2, \ldots$ corresponding to $\Sigma, \Pi, \Delta, \ldots$ states), and $I$ represents the nuclear spin quantum number of the constituent isotope. When $I$ is an integer, $M = 1$, whereas for half-integer $I$, $M = 0$.
Figure~\ref{fig:factorLineAlter} presents the first negative band system of $\mathrm{N}_2^+$ ($\mathrm{B}^2\Sigma_u^+ \rightarrow \mathrm{X}^2\Sigma_g^+$), corresponding to the first two rows in Table~\ref{tab:factorLineAlter}. It is observed that, after incorporating the line alternation factor (blue dashed line), the unbroadened emission coefficient exhibits a pronounced alternating pattern as a function of wavelength, which is directly correlated with the rotational quantum number.

\begin{table}[!htb]
    \caption{\enspace Line alternation factor $L_{\rm alter}$ of diatomic molecules.}
    \footnotesize
    \setlength{\tabcolsep}{4pt}
    \renewcommand{\arraystretch}{1.5}
    \centering
    \begin{tabular}{lcccc}
        \hline
        Type & Level symmetry & $L_{alter}$ & Conditions \\
        \hline
        Homonuclear diatom ($L = 0$) & symmetric & $\left(2I + 1\right)\left(I + M\right)$ & even $J: {\Sigma}_g^+, {\Sigma}_u^-$, odd $J: {\Sigma}_g^-, {\Sigma}_u^+$ \\
        Homonuclear diatom ($L = 0$) & antisymmetric & $\left(2I + 1\right)\left(I + 1 - M\right)$ & even $J: {\Sigma}_g^-, {\Sigma}_u^+$, odd $J: {\Sigma}_g^+, {\Sigma}_u^-$ \\
        Homonuclear diatom ($L > 0$) & / & ${{\left(2I + 1\right)}^2}/2$ & / \\
        Heteronuclear diatom ($L \geq 0$) & / & ${\left(2I_1 + 1\right)}{\left(2I_2 + 1\right)}$ & / \\
        \hline
    \end{tabular}
    \label{tab:factorLineAlter}
\end{table}

\begin{figure}[htb]
    \centering
    \includegraphics[width=1.0\linewidth]{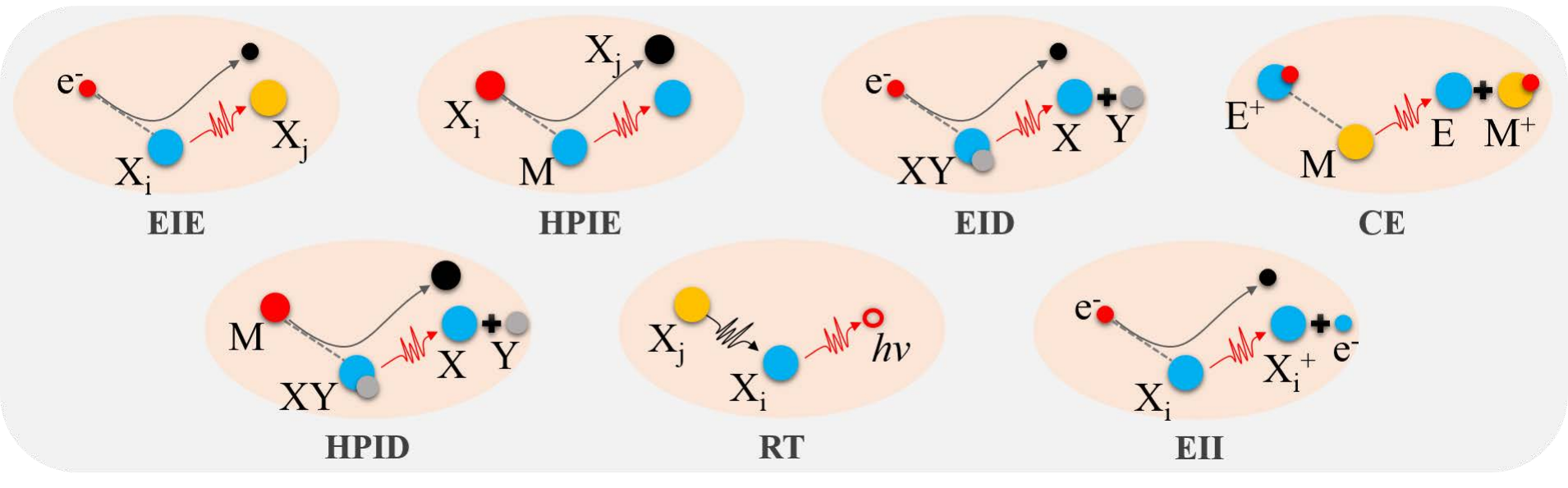}
    \caption{\enspace Schematic of the collisional-radiative mechanisms considered in this study.}
    \label{fig:sketchCRmodel}
\end{figure}

Under nonequilibrium conditions, state-specific modeling is required to resolve the population distributions. The CR model accounts for the elementary processes governing population evolution and establishes a quantitative relationship (master equation) between reaction mechanisms and population kinetics. Specifically, the following mechanisms are considered: heavy-particle impact excitation (HPIE), electron impact excitation (EIE), heavy-particle impact dissociation (HPID), electron impact dissociation (EID), bound-bound radiative transitions (RT), electron impact ionization (EII), and charge exchange (CE), i.e.,
\begin{equation}
    \begin{aligned}
        \frac{d N_i}{dt} &= {\left(\frac{d N_i}{dt}\right)}_{in} - {\left(\frac{d N_i}{dt}\right)}_{out} \\
        &= {\left(\frac{d N_i}{dt}\right)}_{HPIE} + {\left(\frac{d N_i}{dt}\right)}_{EIE} + {\left(\frac{d N_i}{dt}\right)}_{HPID} + {\left(\frac{d N_i}{dt}\right)}_{EID} + {\left(\frac{d N_i}{dt}\right)}_{RT} \\
        &+ {\left(\frac{d N_i}{dt}\right)}_{EII} + {\left(\frac{d N_i}{dt}\right)}_{CE},
    \end{aligned}
\end{equation}
where the subscripts “in” and “out” represent source and sink terms associated with processes that increase and decrease the population of electronic state $i$, respectively (see Figure~\ref{fig:sketchCRmodel}). It is seen that the master equation, as formulated above, involves time derivatives and, when coupled directly with the flow governing equations, imposes a substantial computational burden. 
In practical engineering applications, a quasi-steady-state (QSS) approximation, as proposed by Park \cite{park1989nonequilibrium}, is commonly employed to decouple the master equation from the flow solution, thereby simplifying the computation. The central assumption of this approach is that the time rate of change of excited-state populations is much smaller than the corresponding incoming and outgoing rates. Under this assumption, the time-derivative term on the left-hand side of the master equation can be neglected, yielding the QSS form of the population equations:
\begin{equation}
    \begin{aligned}
        0 = &{\left(\frac{d N_i}{dt}\right)}_{HPIE} + {\left(\frac{d N_i}{dt}\right)}_{EIE} + {\left(\frac{d N_i}{dt}\right)}_{HPID} + {\left(\frac{d N_i}{dt}\right)}_{EID} + {\left(\frac{d N_i}{dt}\right)}_{RT} + {\left(\frac{d N_i}{dt}\right)}_{EII} + {\left(\frac{d N_i}{dt}\right)}_{CE}.
    \end{aligned}
    \label{eq:CRmodel}
\end{equation}
In this regime, the electronic state populations of atoms and molecules can be determined as a post-processing step, using the flowfield solution obtained from the governing equations. Note that the QSS assumption is valid better for excited states \cite{park1989nonequilibrium}, and the ground-state population should be determined from the total number density $N_{\text{total}}$ to ensure closure of the QSS equations:
\begin{equation}
    N_{ground} = N_{i=0} = N_{total} - \sum_{i=1}^{max} N_i.
\end{equation}

The HPIE source term on the right-hand side of Eq.~\eqref{eq:CRmodel} are expressed as follows:
\begin{equation}
    {\left(\frac{d N_i}{dt}\right)}_{HPIE} = \sum_M{\sum_{j\neq i}{\frac{k_{HPIE}(j,i)}{N_A} N_j N_M}} - \sum_M{\sum_{j\neq i}{\frac{k_{HPIE}(i,j)}{N_A} N_i N_M}},
\end{equation}
where $N_A$ is Avogadro's number, and the heavy particles, denoted by the subscript $M$, include all species except electrons. The terms $k_{HPIE}(j,i)$ and $k_{HPIE}(i,j)$ denote the forward and backward rate coefficients ($\rm cm^3/mol/s$) for HPIE, 
\begin{equation}
    k_{HPIE}(i,j) = C_a T_{a,f}^{n_a} exp\left(-\frac{E_a}{T_{a,f}}\right),
    \label{eq:rateFormula_HPIE}
\end{equation}
where $C_a$, $n_a$, and $E_a$ are the Arrhenius coefficients. The forward reaction rate is evaluated at a characteristic temperature $T_{a,f}$, while the backward rate is evaluated at $T_{a,b}$, in accordance with the principle of detailed balance:
\begin{equation}
    k_{HPIE}(j,i) = {\left[k_{HPIE}(i,j)\frac{Q_i}{Q_j}\right]}_{T_{a,b}},
    \label{eq:backRateFormula_HPIE}
\end{equation}
where $Q_i$ and $Q_j$ are the rovibrational partition functions of the lower and upper electronic states, respectively.
The forward reaction rate coefficients for the following mechanisms are also formulated using an Arrhenius expression, consistent with Eq. \eqref{eq:rateFormula_HPIE}. 

Additionally,
\begin{equation}
    {\left(\frac{d N_i}{dt}\right)}_{EIE} = \sum_{j\neq i}{\frac{k_{EIE}(j,i)}{N_A} N_j N_{e^-}} - \sum_{j\neq i}{\frac{k_{EIE}(i,j)}{N_A} N_i N_{e^-}}.
\end{equation}
The backward rate coefficient for EIE is also expressed as in Eq. \eqref{eq:backRateFormula_HPIE}.
\begin{equation}
    {\left(\frac{d N_i}{dt}\right)}_{HPID} = \sum_M{\frac{k_{HPID}(d,i)}{N_A} N_X N_Y N_M} - \sum_M{\frac{k_{HPID}(i,d)}{N_A} N_i N_M},
\end{equation}
where indice $d$ denotes the dissociation product ($X$ and $Y$). Special attention is paid to the calculation of the backward reaction rate, namely
\begin{equation}
    k_{HPID}(d,i) = {\left[k_{HPID}(i,d)\frac{Q_i}{Q_X Q_Y}\right]}_{T_{a,b}} Q_t exp\left(-\frac{D_{0,i}}{T_{a,b}}\right),
\end{equation}
where $D_{0,i}$ is the dissociation energy of electronic state $i$ at the vibrational ground level (K), and $Q_t$ is the translational partition function, for which the approximate expression reported by Vincenti and Kruger \cite{vincenti1966introduction} is adopted.

\begin{equation}
    {\left(\frac{d N_i}{dt}\right)}_{EID} = \frac{k_{EID}(d,i)}{N_A} N_X N_Y N_{e^-} - \frac{k_{EID}(i,d)}{N_A} N_i N_{e^-}.
\end{equation}
The backward reaction rates $k_{EID}(d,i)$ are evaluated in the same manner as those used for HPID.
\begin{equation}
    {\left(\frac{d N_i}{dt}\right)}_{RT} = \sum_{j>i}{\Lambda_{j,i}(\lambda) k_{RT}(j,i) N_j} - \sum_{j<i}{\Lambda_{i,j}(\lambda) k_{RT}(i,j) N_i}.
\end{equation}
In this work, the vacuum-ultraviolet (VUV) wavelength range is assumed to be optically thick (i.e., $\Lambda(\lambda) = 0$ for $\lambda < 200~\rm nm$), since radiation in this band is readily absorbed by $\rm O_2$, while all other wavelength ranges are treated as optically thin ($\Lambda(\lambda) = 1$ for $\lambda > 200~\rm nm$). Photo-induced transitions (the inverse of spontaneous radiative transitions) are neglected by assigning an effectively zero ($1\times10^{-100}$) rate coefficient.
\begin{equation}
    {\left(\frac{d N_i}{dt}\right)}_{EII} = \frac{k_{EII}(c,i)}{N_A} N_{i^+} N_{e^-} N_{e^-} - \frac{k_{EII}(i,c)}{N_A} N_{i} N_{e^-},
\end{equation}
where $c$ denotes the ionization products ($N_{i^+}$ and $e^-$). Since only atomic EII is considered in this study, the backward rate for EII is expressed as:
\begin{equation}
    k_{EII}(c,i) = {\left[k_{EII}(i,c)\frac{Q_i}{Q_{i^+} Q_{e^-}}\right]}_{T_{a,b}}.
\end{equation}

Moreover,
\begin{equation}
    {\left(\frac{d N_i}{dt}\right)}_{CE} = \frac{k_{CE}(E,i)}{N_A} N_{M^+} N_{E} - \frac{k_{CE}(i,E)}{N_A} N_{i} N_{M},
\end{equation}
\begin{equation}
    k_{CE}(E,i) = {\left[k_{CE}(i,E)\frac{Q_{E^+} Q_M}{Q_E Q_{M^+}}\right]}_{T_{a,b}}.
\end{equation}

Substituting the above expressions of source terms into the QSS-form equations and rearranging yields:
\begin{equation}
    \boldsymbol{R}\boldsymbol{n} = \boldsymbol{b},
\end{equation}
where $\boldsymbol{n}$ is the unknown vector of number densities for all electronic states of a given radiating species (e.g., $\rm N_2$ or $\rm N_2^+$) in the CR model $\boldsymbol{n} = \left\{N_1, N_2, ..., N_l\right\}$, $\boldsymbol{R}$ is the $l\times l$ coefficient matrix, and $\boldsymbol{b}$ is the $l$-dimensional source vector.
If the total number of electronic states for this radiating species is $l^*$, then the system can be expressed as:
\begin{equation}
    \begin{aligned}
        &\boldsymbol{R}(i=1, j) = \left\{\begin{aligned}
            &1, \enspace j < l \\
            &1 + \sum_{p=l+1}^{l^*} \frac{Q_p}{Q_l}, j = l
            \end{aligned}\right. \\
        &\boldsymbol{R}(i>1, j=i) = \left\{\begin{aligned} \\
            &- \sum_{m\neq i}\sum_M{\frac{k_{HPIE}(i,m)}{N_A} N_M} - \sum_{m\neq i}{\frac{k_{EIE}(i,m)}{N_A} N_{e^-}} \\
            &- \sum_M\frac{k_{HPID}(i,d)}{N_A} N_M - \frac{k_{EID}(i,d)}{N_A} N_{e^-} \\
            &- \sum_{m<i}{\Lambda_{i,m}(\lambda) k_{RT}(i,m)} \\
            &- \frac{k_{EII}(i,c)}{N_A} N_{e^-} - \frac{k_{CE}(i,E)}{N_A} N_{M}\end{aligned}\right. \\
        &\boldsymbol{R}(i>1, j>i) = \sum_M\frac{k_{HPIE}(j,i)}{N_A} N_M + \frac{k_{EIE}(j,i)}{N_A} N_{e^-} + \Lambda_{j,i}(\lambda) k_{RT}(j,i) \\
        &\boldsymbol{R}(i>1, j<i) = \sum_M\frac{k_{HPIE}(j,i)}{N_A} N_M + \frac{k_{EIE}(j,i)}{N_A} N_{e^-}.
    \end{aligned}
\end{equation}
The source vector is expressed as:
\begin{equation}
    \begin{aligned}
        &\boldsymbol{b}(i=1) = N_{total} \\
        &\boldsymbol{b}(i>1) = \left\{\begin{aligned}
            & - \sum_M{\frac{k_{HPID}(d,i)}{N_A} (N_X N_Y N_M)_{total}} - \frac{k_{EID}(d,i)}{N_A} (N_X N_Y N_{e^-})_{total} \\
            & - \frac{k_{EII}(c,i)}{N_A} (N_{i^+} N_{e^-} N_{e^-})_{total} - \frac{k_{CE}(E,i)}{N_A} (N_{M^+} N_{E})_{total}
        \end{aligned}\right.
    \end{aligned}
\end{equation}
The unknown population vector, $\boldsymbol{n}$, can then be computed by performing an LU decomposition of the coefficient matrix $\boldsymbol{R}$, which is numerically more robust than directly inverting the matrix.

\subsection{Radiative module}
Radiative energy propagating through a medium undergoes physical processes such as absorption and scattering. The conservation of radiative energy in the presence of these mechanisms is governed by the RTE \cite{modest2003radiative}:
\begin{equation}
    \frac{1}{c} \frac{\partial I_\lambda}{\partial t} + \frac{\partial I_\lambda}{\partial s} = j_\lambda - \kappa_\lambda I_\lambda - \sigma_{s \lambda} I_\lambda + \frac{\sigma_{s \lambda}}{4 \pi} \int_{4 \pi} I_\lambda\left(\hat{\mathbf{s}}_i\right) \Phi_\lambda\left(\hat{\mathbf{s}}_i, \hat{\mathbf{s}}\right) d \Omega_i,
\end{equation}
where $I_\lambda$ is the spectral radiative intensity at wavelength $\lambda$ and $\sigma_{s\lambda}$ represents the scattering coefficient. The term $\Phi_\lambda\left(\hat{\mathbf{s}}_i, \hat{\mathbf{s}}\right)$ is the scattering phase function, which characterizes the probability of photons being scattered from the incident direction $\hat{\mathbf{s}}_i$ into the direction $\hat{\mathbf{s}}$. The differential solid angle associated with radiation incident along $\hat{\mathbf{s}}_i$ is denoted by $d\Omega_i$.

In the subsequent analysis, the following assumptions are adopted:
\begin{enumerate}[leftmargin=*]
    \item Radiation propagates along straight paths, i.e., the refractive index of the medium is constant;
    \item The flow velocity of the medium is much smaller than the speed of light. This assumption is valid for most engineering applications, including the present study, with notable exceptions such as ultrashort-pulse lasers, where pulse durations on the order of picoseconds ($10^{-12}~\mathrm{s}$) or even femtoseconds ($10^{-15}~\mathrm{s}$) render the temporal derivative of radiative intensity non-negligible relative to the speed of light;
    \item The medium is in local thermodynamic equilibrium;
    \item Scattering processes are neglected.
\end{enumerate}
Under these assumptions, the RTE can be simplified as:
\begin{equation}
    \begin{aligned}
        &\frac{d I_\lambda}{d s} = \kappa_\lambda B_\lambda - \kappa_\lambda I_\lambda, \\
        & B_{\lambda} = \frac{2hc^{2}}{1\times10^{4}\left({\lambda\times10^{-7}}\right)^{5}\left[exp\left(\frac{hc}{\lambda\times10^{-7} k T_{ee}}\right)-1\right]}  \overset{\rm Boltzmann}{=} \frac{2hc^{2}}{1\times10^{4}\left({\lambda\times10^{-7}}\right)^{5}\left(\frac{N_l}{N_u}\frac{g_u}{g_l}-1\right)}.
    \end{aligned}
    \label{eq:funcBlackBodyRad}
\end{equation}
In this formulation, solving the RTE requires prior evaluation of two wavelength-dependent quantities: the emission coefficient $j_\lambda (= \kappa_\lambda B_\lambda)$ and the absorption coefficient $\kappa_\lambda$. Here, $B_{\lambda}$ denotes the Planck blackbody radiation function expressed in terms of wavelength ($\mathrm{W/cm^{2}/sr/\mu m}$) \cite{whiting1996neqair96}, where a factor of $\times 10^{4}$ is introduced to convert the wavelength unit from cm to $\mathrm{\mu m}$. 
$N_{l}$ and $N_{u}$ represent the number densities ($\mathrm{cm^{-3}}$) of the lower and upper energy levels, respectively, and $g_{l}$ and $g_{u}$ denote their corresponding degeneracies.

The primary radiating species in air mixture consist of atoms and diatomic molecules. In this work, the calculation of emission and absorption coefficients includes atomic bound-bound (B-B), bound-free (B-F), and free-free (F-F) transitions, as well as molecular B-B transitions. 
Contributions from the molecular continuum, namely B-F transitions (photo-dissociation and photo-ionization) and F-F transitions (bremsstrahlung), could be negligible at high temperatures ($\ge 7000~\rm K$) \cite{chauveau2003radiative} and are therefore not included in the present implementation.

\subsubsection{Atoms}
The spontaneous emission coefficient for atomic B-B transitions, $j^{b-b}_{\lambda}$ (in $\rm W/cm^{3}/sr/{\mu}m$), is given by \cite{potter2011modelling}:
\begin{equation}
    j^{b-b}_{\lambda} = {\frac{hc}{4\pi}}{\frac{N_{u}A_{ul}}{\lambda\times10^{-7}}}{\phi_{\lambda}},
\end{equation}
where $A_{ul}$ is the Einstein coefficient for spontaneous emission (s$^{-1}$; typically available from the NIST Atomic Spectra Database \cite{NIST_ASD}), $\lambda$ is the central wavelength (in unit of nm unless otherwise specified), and $\phi_\lambda$ is the line shape function expressed in wavelength units ($\rm \mu m^{-1}$). The factor $10^{-7}$ converts wavelength from nm to cm. 
Induced emission is generally combined with absorption, and their net effect is represented by the absorption coefficient $\kappa$ (cm$^{-1}$):
\begin{equation}
    \begin{split}
        {\kappa}^{b-b}_{\lambda} = \frac{j^{b-b}_{\lambda}}{B_{\lambda}}.
    \end{split}
\end{equation}

\begin{figure}[!htb]
    \centering
    \includegraphics[width=0.4\linewidth]{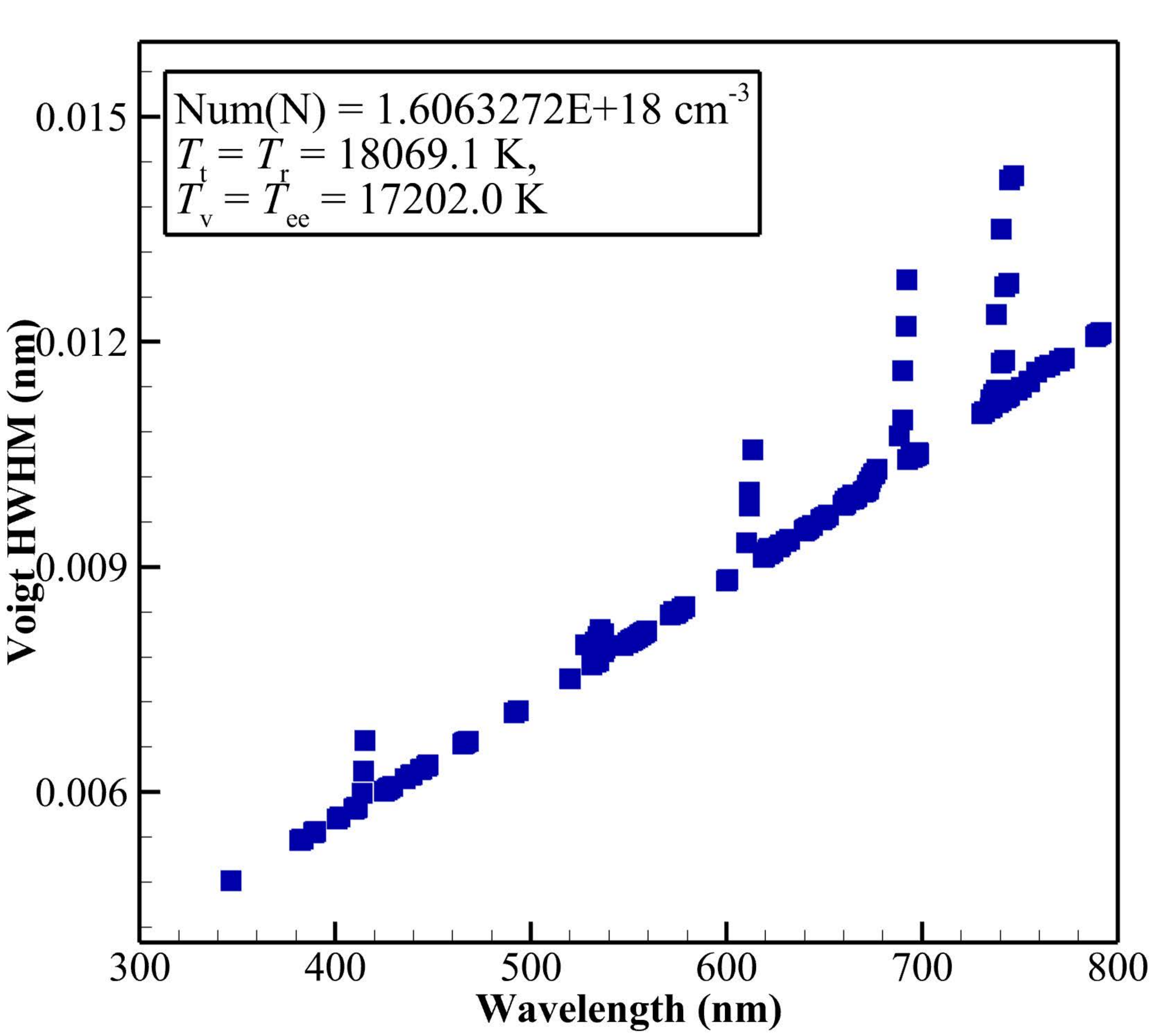}
    \caption{\enspace Voigt HWHM for atomic N in the 300-800 nm wavelength range under a specified condition.}
    \label{fig:voigtHWHM}
\end{figure}

For radiating atoms in excited states, each radiation line has a central wavelength uniquely determined by $\lambda = \left(hc\times10^{7}\right)/{\Delta}E_{ul}$, where $\Delta E_{ul}$ (J) is the energy released via radiation. Consequently, the emitted photons form a spectrum composed of discrete spectral lines.
However, due to thermal motion, collisions, and other broadening effects, the actual spectral line is distributed over a range of wavelengths centered around $\lambda$, corresponding to the broadening of an otherwise sharp line into a finite-width profile. To account for these broadening mechanisms, a line shape function $\phi_\lambda$ is introduced. At relatively high pressures and low temperatures, collisional broadening dominates, and the Lorentzian line shape is appropriate; at high temperatures and low pressures, Doppler broadening prevails, and a Gaussian profile is used.
In general, to accommodate different thermochemical nonequilibrium conditions, a combination of broadening mechanisms is required. The Voigt profile \cite{johnston2006nonequilibrium} is commonly employed to define $\phi_\lambda$ under such conditions:
\begin{equation}
    \begin{split}
        & \phi_{\lambda} = \frac{1\times10^{3}}{2{\lambda}^{V}\left[1.065+0.447\frac{\lambda^L}{\lambda^V}+0.058\left(\frac{\lambda^L}{\lambda^V}\right)^{2}\right]} \cdot \\
        &~ \left\{
        \begin{aligned}
            \left(1-\frac{\lambda^L}{\lambda^V}\right)exp\left[-2.772\left(\frac{\Delta{\lambda}}{2{\lambda}^{V}}\right)^{2}\right] + \frac{\lambda^L}{\lambda^V\left[1+4\left(\frac{\Delta{\lambda}}{2{\lambda}^{V}}\right)^{2}\right]} \\
            + 0.016\frac{\lambda^L}{\lambda^V}\left(1-\frac{\lambda^L}{\lambda^V}\right)\left[exp\left(-0.4\left(\frac{\Delta{\lambda}}{2{\lambda}^{V}}\right)^{2.25}\right)-\frac{1}{1+0.1\left(\frac{\Delta{\lambda}}{2{\lambda}^{V}}\right)^{2.25}}\right]
        \end{aligned}
        \right\},
    \end{split}
\end{equation}
where $\Delta \lambda$ represents the wavelength difference between a spectral grid node and the central wavelength and is always taken as a positive value. The factor $10^3$ converts the Voigt half-width at half-maximum (HWHM), $\lambda^V$, from nm to $\mu$m in the denominator. The Voigt HWHM is defined as:
\begin{equation}
    \begin{split}
        & \lambda^{V} = \left\{
        \begin{aligned}
            1 - 0.18121\left(1-d^{2}\right) \\
            - \left[0.023665 exp\left(0.6d\right)+0.00418 exp\left(-1.9d\right)\right]sin\left(\pi d\right)
        \end{aligned}
        \right\}
        \left({\lambda^L} + {\lambda^G}\right), \\
        & d = \frac{{\lambda^L} - {\lambda^G}}{{\lambda^L} + {\lambda^G}},
    \end{split}
\end{equation}
where $\lambda^L$ and $\lambda^G$ are the Lorentz and Doppler HWHMs, respectively. 
Figure~\ref{fig:voigtHWHM} presents the Voigt HWHM for atomic nitrogen under a representative flow condition considered in this study, providing a reference for the order of magnitude of this parameter.

Doppler broadening arises from the thermal motion of radiating atoms. When an atom emits electromagnetic radiation due to an electronic transition, the atom itself moves randomly in all directions. If the radiation propagates in the same direction as the atomic motion, the observed wavelength is shortened and the frequency is shifted higher; in the opposite direction, the wavelength lengthens (frequency decreases).
Under the assumption of a Maxwellian velocity distribution, the resulting line shape is Gaussian, with a HWHM given by \cite{drake2007springer}:
\begin{equation}
    {\lambda}^{G} = \frac{7.16233\times10^{-7}}{2} \lambda {(\frac{T_t}{\bar{M}})}^{0.5},
\end{equation}
where $T_t$ is the translational temperature (K) and $\bar{M}$ is the molar mass (g/mole).

Pressure broadening, on the other hand, results from collisions between radiating atoms and surrounding species. This effect produces a Lorentzian line shape, with the HWHM arising from the combined contributions of four mechanisms \cite{arnold1979line}: natural broadening, Stark broadening, resonance broadening, and Van der Waals broadening. The Lorentzian HWHM is expressed as:
\begin{equation}
    {\lambda}^{L} = {\lambda}^{N} + {\lambda}^{S} + {\lambda}^{R} + {\lambda}^{W}.
\end{equation}

Natural broadening arises from the Heisenberg uncertainty principle. Due to the fundamental quantum uncertainty, the radiating photon's energy is not exactly fixed but fluctuates around a central value, resulting in a finite line width. The characteristic timescale of natural broadening is generally much longer than those of other broadening mechanisms; therefore, it is often treated as a constant in engineering applications. In the present work, natural broadening is set as \cite{arnold1979line}: ${\lambda}^{N} = 5.9 \times 10^{-6}~\rm nm$.
\begin{figure}[!htb]
    \centering
    \begin{subfigure}{0.48\textwidth}
        \includegraphics[width=\linewidth]{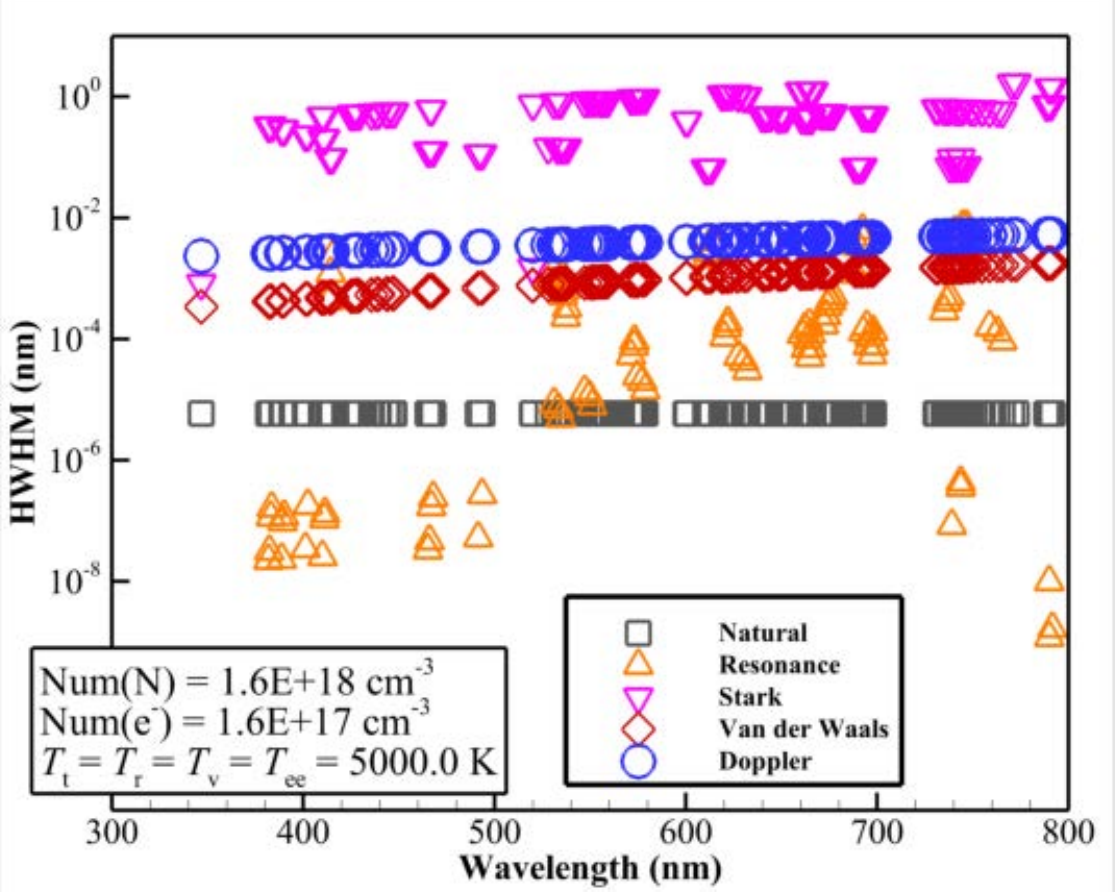}
        \caption{5000 K, $N_{e^{-}} = 1.6\times10^{17}~\rm cm^{-3}$}
    \end{subfigure}
    \begin{subfigure}{0.48\textwidth}
        \includegraphics[width=\linewidth]{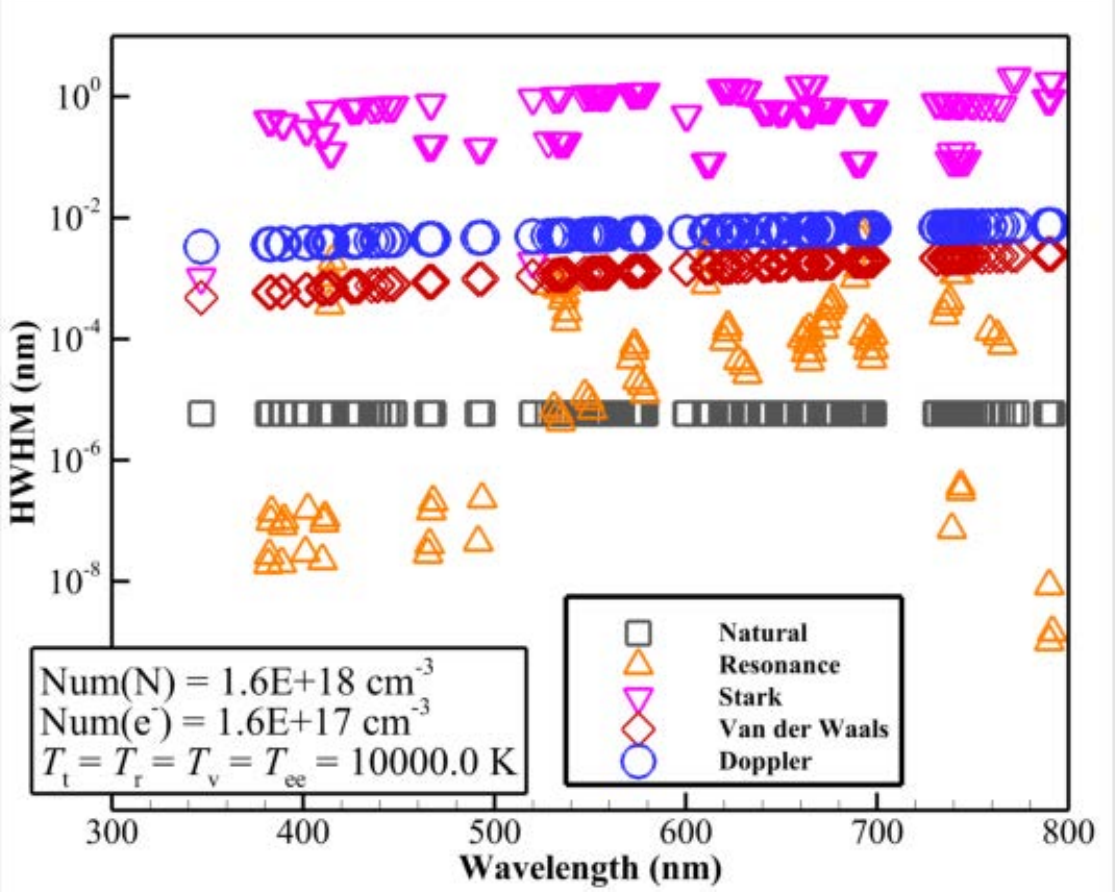}
        \caption{10000 K, $N_{e^{-}} = 1.6\times10^{17}~\rm cm^{-3}$}
    \end{subfigure}
    
    \begin{subfigure}{0.48\textwidth}
        \includegraphics[width=\linewidth]{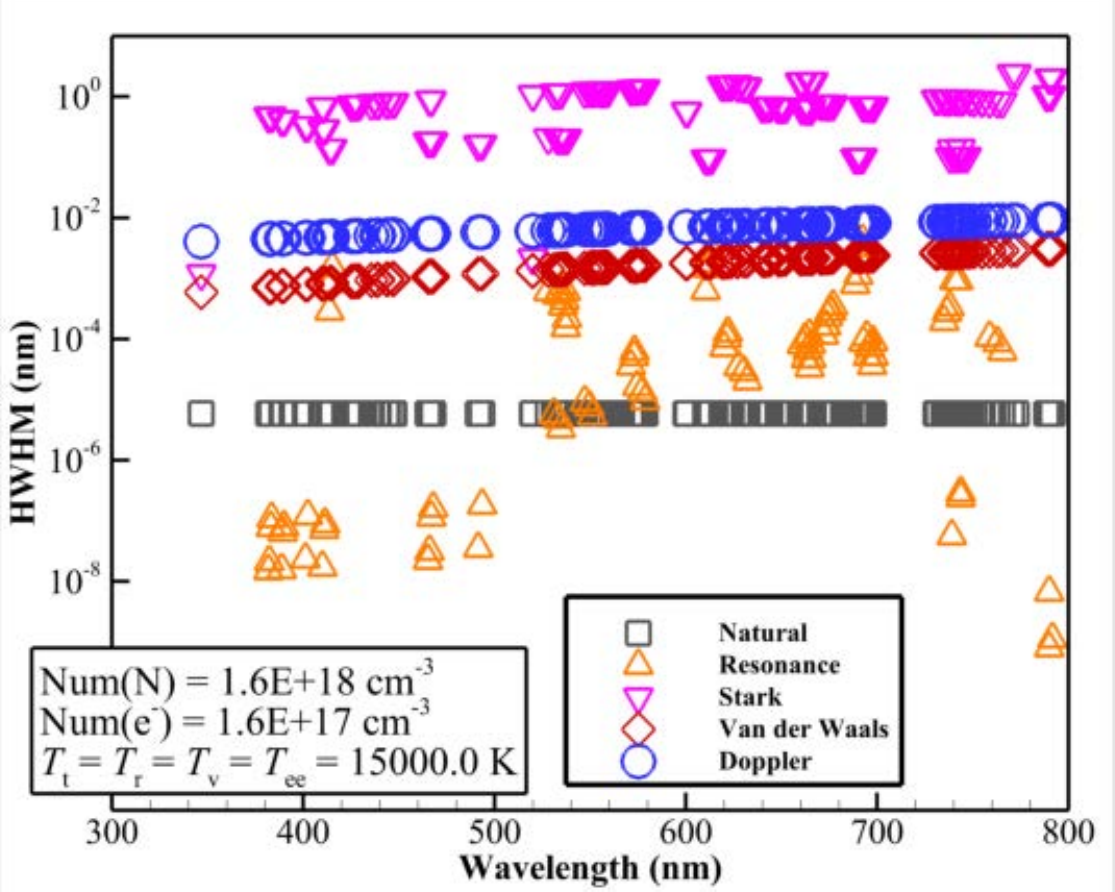}
        \caption{15000 K, $N_{e^{-}} = 1.6\times10^{17}~\rm cm^{-3}$}
    \end{subfigure}
    \begin{subfigure}{0.48\textwidth}
        \includegraphics[width=\linewidth]{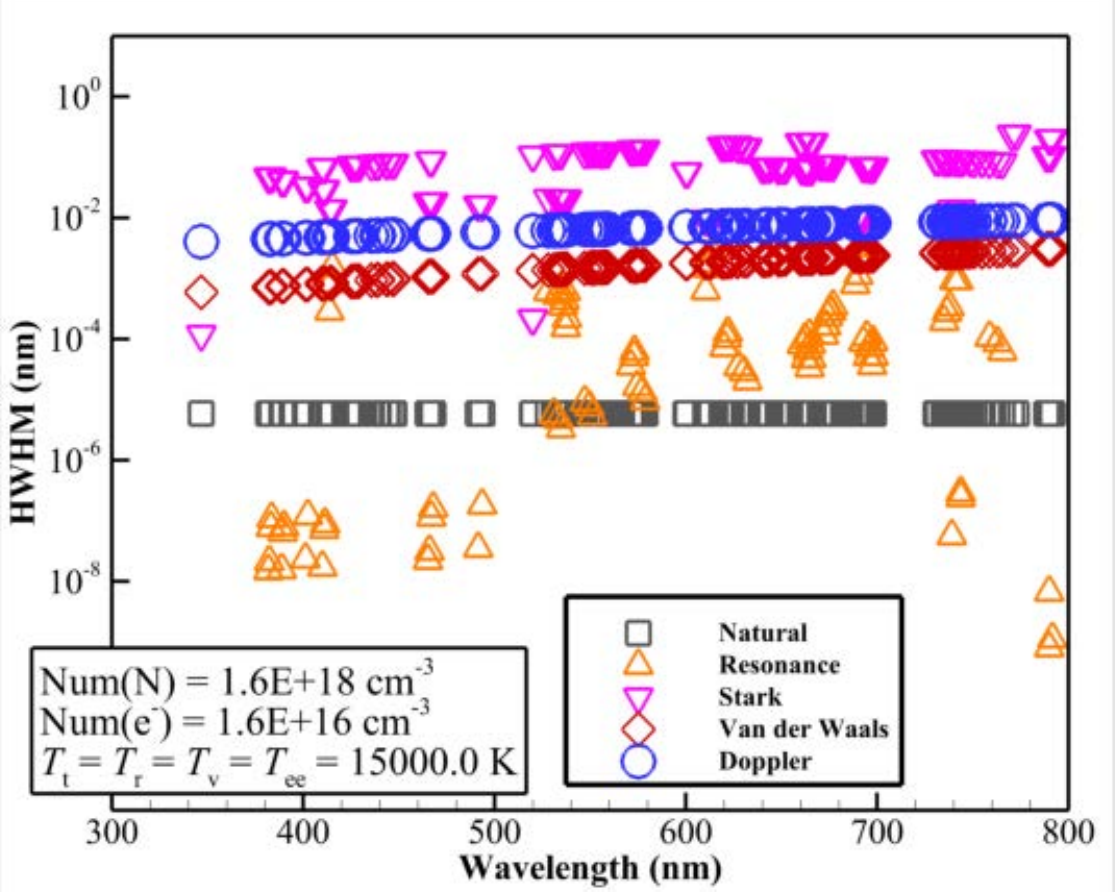}
        \caption{15000 K, $N_{e^{-}} = 1.6\times10^{16}~\rm cm^{-3}$}
    \end{subfigure}
    \caption{\enspace Comparison of HWHMs under thermal equilibrium at different temperatures and electron number densities.}
    \label{fig:HWHMs}
\end{figure}

Stark broadening results from collisions between radiating atoms and charged particles, including ions and free electrons. A more detailed discussion can be found in Johnston's dissertation \cite{johnston2006nonequilibrium}. In this work, the most recent fitted broadening coefficient \cite{johnston2006nonequilibrium}, ${\lambda}^{S,0}$, which closely matches detailed calculations, is adopted. The corresponding HWHM is expressed as:
\begin{equation}
    \begin{split}
        & {\lambda}^{S} = {\lambda}^{S,0}{\left(\frac{T_{ee}}{10000}\right)}^{0.33}{\left(\frac{{N}_{e^-}}{1\times10^{16}}\right)}, \\
        & {\lambda}^{S,0} = \frac{1.69\times10^{3}{\lambda}^{2}}{(\Delta E_{ion})^{2.623}}, \\
        & {\Delta E_{\rm ionize}} = max\left({{E}_{\rm ionize}-{E}_{u}, \frac{{E}_{\rm ionize}^{H}}{25}}\right),
    \end{split}
\end{equation}
where $E_{\rm ionize}^{\rm H}$ is the ionization energy of hydrogen (cm$^{-1}$), $E_u$ is the energy of the upper level corresponding to the spectral line with central wavelength $\lambda$ (cm$^{-1}$), and $\max()$ indicates taking the larger of the two arguments.

Resonance broadening arises from collisions between a radiating atom and identical atoms, and it occurs only for spectral lines corresponding to transitions from excited states with an electric dipole moment to the ground state (resonance lines). The HWHM for resonance broadening is expressed as \cite{drake2007springer}:
\begin{equation}
    \begin{split}
        & {\lambda}^{R} = {\lambda}^{R,0}{\left(\frac{g_{l}}{g_{u}}\right)}^{0.5}{\lambda}^{2}, \\
        & {\lambda}^{R,0} = {4.3\times10^{-28}}{\lambda}_{r}{f_{lu,r}}{N_{i=0}}, \\
        & {f_{lu,r}} = \frac{m_ec{\left(\lambda\times10^{-7}\right)}^{2}}{8{\pi}^{2}e^{2}}\frac{g_{u,r}}{g_{l,r}}A_{ul,r},
    \end{split}
\end{equation}
where $f_{lu}$ is the absorption oscillator strength (dimensionless), the subscript $r$ indicates a resonance line, $m_e$ and $e$ are the electron mass (g) and charge (esu), respectively, and the factor $10^{-7}$ converts wavelength from nm to cm.

Van der Waals broadening arises from the interaction between the permanent or induced dipole of a radiating atom and the induced dipole of surrounding ground-state atoms. The HWHM for Van der Waals broadening is expressed as:
\begin{equation}
    \begin{split}
        & {\lambda}^{W} = {\lambda}^{W,0}{\lambda}^{2}, \\
        & {\lambda}^{W,0} = {6.65\times10^{-29}}N{\left(\frac{2T_t}{\bar{M}}\right)}^{0.5} + {2.925\times10^{-29}}{N_{others}}{\left(\frac{2T_t}{\bar{M}_{others}}\right)}^{0.5},
    \end{split}
\end{equation}
where $N$ and $N_{\rm others}$ are the number densities of the radiating atom and other species (cm$^{-3}$), respectively.
Figure~\ref{fig:HWHMs} shows the HWHMs of different broadening mechanisms under thermal equilibrium at different temperatures and electron number densities.
It can be observed that the Stark HWHM is consistently much larger than the other mechanisms.

During the process in which a bound-state electron in an atom ($N_i$) is excited and subsequently freed from the atomic nucleus, the absorption coefficient (neglecting induced emission) can be expressed as \cite{johnston2006nonequilibrium}:
\begin{equation}
    {\kappa}^{b-f}_{\lambda,i} = N_{i}{\sigma}^{b-f}_{\lambda,i},
    \label{eq:abCoeBF}
\end{equation}
where $\sigma^{b-f}_{\lambda,i}$ is the photoionization cross section (cm$^2$) for absorption of a photon with wavelength $\lambda$ by an atom in bound state $i$, resulting in a free electron. In this work, atomic energy levels and photoionization cross sections are taken from the TOPbase database \cite{TOPbase}.
The corresponding emission coefficient is proportional to the product of the resulting ion number density $N_+$ and electron number density $N_{e^-}$:
\begin{equation}
    {j}^{b-f}_{\lambda,i} = N_{+}N_{e^-}{\xi}^{b-f}_{\lambda,i},
    \label{eq:emssCoeBF}
\end{equation}
where $\xi^{b-f}_{\lambda,i}$ is a physical quantity independent of the species number density. According to Kirchhoff's law, the absorption and emission coefficients under LTE satisfy:
\begin{equation}
    j^{b-f}_{eq} = B_{\lambda}{\kappa}^{b-f}_{eq} = B_{\lambda}N_{i,eq}{\sigma}^{b-f}_{\lambda,i}.
    \label{eq:emssCoeBF_LTE_cross}
\end{equation}
The LTE bound-state number density $N_{i,\rm eq}$ can be calculated using the Saha-Boltzmann relation:
\begin{equation}
    N_{i,eq} = N_{+,eq}N_{e^{-},eq}\left\{1\times{10^{6}}\left(\frac{h^2}{2\pi m\times{10^{-3}}kT_{ee}}\right)^{1.5}\frac{g_{i}exp\left[\frac{-hc\left(E_{i}-E_{ionize}\right)}{kT_{ee}}\right]}{2Q_{+}}\right\}.
\end{equation}
Since $\xi^{b-f}_{\lambda,i}$ is independent of whether the LTE assumption is valid, substituting $N_{i,\rm eq}$ into the emission expression allows one to solve for $\xi^{b-f}_{\lambda,i}$ explicitly. Finally, substituting both $\sigma^{b-f}_{\lambda,i}$ and $\xi^{b-f}_{\lambda,i}$ back into the general B-F emission formula gives the final form:
\begin{equation}
    \begin{split}
        {j}^{b-f}_{\lambda,i} = N_{+}N_{e^{-}}\frac{2hc^{2}}{1\times10^{4}\left({\lambda\times10^{-7}}\right)^{5}\left[exp\left(\frac{hc}{\lambda\times10^{-7} k T_{ee}}\right)-1\right]}{\sigma}^{b-f}_{\lambda,i} \cdot \\
        \left\{1\times{10^{6}}\left(\frac{h^2}{2\pi m\times{10^{-3}}kT_{ee}}\right)^{1.5}\frac{g_{i}exp\left[\frac{-hc\left(E_{i}-E_{ionize}\right)}{kT_{ee}}\right]}{2Q_{+}}\right\}.
    \end{split}
\end{equation}


Furthermore, F-F transitions refer to the process in which a free electron, when passing near a charged particle $P^{+l}$ ($l = 0, 1, 2$; for nonequilibrium air radiation, considering ions with up to two lost electrons is usually sufficient), is decelerated by the Coulomb field of the particle and emits a photon carrying part of its kinetic energy. This process is commonly referred to as Bremsstrahlung, or 'braking radiation,' due to the electron's deceleration.
For all non-hydrogenic atomic species, the emission and absorption coefficients are obtained by scaling the corresponding hydrogen coefficients, following the hydrogenic atom approximation. Under this assumption, the volumetric absorption coefficient for F-F (electron-ion) transitions is expressed as \cite{whiting1996neqair96}:
\begin{equation}
        \begin{split}
            & {\kappa}^{f-f}_{\lambda} = N_{a}{\sigma}^{f-f}_{\lambda}, \\
            & {\sigma}^{f-f}_{\lambda} = \frac{N_{+}N_{e^-}}{N_{a}}{\sigma}^{H,f-f}_{\lambda}\left[1+D\left(\sigma,T_{ee}\right)\right],
        \end{split}
\end{equation}
where $\sigma_\lambda^{f-f}$ is the absorption cross section of a single particle, $N_a$ is the number density of the charged particle $P^{+l}$ ($l = 0, 1, 2$), and $\sigma_\lambda^{H,f-f}$ is the corresponding hydrogenic absorption cross section. In this work, $\sigma_\lambda^{H,f-f}$ is obtained by fitting the tabulated data in Peach \cite{peach1970continuous}, yielding an approximate expression:
\begin{equation}
    \begin{split}
        & {z^4{\sigma}^3{\sigma}^{H,f-f}_{\lambda}\times10^{40}} = \frac{A_{1}(T_{ee})}{\sigma} + A_{2}(T_{ee}), \\
        & A_{1}(T_{ee}) = c_1 + c_{2}exp(-c_{3}T_{ee}), \\
        & A_{2}(T_{ee}) = c_4 + c_{5}exp(-c_{6}T_{ee}) + c_{7}exp(-c_{8}T_{ee}),
    \end{split}
\end{equation}
where $c_i$ are fitting coefficients (cm$^5$) listed in Table~\ref{tab:fitCoeFF} for atomic N and O, $z = l + 1$, and $\sigma$ is a dimensionless wavelength parameter
\begin{equation}
    \sigma = {\lambda}_{ground}^{H}/{\lambda} = 91.1267/{\lambda},
\end{equation}
where $\lambda_{\rm ground}^{H}$ corresponds to the hydrogen atom B-F transition wavelength (nm).
The factor $\frac{N_+ N_{e^-}}{N_a}$ can be obtained by solving the Saha-Boltzmann and Boltzmann equations simultaneously
\begin{equation}
    \frac{N_{+}N_{e^-}}{N_{a}} = \frac{1}{1\times{10^{6}}\left(\frac{h^2}{2\pi m\times{10^{-3}}kT_{ee}}\right)^{1.5}}\frac{2Q_{a}}{Q_{a,-}}exp\left(\frac{-hcE_{\rm ionize,a}}{kT_{ee}}\right),
\end{equation}
where $N_+$ is the number density of the next higher ionization state $P^{+(l+1)}$ and $Q_{a,-}$ is the partition function of the lower ionization state $P^{+(l-1)}$. The normalization factor $D(\sigma,T_{ee})$ is obtained by direct interpolation of the tabulated data reported by Peach \cite{peach1970continuous}.

\begin{table}[htb]
    \caption{\enspace Fitting parameters for the absorption coefficients of F-F transitions of atomic N and O.}
    \footnotesize
    \setlength{\tabcolsep}{4pt}
    \renewcommand{\arraystretch}{1.5}
    \centering
    \begin{tabular}{lcccccccc}
        \hline
        Atom & $c_1$ & $c_2$ & $c_3$ & $c_4$ & $c_5$ & $c_6$ & $c_7$ & $c_8$\\
        \hline
        N & 0.006805 & -0.00934 & 0.000116 & 0.508 & 1.046 & 5.77$\times10^{-5}$ & 1.85 & 0.000343\\
        O & 0.001905 & -0.005105 & 8.176$\times10^{-5}$ & 0.4554 & 1.0238 & 5.329$\times10^{-5}$ & 1.7587 & 0.003121\\
        \hline
    \end{tabular}
    \label{tab:fitCoeFF}
\end{table}

Under the LTE assumption, the emission coefficient for F-F transitions is obtained from Kirchhoff's law:
\begin{equation}
    j^{f-f}_{\lambda} = B_{\lambda}{\kappa}^{f-f}_{\lambda}.
\end{equation}

\begin{figure}[!htb]
    \centering
    \begin{subfigure}{0.48\textwidth}
        \includegraphics[width=\linewidth]{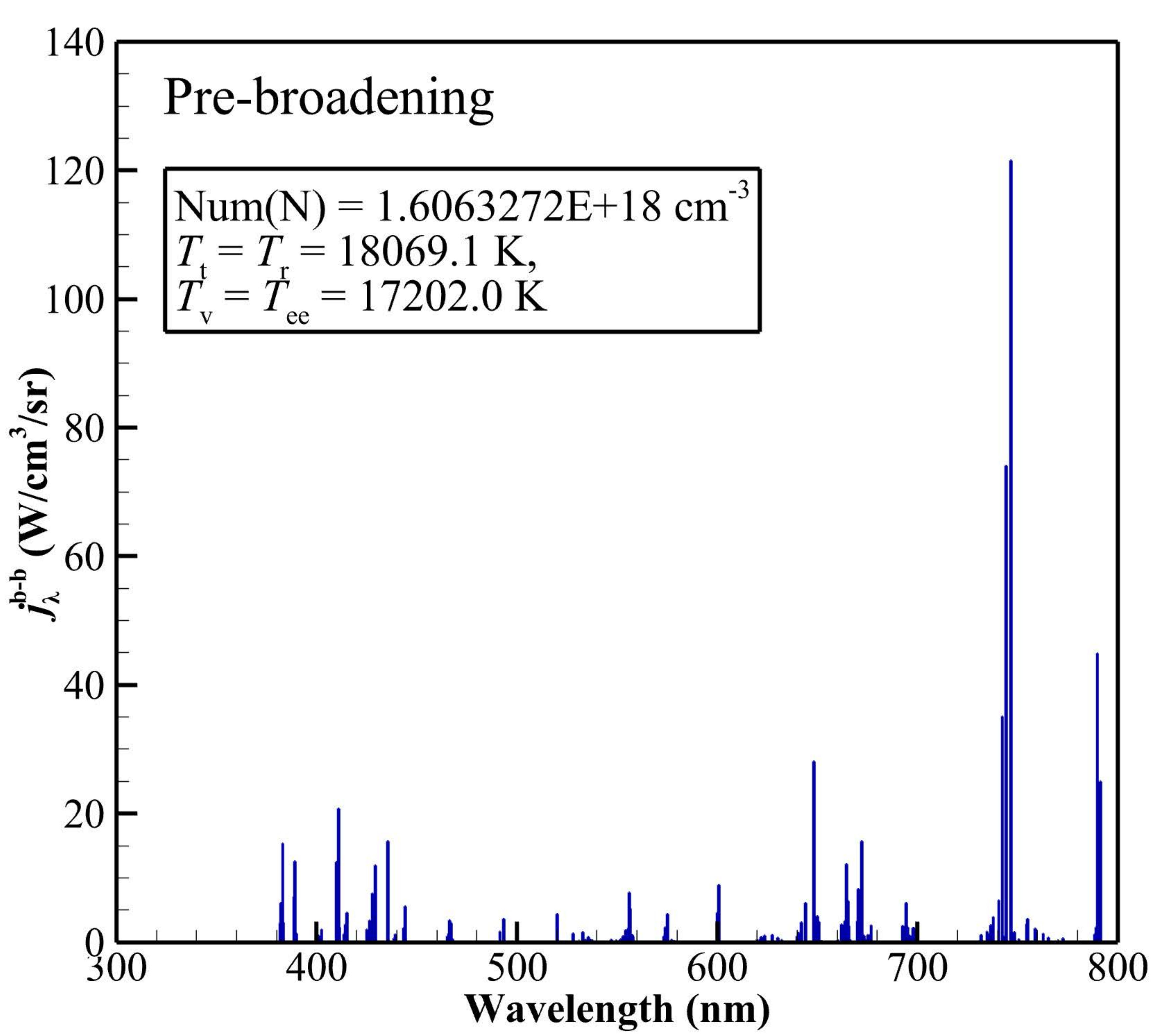}
        \caption{B-B transitions before broadening}
    \end{subfigure}
    \begin{subfigure}{0.48\textwidth}
        \includegraphics[width=\linewidth]{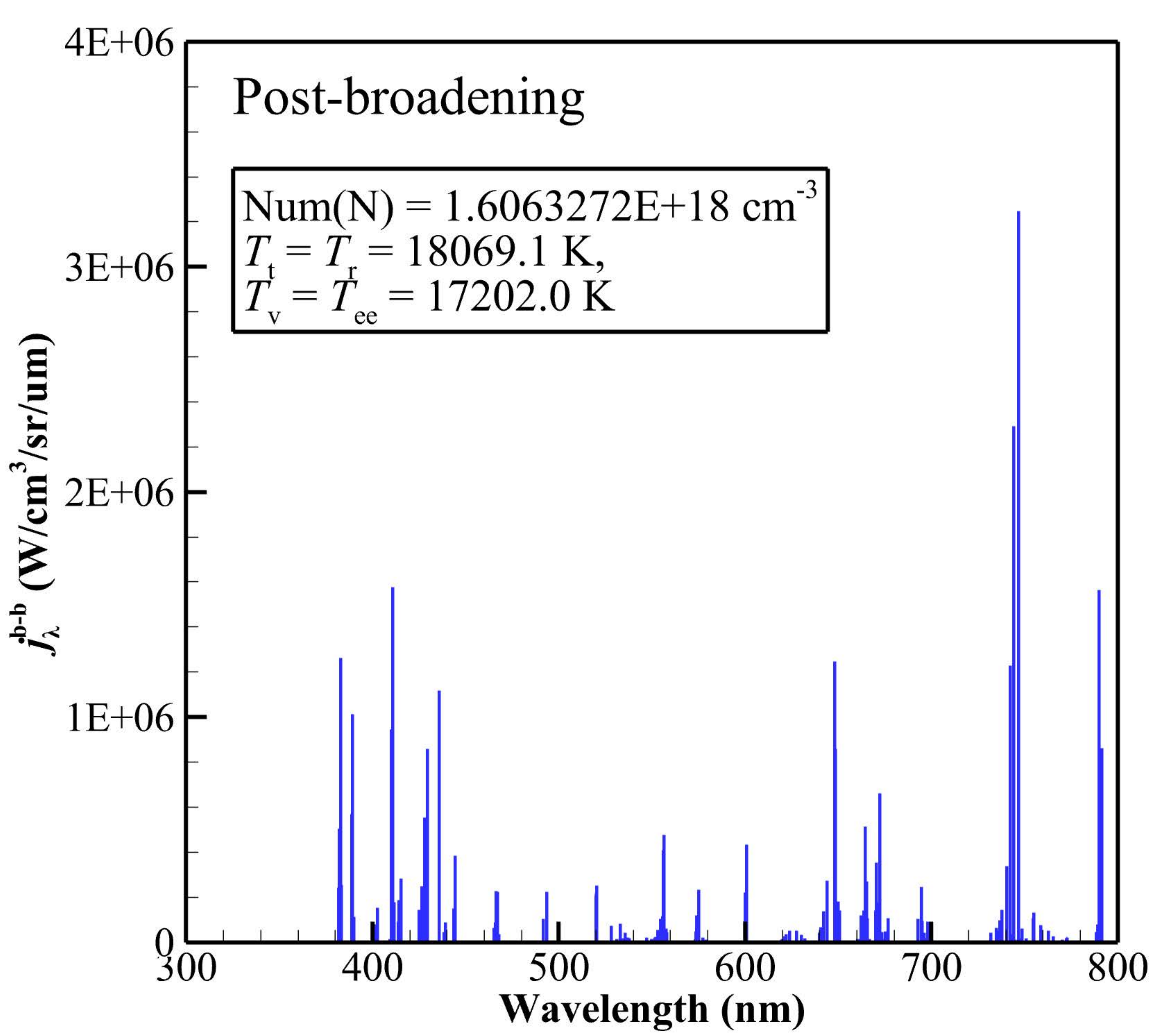}
        \caption{B-B transitions after broadening}
    \end{subfigure}
    
    \begin{subfigure}{0.48\textwidth}
        \includegraphics[width=\linewidth]{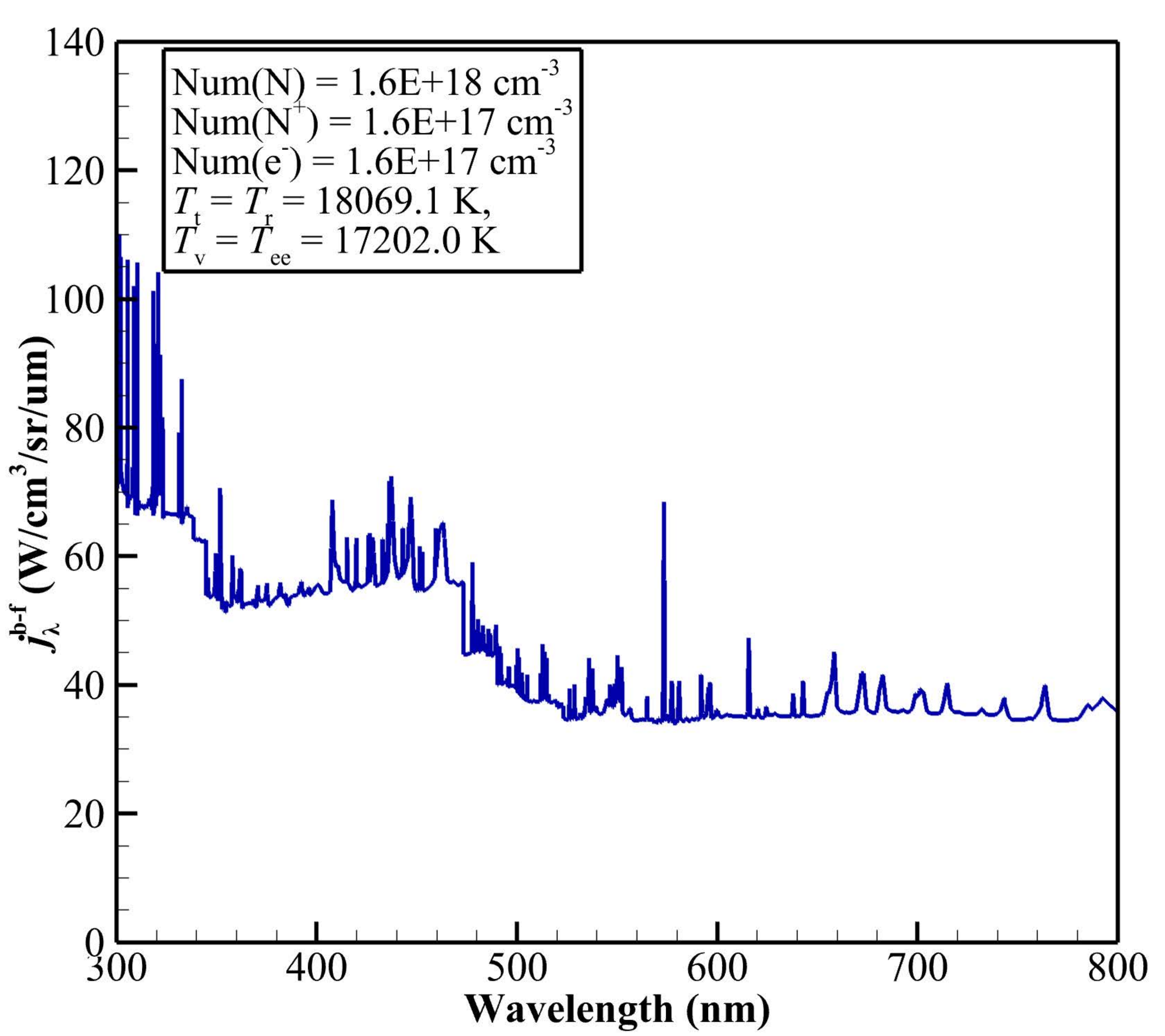}
        \caption{B-F transitions}
    \end{subfigure}
    \begin{subfigure}{0.48\textwidth}
        \includegraphics[width=\linewidth]{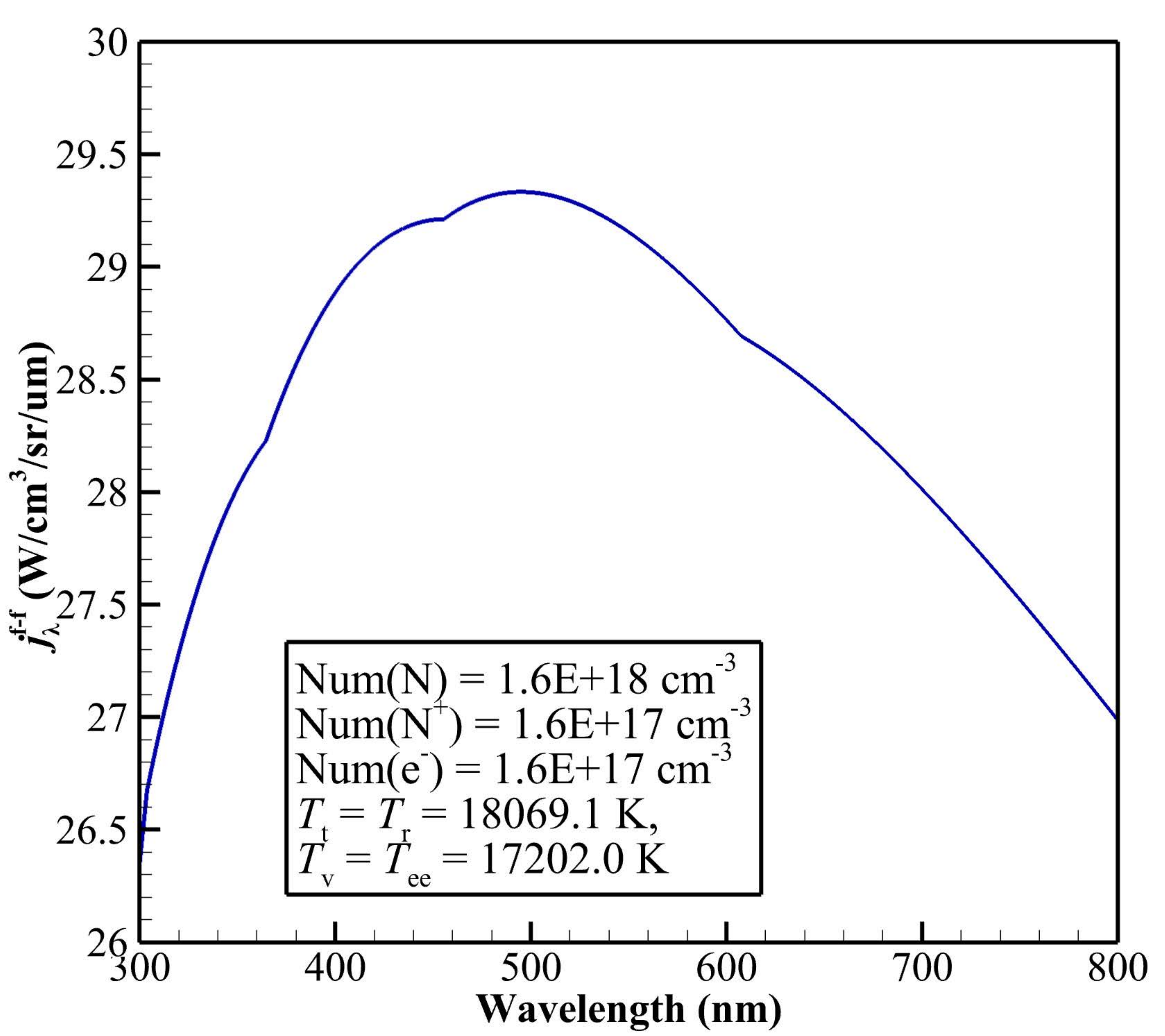}
        \caption{F-F transitions}
    \end{subfigure}
    \caption{Emission coefficients of atomic N from different transition types under the illustrated conditions.}
    \label{fig:coeEmiss_N_BB-BF-FF}
\end{figure}

Figure~\ref{fig:coeEmiss_N_BB-BF-FF} shows the emission coefficients of atomic nitrogen under the illustrated nonequilibrium conditions for the three radiative mechanisms. Overall, the hierarchy is: B-B transitions (after line broadening) $\gg$ B-F transitions $>$ F-F transitions, which corresponds to typical atomic radiation for reentry speeds below 10 km/s. As the reentry speed and ionization level increase, the contributions from B-F and F-F transitions become more significant and may become comparable to, or even exceed, those of B-B transitions in the total atomic radiative emission.

\subsubsection{Molecules}
Molecular radiation involves multiple types of internal energy transitions, including vibrational, rotational, electronic, and combined transitions. Transitions emit photons at specific wavelengths, producing discrete spectral lines on the spectral grid. 
The central wavelength of each line is determined by the energy difference between the corresponding upper and lower states. Collectively, these lines form the overall molecular emission band.
Molecular radiation calculations are generally performed using the following approaches. The first is the LBL method, in which each individual transition is explicitly computed and its contribution to the overall band is evaluated. This approach offers high accuracy but comes with significant computational cost. The second approach involves band-model methods, where transitions are grouped into clusters that collectively contribute to the spectral band, rather than being treated individually. This reduces computational cost and can be further categorized into narrow-band and wide-band models.
In this work, the LBL method is adopted to achieve higher accuracy.

The spontaneous emission coefficient for molecular B-B transitions is expressed analogously to atomic transitions:
\begin{equation}
    j^{b-b}_{\lambda} = {\frac{hc}{4\pi}}{\frac{N_{u}A_{ul}^{M}}{\lambda\times10^{-7}}}{\phi_{\lambda}},
\end{equation}
where $N_u$ is the number density (cm$^{-3}$) of molecules in the upper state, uniquely determined by the combination of electronic, vibrational, and rotational quantum numbers. The lineshape function $\phi_\lambda$ is defined and computed in the same manner as for atomic B-B transitions. $A_{ul}^M$ is the molecular Einstein coefficient (s$^{-1}$) \cite{whiting1996neqair96}, which can be expressed as:
\begin{equation}
    A_{ul}^{M} = \frac{64{\pi}^4\times10^{-7}}{3h{\left({\lambda}\times10^{-7}\right)}^3}\left({e{a_0}Re}\right)^{2}{F_{V_uV_l}}\frac{S_{J_uJ_l}}{2J_u + 1},
\end{equation}
where $a_0 = 5.291772083\times10^{-9}$ cm is the Bohr radius, $Re$ is the electronic transition moment (dimensionless), $F_{V_uV_l}$ is the Franck-Condon factor (dimensionless) representing vibrational coupling between upper ($V_u$) and lower ($V_l$) states, and $S_{J_uJ_l}$ is the H\"onl-London factor describing the rotational contribution for upper ($J_u$) and lower ($J_l$) states.
For diatomic molecules and ions in air, $Re$ and $F_{V_uV_l}$ can be obtained from the literature \cite{gilmore1992franck,liang2021radiative}. Since these two factors depend only on the electronic and vibrational states and are independent of rotation, they can be combined into a single term $Re'= {\left({Re^2}F_{V_uV_l}\right)}^{0.5}$, giving a simplified expression for the molecular Einstein coefficient $A_{ul}^{M}$:
\begin{equation}
    \begin{split}
        & A_{ul}^{M} = \frac{64{\pi}^4\times10^{-7}}{3h{\left({\lambda}\times10^{-7}\right)}^3}\left({e{a_0}Re'}\right)^{2}\frac{S_{J_uJ_l}}{2J_u + 1}.
    \end{split}
\end{equation}
This formulation allows the rotational contribution to be treated separately while retaining the accurate vibronic structure in the emission spectrum.

The central wavelength of each molecular spectral line is determined by the energy difference between the corresponding upper and lower states. The energy of a rovibronic state can be expressed as the sum of electronic, vibrational, and rotational contributions:
\begin{equation}
    \begin{split}
        & E_{i,V,J} = {E_i} + E_{V} + {E_J}, \\
        & E_V = {\omega_{e}}{\left(V + 0.5\right)} - {\omega_{e}x_e}{\left(V + 0.5\right)}^{2} + {\omega_{e}y_e}{\left(V + 0.5\right)}^{3},
    \end{split}
\end{equation}
where $E_V$ is the vibrational energy, with ${\omega_e}$, ${\omega_e x_e}$, and ${\omega_e y_e}$ representing the vibrational constants, and $E_J$ is the rotational energy that strongly depends on the Hund's coupling case of the molecule \cite{herzberg2013molecular}.
\begin{figure}[htb]
    \centering
    \includegraphics[width=0.6\linewidth]{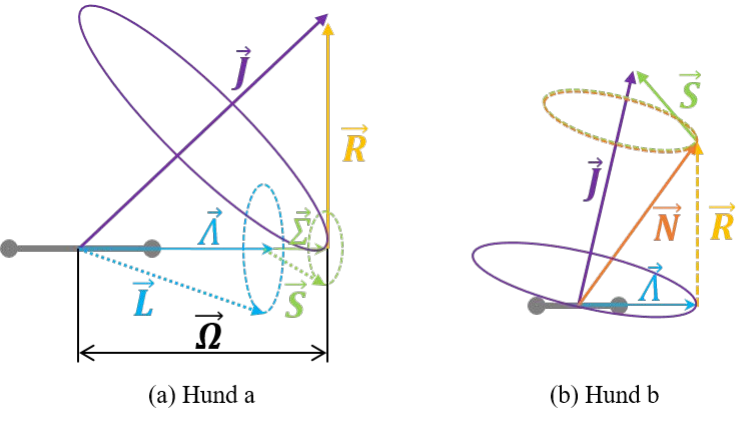}
    \caption{\enspace Vector diagrams for Hund's cases (a) and (b) \cite{herzberg2013molecular}. Solid ellipses represent nutation of the graphical axis, whiel dashed ellipses indicate precession. In case (a), precession is much faster than nutation, whereas in case (b), nutation dominates over precession.}
    \label{fig:sketchHund_A_B}
\end{figure}

Hund's coupling cases describe the way in which electronic spin angular momentum, electronic orbital angular momentum, and nuclear rotational angular momentum interact, reflecting the coupling between molecular rotation and electronic motion. Based on the relative strength of these interactions, Hund classified them into five distinct coupling types (a-e).
In Hund's case (a), the coupling between nuclear rotation and electronic motion (including spin and orbital contributions) is assumed to be weak, whereas the electronic motion itself is strongly coupled to the internuclear axis. In this case, the electronic angular momentum $\vec{\Omega}$ and the nuclear rotational angular momentum $\vec{R}$ combine to form the total angular momentum $\vec{J}$ (see Figure~\ref{fig:sketchHund_A_B}a). The vector $\vec{J}$ remains constant in both magnitude and direction, while $\vec{\Omega}$ and $\vec{R}$ precess around it. Simultaneously, the orbital angular momentum $\vec{L}$ and spin angular momentum $\vec{S}$ precess around the internuclear axis. Hund's case (a) corresponds to the situation where this precession is much faster than the overall rotational motion (nutation). The corresponding rotational energy can be expressed as:
\begin{equation}
    \begin{split}
        & E_J^{(a)} = B_v{J\left(J+1\right)} - D_v{J^2{\left(J+1\right)}^2}, \\
        & B_v = B_e - \alpha_e\left(V + \frac{1}{2}\right), \\
        & D_v = D_e + \beta_e\left(V + \frac{1}{2}\right),
    \end{split}
\end{equation}
where the constants $B_e$, $\alpha_e$, $D_e$, and $\beta_e$ represent the rotational constants. In this work, all these constants are adopted from the NIST database \cite{NIST_MCD}.
In Hund's case (b), the coupling between the electronic motion (particularly the spin vector) and the internuclear axis is assumed to be weak or negligible. In this case, the component of the electronic orbital angular momentum along the internuclear axis, $\vec{\Lambda}$ (when nonzero), combines with the rotational angular momentum $\vec{R}$ to form a resultant vector denoted by $\vec{N}$ (see Figure~\ref{fig:sketchHund_A_B}b). This vector characterizes the total angular momentum excluding spin. The total angular momentum including spin is then given by $\vec{J} = \vec{N} + \vec{S}$. The rotation of the molecule generates a small magnetic moment along $\vec{N}$, which in turn induces a weak magnetic coupling between $\vec{N}$ and $\vec{S}$. This interaction, along with other minor effects, leads to small splittings among levels with the same $\vec{N}$ but different $\vec{J}$, with the magnitude of the splitting increasing with $\vec{N}$. The rotational energy in Hund's case (b) (doublets: $S = -1/2, +1/2$, triplets: $S = -1, 0, +1$, etc.) can thus be expressed as:
\begin{equation}
    \begin{split}
        & E_J^{(b)} = B_v{N\left(N+1\right)} - D_v{N^2{\left(N+1\right)}^2} + S{\gamma}\left(N+\frac{1}{2}-S\right), \\
        & N = J - S,
    \end{split}
\end{equation}
where $\gamma$ denotes the spin-rotation coupling constant.
In Hund's case (c), the interaction between the orbital and spin angular momenta is assumed to be stronger than their respective couplings with the internuclear axis.
In Hund's case (d), the orbital angular momentum is weakly coupled to the internuclear axis but more strongly coupled to the rotational axis.
In Hund's case (e), the orbital and spin angular momenta are strongly coupled, whereas other coupling conditions are similar to those in case (d).
Compared with the coupling types in Hund's cases (a) and (b), instances corresponding to cases (c)-(e) are rarely observed, particularly for the air species considered in the present study. Therefore, these cases are not considered further in this work.

\begin{table}[htb]
    \caption{\enspace H\"onl-London factors for Hund's case (a).}
    \footnotesize
    \setlength{\tabcolsep}{4pt}
    \renewcommand{\arraystretch}{1.5}
    \centering
    \begin{tabular}{lccc}
        \hline
        Branch & $\Lambda_u = \Lambda_l = 0$ & $\Delta \Lambda = 0$ & $\Delta \Lambda = \pm 1$\\
        \hline
        P ($\Delta J = -1$) & $J_u + 1$ & $\frac{\left(J_u+1+\Lambda_u\right)\left(J_u+1-\Lambda_u\right)}{J_u + 1}$ & $\frac{\left(J_u+1\mp\Lambda_u\right)\left(J_u+2\mp\Lambda_u\right)}{2\left(J_u + 1\right)}$\\
        Q ($\Delta J = 0$) & $0$ & $\frac{\left(2J_u +1\right){\Lambda_u}^2}{J_u\left(J_u+1\right)}$ & $\frac{\left(J_u\pm\Lambda_u\right)\left(J_u+1\mp\Lambda_u\right)\left(2J_u +1\right)}{2J_u\left(J_u + 1\right)}$\\
        R ($\Delta J = +1$) & $J_u$ & $\frac{\left(J_u+\Lambda_u\right)\left(J_u-\Lambda_u\right)}{J_u}$ & $\frac{\left(J_u\pm\Lambda_u\right)\left(J_u-1\pm\Lambda_u\right)}{2J_u}$\\
        \hline
    \end{tabular}
    \label{tab:HLfactor_HundA}
\end{table}

\begin{table}[htb]
    \caption{\enspace H\"onl-London factors for Hund's case (b).}
    \footnotesize
    \setlength{\tabcolsep}{4pt}
    \renewcommand{\arraystretch}{1.5}
    \centering
    \begin{tabular}{lcc}
        \hline
        Branch & $S_{J_uJ_l}$ \\
        \hline
        P ($\Delta J = -1$) & $\frac{{J_l}^2 - 1/4}{J_l}$ \\
        Q ($\Delta J = 0$) & $\frac{2{J_l} + 1}{4J_l\left(J_l + 1\right)}$ \\
        R ($\Delta J = +1$) & $\frac{{\left(J_l+1\right)}^2 - 1/4}{J_l + 1}$ \\
        \hline
    \end{tabular}
    \label{tab:HLfactor_HundB}
\end{table}

The five coupling types defined by Hund represent several idealized limiting cases, in which certain interactions of relatively small magnitude are neglected: for example, the interaction between nuclear rotation and orbital angular momentum is ignored in cases (a) and (b). However, as molecular rotation increases, transitions from one coupling scheme to another may occur. For instance, in the case of slow rotation, the angular momentum vectors coupled to the internuclear axis may decouple from the axis as the rotation accelerates. 
These decoupling phenomena can be broadly categorized into three types: spin decoupling (transition from Hund's case (a) to (b)), $\Lambda$-type doublet splitting (initiation of the transition from Hund's case (a) or (b) to (d)), and $L$-uncoupling (complete transition from Hund's case (b) to (d)). The present work primarily focuses on the first type. Considering Hund's case (a) (assuming slow nuclear rotation and weak interaction with electronic motion), as the rotational speed increases, the molecular rotation gradually approaches and eventually surpasses the precession rate of the electronic spin angular momentum around the internuclear axis. At this point, the spin vector can be considered effectively decoupled from the internuclear axis. For doublet states ($S = -1/2, +1/2$) corresponding to this scenario, the rotational energy can be calculated as:
\begin{equation}
    \begin{split}
        & E_J^{(a\rightarrow b),doublet} = B_v\left\{{\left(J+\frac{1}{2}\right)}^2 - {\Lambda^2} - S{\left[4{\left(J+\frac{1}{2}\right)}^2+Y\left(Y-4\right){\Lambda^2}\right]}^{0.5}\right\} - D_v{\left(J+\frac{1}{2}-S\right)}^4, \\
        & Y = A/B_v,
    \end{split}
\end{equation}
where $A$ denotes the spin-orbit coupling constant, reflecting the strength of the interaction between the spin and orbital angular momenta.
For triplet states ($S = -1, 0, +1$) corresponding to the $(a\rightarrow b)$ transition, the rotational energy is expressed as:
\begin{equation}
    \begin{split}
        & E_J^{(a\rightarrow b),triplet} = B_v\left[J\left(J+1\right) - S(Z_1)^{0.5} + Z_2\left(4-6|S|\right)\right] - D_v{\left(J+\frac{1}{2}-S\right)}^4, \\
        & Z_1 = {\Lambda^2}Y\left(Y-4\right) + \frac{4}{3} + 4J\left(J+1\right), \\
        & Z_2 = \frac{1}{3Z_1}\left[{\Lambda^2}Y\left(Y-1\right) - \frac{4}{9} - 2J\left(J+1\right)\right].
    \end{split}
\end{equation}

The values of the H\"onl-London factors are closely related to the type of transition. In Hund's case (a), the factors are given in Table~\ref{tab:HLfactor_HundA} \cite{arnold1969line}, where $\Delta J = J_u - J_l$ and $\Delta \Lambda = \Lambda_u - \Lambda_l$.
In Hund's case (b), the corresponding values are listed in Table~\ref{tab:HLfactor_HundB} \cite{arnold1969line}.
For doublet states in the $(a\rightarrow b)$ transition (in this work, only ${^2 \Sigma} \leftrightarrow {^2 \Pi}$ transitions are considered), the H\"onl-London factors are provided in Table~\ref{tab:HLfactor_HundA-B} \cite{arnold1969line}, where $Y_u$ and $U_u$ are defined as following:
\begin{equation}
    \begin{split}
        & Y_u = A_u/B_{v,u}, \\
        & U_u = {\left[Y_u^2 -4Y_u + {\left(2J_u + 1\right)}^2\right]}^{-0.5}.
    \end{split}
\end{equation}
The value of $\epsilon$ and the sign variable $\eta$ are determined by the branch of the spectral line, as listed in Table~\ref{tab:coesHL_HundA-B}.

\begin{table}[htb]
    \caption{\enspace H\"onl-London factors for doublet states in the Hund's case $(a\rightarrow b)$ transition.}
    \footnotesize
    \setlength{\tabcolsep}{4pt}
    \renewcommand{\arraystretch}{1.5}
    \centering
    \begin{tabular}{lcc}
        \hline
        $\epsilon$ & $S_{J_uJ_l}$ \\
        \hline
        0 & $\frac{{\left(2J_u+1\right)}^2 + \eta{\left(2J_u+1\right)}{U_u}{\left(4J_u^2+4J_u+1-2Y_u\right)}}{16{\left(J_u+1\right)}}$ \\
        1 & $\frac{{\left(2J_u+1\right)}^2 + \eta{\left(2J_u+1\right)}{U_u}{\left(4J_u^2+4J_u-7+2Y_u\right)}}{16{\left(J_u+1\right)}}$ \\
        2 & $\frac{{\left(2J_u+1\right)}{\left[\left(4J_u^2+4J_u-1\right) + \eta{U_u}{\left(8J_u^3+12J_u^2-2J_u+1-2Y_u\right)}\right]}}{16J_u{\left(J_u+1\right)}}$ \\
        3 & $\frac{{\left(2J_u+1\right)}{\left[\left(4J_u^2+4J_u-1\right) + \eta{U_u}{\left(8J_u^3+12J_u^2-2J_u-7+2Y_u\right)}\right]}}{16J_u{\left(J_u+1\right)}}$ \\
        4 & $\frac{{\left(2J_u+1\right)}^2 + \eta{\left(2J_u+1\right)}{U_u}{\left(4J_u^2+4J_u-7+2Y_u\right)}}{16{J_u}}$ \\
        5 & $\frac{{\left(2J_u+1\right)}^2 + \eta{\left(2J_u+1\right)}{U_u}{\left(4J_u^2+4J_u+1-2Y_u\right)}}{16{J_u}}$ \\
        \hline
    \end{tabular}
    \label{tab:HLfactor_HundA-B}
\end{table}

\begin{table}[htb]
    \caption{\enspace Intermediate variables for doublet states in the Hund's case $(a\rightarrow b)$ transition.}
    \footnotesize
    \setlength{\tabcolsep}{4pt}
    \renewcommand{\arraystretch}{1.5}
    \centering
    \begin{tabular}{lccccc}
        \hline
        Branch & $\epsilon ({^2 \Pi} \rightarrow {^2 \Sigma})$ & $\eta ({^2 \Pi} \rightarrow {^2 \Sigma})$ & $\epsilon ({^2 \Sigma} \rightarrow {^2 \Pi})$ & $\eta ({^2 \Sigma} \rightarrow {^2 \Pi})$ \\
        \hline
        $P_2 (\Delta J = -1, S_u = -1/2, S_l = -1/2)$ & 0 & +1 & 4 & +1 \\
        $^O{P_{12}} (\Delta J = -1, S_u = +1/2, S_l = -1/2)$ & 0 & -1 & 5 & -1 \\
        $^Q{P_{21}} (\Delta J = -1, S_u = -1/2, S_l = +1/2)$ & 1 & -1 & 4 & -1 \\
        $P_1 (\Delta J = -1, S_u = +1/2, S_l = +1/2)$ & 1 & +1 & 5 & +1 \\
        $Q_2 (\Delta J = 0, S_u = -1/2, S_l = -1/2)$ & 2 & +1 & 2 & +1 \\
        $^P{Q_{12}} (\Delta J = 0, S_u = +1/2, S_l = -1/2)$ & 2 & -1 & 3 & -1 \\
        $^R{Q_{21}} (\Delta J = 0, S_u = -1/2, S_l = +1/2)$ & 3 & -1 & 2 & -1 \\
        $Q_1 (\Delta J = 0, S_u = +1/2, S_l = +1/2)$ & 3 & +1 & 3 & +1 \\
        $R_2 (\Delta J = +1, S_u = -1/2, S_l = -1/2)$ & 4 & +1 & 0 & +1 \\
        $^Q{R_{12}} (\Delta J = +1, S_u = +1/2, S_l = -1/2)$ & 4 & -1 & 1 & -1 \\
        $^S{R_{21}} (\Delta J = +1, S_u = -1/2, S_l = +1/2)$ & 5 & -1 & 0 & -1 \\
        $R_1 (\Delta J = +1, S_u = +1/2, S_l = +1/2)$ & 5 & +1 & 1 & +1 \\
        \hline
    \end{tabular}
    \label{tab:coesHL_HundA-B}
\end{table}
The H\"onl-London factors for triplet states in the $(a\rightarrow b)$ transition are reported in Tables 3.8 and 3.10 of Kov\'acs \cite{kovacs1972rotational} and are not repeated here.

\subsubsection{Radiative transfer}
Based on the recent work of Johnston and Mazaheri \cite{johnston2018impact}, a ray-tracing approach is employed in RAPRAL to integrate the RTE for the calculation of wall radiative heat flux. This approach provides adequate accuracy in both the strongly compressed region behind the bow shock and the expansion-cooled region. 

The core objective is to determine the radiative properties along the lines of sight (LOS) that converge from the entire flowfield to a target point. These properties include the emission coefficient, absorption coefficient, and escape factor. The RTE is then integrated along each LOS to obtain quantities such as the radiative heat flux at the target location. The implementation procedure is summarized as follows:
1. A sufficient number of LOS (denoted as $n_{LOS}$) are generated in three-dimensional (3D) space using the Fibonacci sphere method. These rays are then projected onto the actual flowfield, and the flow variables (including temperature, pressure, and total number density of species) are interpolated onto the nodes of each LOS.
2. A LBL approach is adopted to compute the radiative contributions at each LOS node. Specifically, atomic radiation processes (B-B, B-F, and F-F transitions) and molecular radiation processes (B-B transitions) are evaluated to obtain quantities such as the emission and absorption coefficients.
3. The RTE is integrated along each LOS to obtain the radiative intensity $I(\Omega,\enspace \lambda)$ at the target location as a function of direction $\Omega$ and wavelength $\lambda$.
4. Finally, the radiative intensity is numerically integrated to yield the radiative heat flux over a specified spectral range at the target point.
With the incorporation of efficient grid-search and interpolation algorithms for retrieving flow variables at LOS nodes, the present approach can be readily extended to more general flowfields on unstructured grids.

\begin{figure}[!htb]
    \centering
    \begin{subfigure}{0.48\textwidth}
        \includegraphics[width=\linewidth]{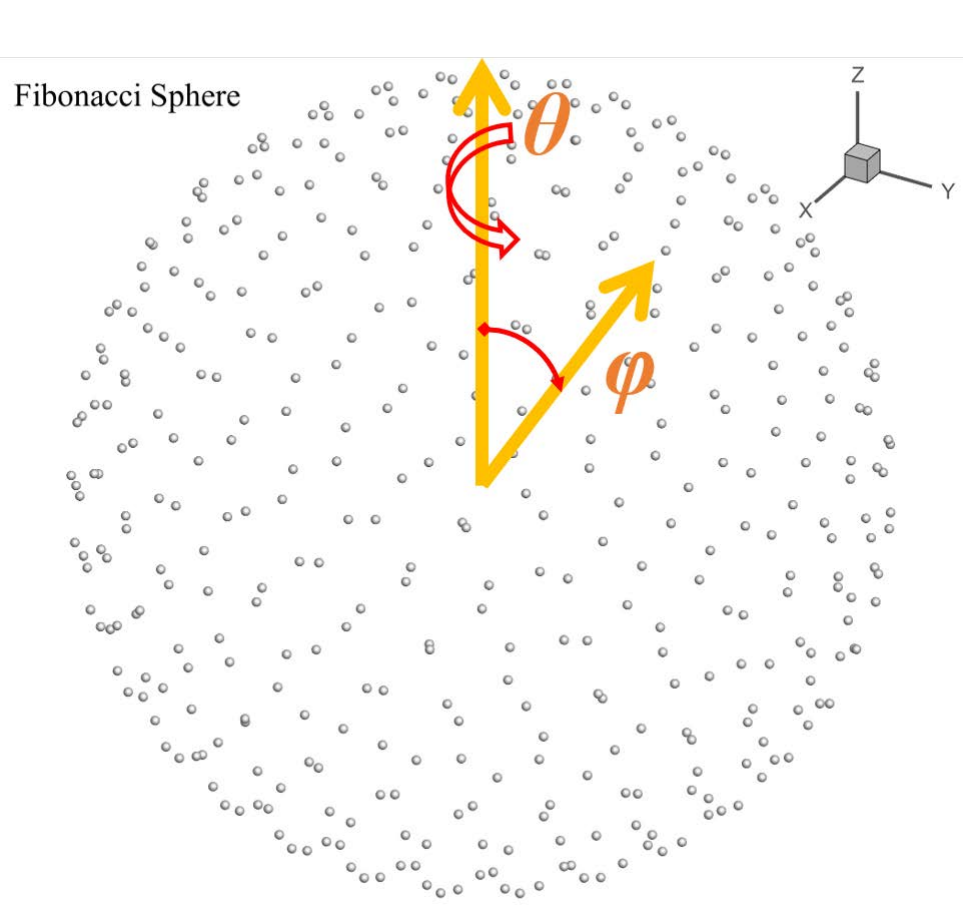}
        \caption{Fibonacci sphere}
    \end{subfigure}
    \begin{subfigure}{0.48\textwidth}
        \includegraphics[width=\linewidth]{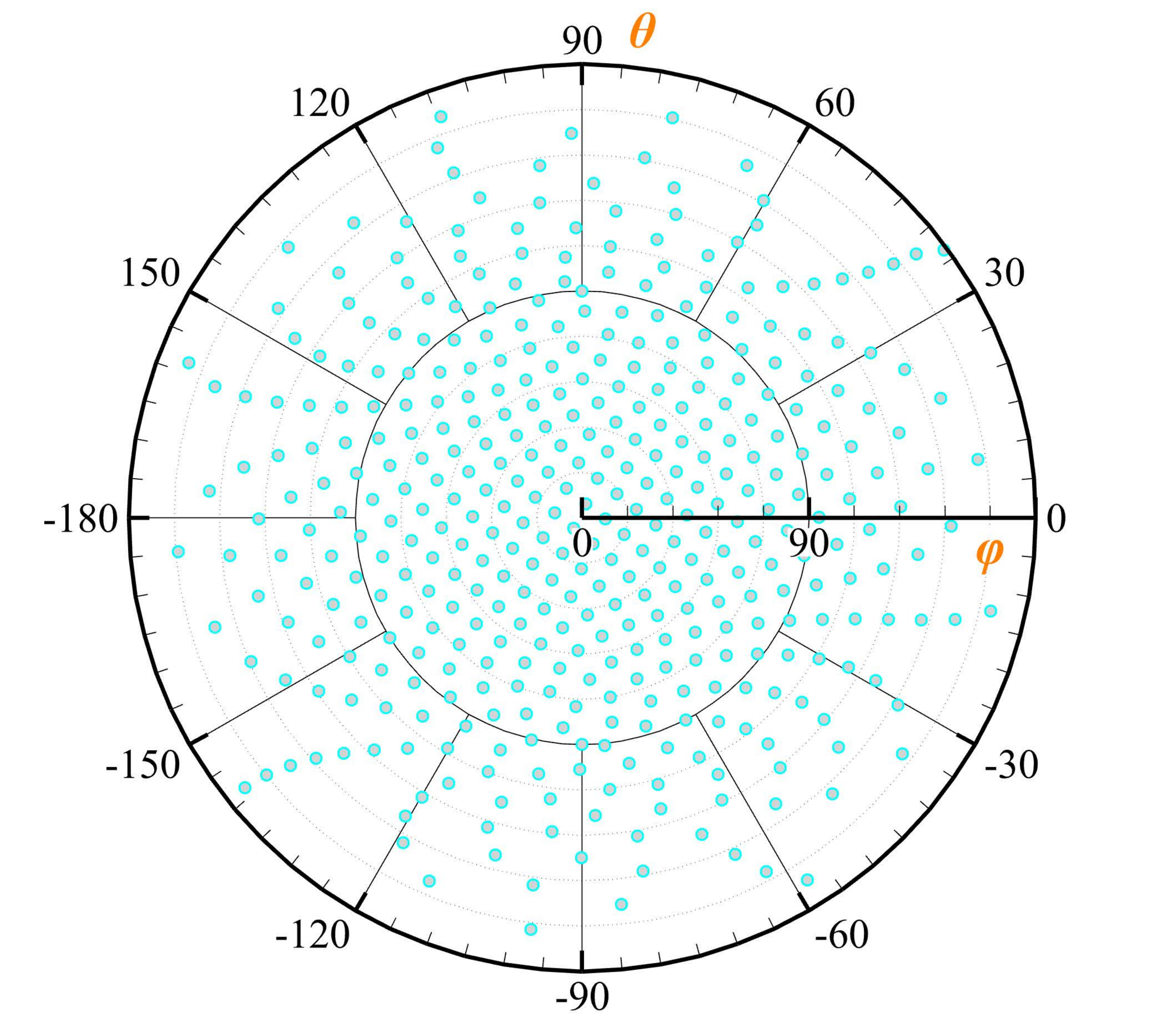}
        \caption{discrete azimuthal angles}
    \end{subfigure}
    \caption{\enspace Schematic of spatial discretization for 3D LOS generation using the Fibonacci sphere method with $n_{LOS} = 400$.}
    \label{fig:DisFibonacci}
\end{figure}

Both two-dimensional (2D) planar flows and axisymmetric flows are, in essence, rational simplifications of the 3D flows, provided that the fundamental physical laws are preserved. Accordingly, in the present work, the LOS are uniformly generated in 3D space and subsequently projected onto the actual flow domain as required.
To avoid the non-uniformity associated with discretizing the azimuthal angle $\theta$ and polar angle $\varphi$ using equal intervals--where excessive clustering occurs near $\varphi=0^{\circ}$ and $180^{\circ}$, while insufficient resolution appears near $\varphi=90^{\circ}$--as well as the high computational cost arising from the need for smaller discretization steps $\Delta \theta$ and $\Delta \varphi$ to ensure convergence of the radiative integration \cite{johnston2018impact}, the Fibonacci sphere method \cite{gonzalez2010measurement} is adopted to discretize the 3D space and determine the LOS directions. This approach ensures a nearly uniform distribution for arbitrary values of $n_{LOS}$. Figure~\ref{fig:DisFibonacci} illustrates the Fibonacci sphere with unit radius and the corresponding angular distribution of the LOS directions.
Taking the target point $O\left(x_0, y_0, z_0\right)$ as an example, the Cartesian coordinates of node $j$ along the $i$-th LOS ($i_{LOS}=1,2,...,n_{LOS}$) are given by:
\begin{equation}
    \begin{split}
        & x_{i, j} = x_0 + r{cos\left(a\right)}{L_j}, \\
        & y_{i, j} = y_0 + w{L_j}, \\
        & z_{i, j} = z_0 + r{sin\left(a\right)}{L_j},
    \end{split}
\end{equation}
where $L_j$ denotes the distance between node $j$ and the target point $O$, and
\begin{equation}
    \begin{split}
        & w = 1 - 2\frac{i_{LOS} - 1}{n_{LOS} - 1}, \\
        & r = {\left(1 - w^2\right)}^{0.5}, \\
        & a = {\left(i_{LOS} - 1\right)}{\left(\sqrt{5} - 1\right)}\pi.
    \end{split}
\end{equation}

\begin{figure}
    \centering
    \begin{subfigure}{0.48\textwidth}
        \includegraphics[width=\linewidth]{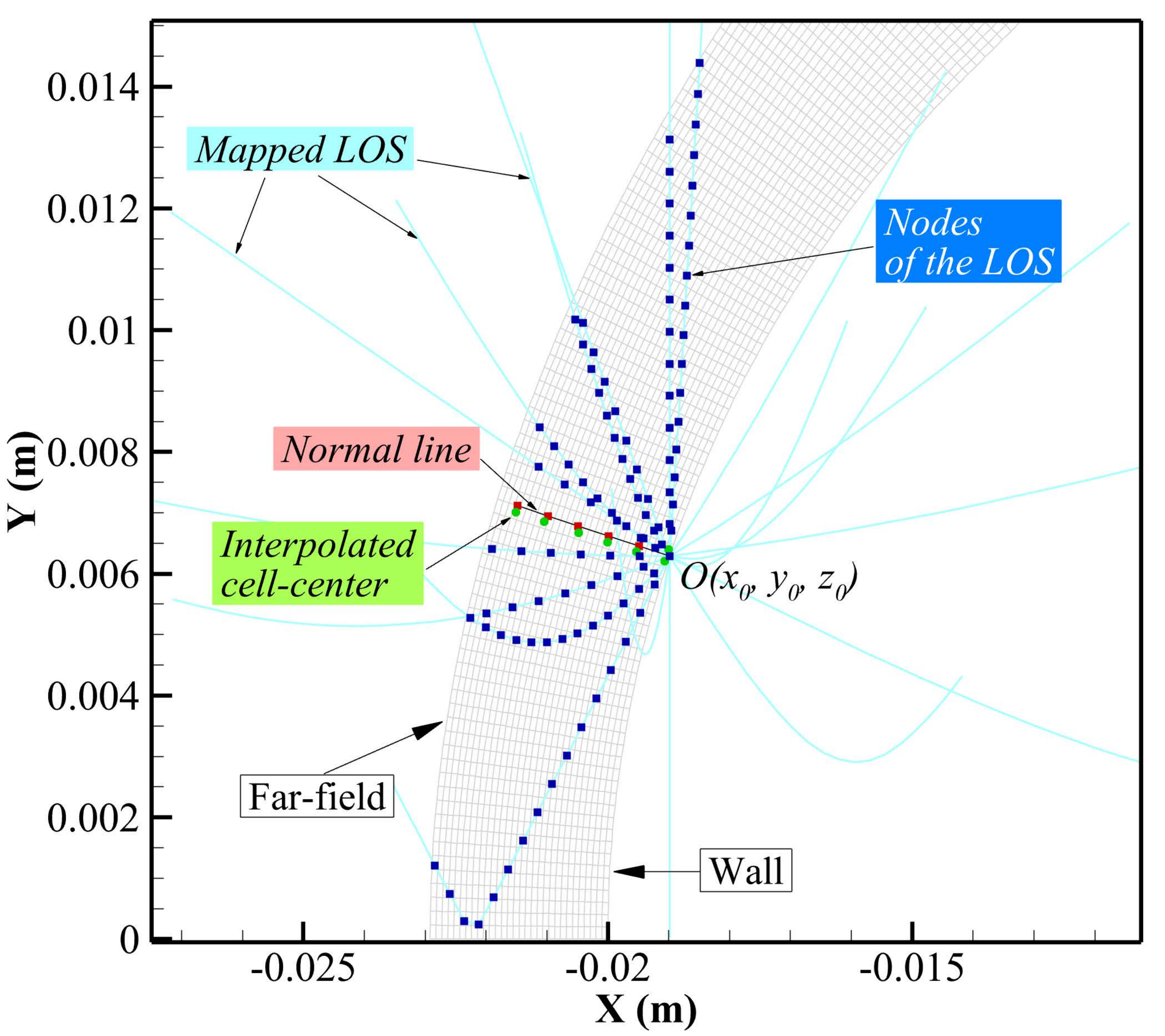}
        \caption{}
    \end{subfigure}
    \begin{subfigure}{0.48\textwidth}
        \includegraphics[width=\linewidth]{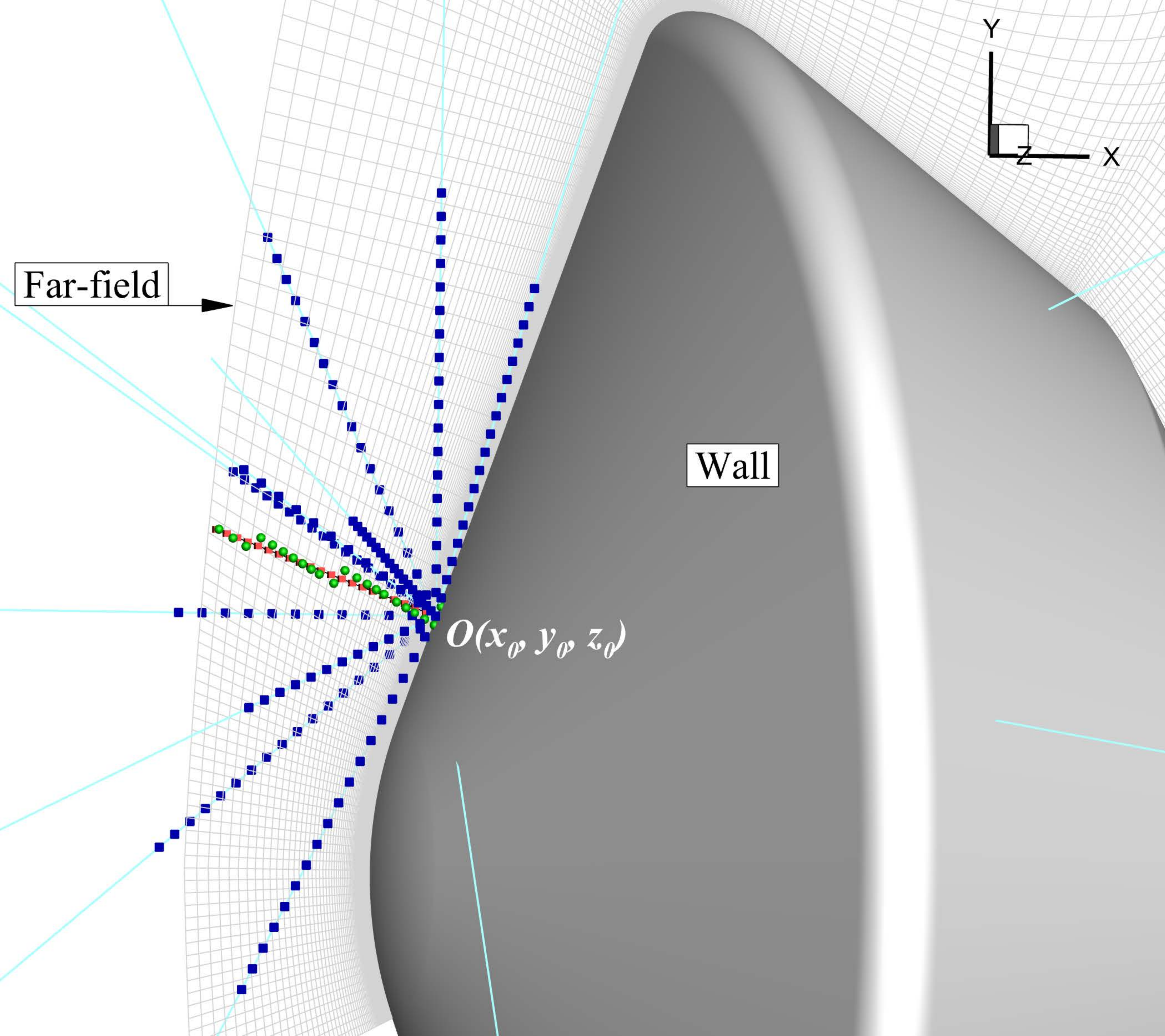}
        \caption{}
    \end{subfigure}
    \caption{\enspace LOS projections (a) onto a 2D axisymmetric flowfield and (b) in a 3D flowfield.}
    \label{fig:LOSinterpolate}
\end{figure}

\begin{figure}
    \centering
    \begin{subfigure}{0.9\textwidth}
        \includegraphics[width=\linewidth]{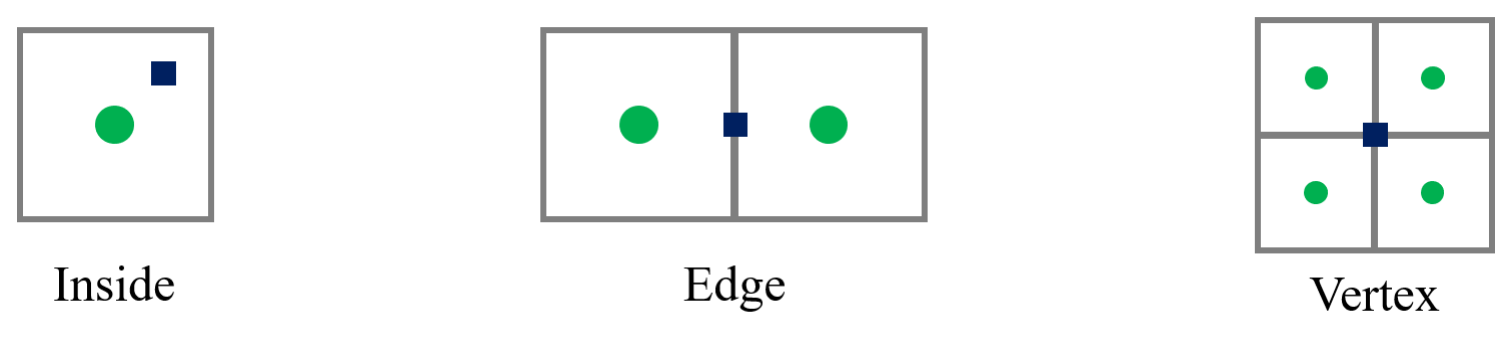}
        \caption{}
    \end{subfigure}
    
    \begin{subfigure}{0.9\textwidth}
        \includegraphics[width=\linewidth]{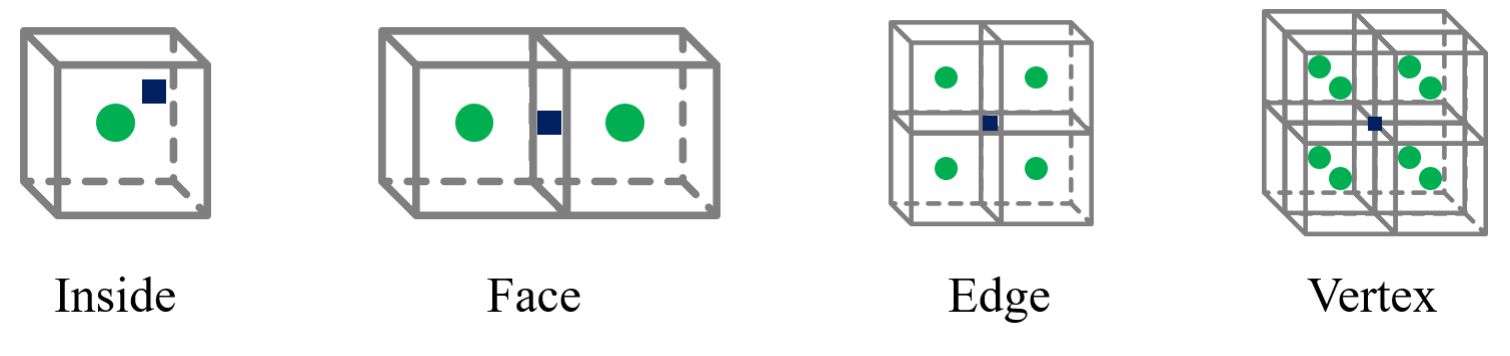}
        \caption{}
    \end{subfigure}
    \caption{\enspace Interpolation of flow variables at LOS nodes (blue squares) based on cell-centered variables (green circles) in (a) 2D and (b) 3D structured grids.}
    \label{fig:AlgorithmInterpolate}
\end{figure}

The interpolation procedure for obtaining flow properties at LOS nodes is illustrated using axisymmetric and 3D flowfields as examples. As shown in Figure~\ref{fig:LOSinterpolate}a, the LOS are first projected onto the axisymmetric plane (light blue lines). Based on the relative positions between the LOS and the flow boundaries (e.g., far-field boundary, wall, and symmetry axis), the valid LOS nodes used for subsequent radiative calculations are identified and filtered (dark blue/red squares).
Subsequently, a grid-search algorithm is employed to determine the surrounding computational cells for each LOS node, where the green dots represent the cell centers used for interpolation onto the LOS nodes. For 3D flowfields, no projection is required, and the corresponding computational cells are directly identified through the grid-search procedure (Figure~\ref{fig:LOSinterpolate}b).
Finally, appropriate interpolation schemes (e.g., algebraic averaging or inverse distance weighting) are applied to reconstruct the flow properties at each LOS node from the values stored in the computational grid (Figure~\ref{fig:AlgorithmInterpolate}).

\begin{figure}
    \centering
    \includegraphics[width=0.7\linewidth]{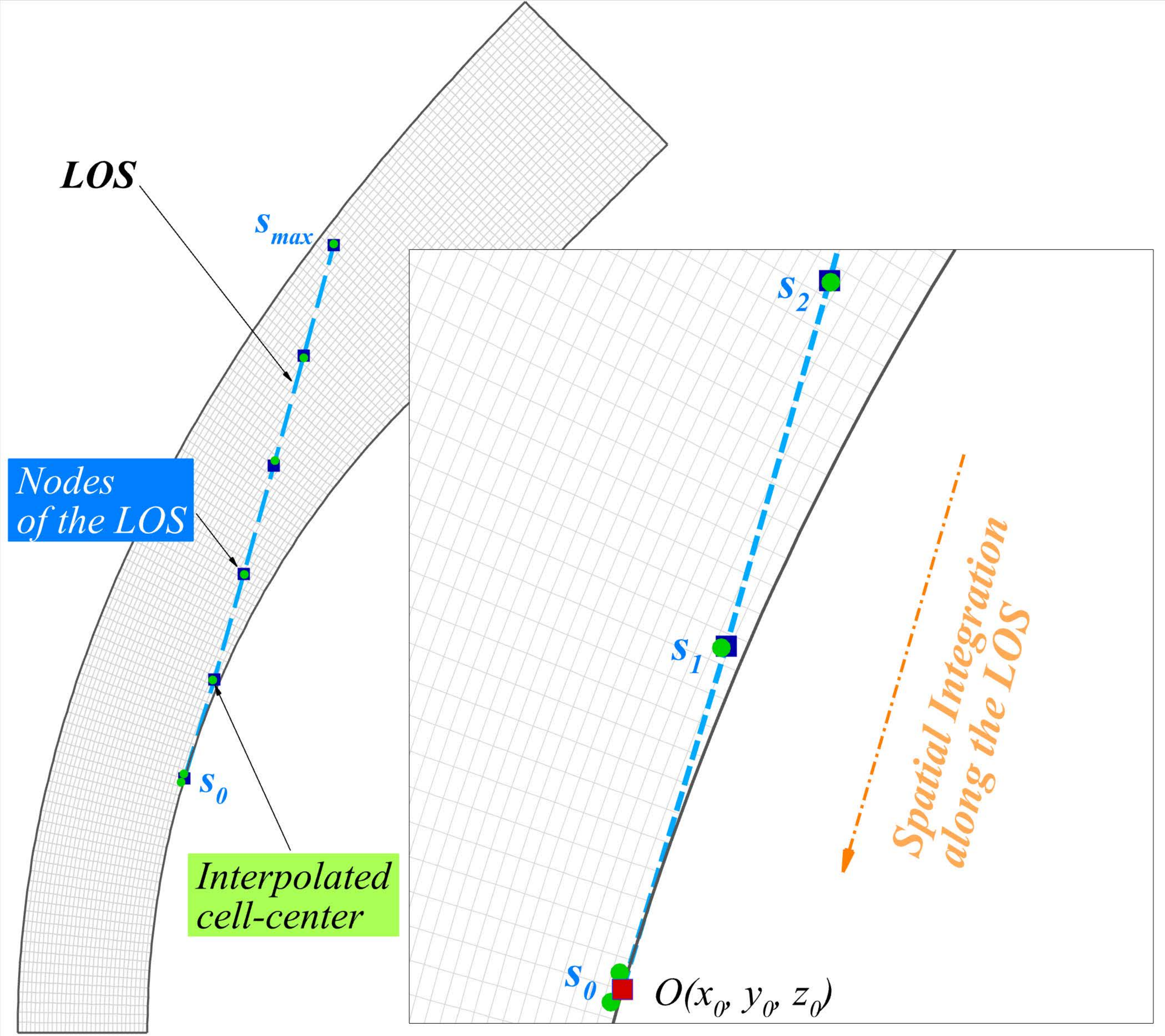}
    \caption{\enspace Schematic of integrating the RTE along LOS in the ray-tracing method, with the spacing between LOS nodes intentionally exaggerated for clarity.}
    \label{fig:LOSintegral}
\end{figure}

After obtaining the flow properties at each LOS node, the spectral radiative quantities of atomic and molecular sources are evaluated using the methods described previously. These quantities are then substituted into the RTE and integrated to obtain the radiative intensity at the target point:
\begin{equation}
    I_{\lambda, 0} = I_\lambda\left(s_{\max }\right) e^{-\tau_\lambda\left(s_{\max }\right)}+\int_{s^{\prime}=s_{\max }}^{s^{\prime}=s_0} j_\lambda\left(s^{\prime}\right) e^{-\tau_\lambda\left(s^{\prime}\right)} \mathrm{d} s^{\prime},
\end{equation}
where $s_0$ and $s_\text{max}$ denote the target point on the LOS and the intersection with the farthest boundary (e.g., a supersonic inflow boundary), respectively (see Figure~\ref{fig:LOSintegral}). If the farthest boundary corresponds to a solid surface, $I_\lambda\left(s_{\max }\right)$ is given by the blackbody radiation intensity; otherwise, $I_\lambda\left(s_{\max }\right) = 0$. The optical thickness $\tau_\lambda$ represents the integral of the absorption coefficient along the LOS between node $i(s_i)$ and the target point:
\begin{equation}
    \tau_\lambda(s_i) = \int_{s^{\prime}=s_{i}}^{s^{\prime}=s_0} \kappa_\lambda\left(s^{\prime}\right) \mathrm{d} s^{\prime}.
\end{equation}
In the evaluation of the two preceding integral expressions, the approximate method proposed by Johnston and Mazaheri \cite{johnston2018impact} is adopted to balance computational accuracy and cost. Specifically, for two adjacent nodes along the LOS, i.e., $\left[s_i, s_{i+1}\right]$ (for clarity, the indexing here is reversed relative to the previous definition, such that the farthest point from the target is denoted as $s_0$ and the target point corresponds to $s_{max}$), the emission coefficient is assumed to vary linearly, while the absorption coefficient is treated as constant and equal to the value at the node farther from the target point.
\begin{equation}
    \left.\begin{aligned}
        & \kappa_{\lambda}(s) = \kappa_{\lambda, i}, \\
        & j_{\lambda}(s) = j_{\lambda, i} + \frac{j_{\lambda, i+1}-j_{\lambda, i}}{\Delta s}(s-s_i)
    \end{aligned}\right\}, s \in \left[s_i, s_{i+1}\right], {\Delta s} = s_{i+1} - s_i.
\end{equation}
Under these assumptions, integrating the RTE over the interval $\left[s_i, s_{i+1}\right]$ yields
\begin{equation}
    I_{\lambda,i+1} = I_{\lambda,i}exp(-\kappa_{\lambda,i} \Delta s) + exp(-\kappa_{\lambda,i}s_{i+1})\int_{s_i}^{s_{i+1}}{j_{\lambda}(s)exp(\kappa_{\lambda,i} s)}ds.
\end{equation}
Let $x = s - s_i, x \in [0, \Delta s]$, then
\begin{equation}
    \begin{aligned}
        & j_{\lambda}(s) = j_{\lambda, i} + \frac{j_{\lambda, i+1}-j_{\lambda, i}}{\Delta s}x, \\
        & exp(\kappa_{\lambda,i} s) = exp(\kappa_{\lambda,i} s_i)exp(\kappa_{\lambda,i} x).
    \end{aligned}
\end{equation}
Substituting into the above integral expression gives:
\begin{equation}
    \begin{aligned}
        & exp(-\kappa_{\lambda,i}s_{i+1})\int_{s_i}^{s_{i+1}}{j_{\lambda}(s)exp(\kappa_{\lambda,i} s)}ds \\
        & = exp(-\kappa_{\lambda,i}\Delta s)\int_0^{\Delta s}{\left[j_{\lambda,i}+\frac{j_{\lambda, i+1}-j_{\lambda, i}}{\Delta s}x\right]exp(\kappa_{\lambda,i} x)}dx \\
        & = j_{\lambda, i}\frac{1-exp(-\kappa_{\lambda,i}\Delta s)}{\kappa_{\lambda,i}} + (j_{\lambda,i+1}-j_{\lambda,i}){\left(\frac{1}{\kappa_{\lambda,i}}-\frac{1-exp(-\kappa_{\lambda,i}\Delta s)}{\kappa_{\lambda,i}^2 \Delta s}\right)}
    \end{aligned}
\end{equation}
By introducing the local optical thickness $\Delta \tau_{\lambda,i} = \kappa_{\lambda,i}\Delta s$ and substituting into the integral and rearranging, the final expression becomes:
\begin{equation}
    I_{\lambda,i+1} = I_{\lambda,i}exp(-\Delta \tau_{\lambda,i}) + \alpha_{\lambda,i}j_{\lambda,i} + \beta_{\lambda,i}j_{\lambda,i+1},
\end{equation}
where
\begin{equation}
    \begin{aligned}
        & \alpha_{\lambda,i} = \frac{1}{\kappa_{\lambda,i}}\left(1-\frac{1-exp(-\Delta \tau_{\lambda,i})}{\Delta \tau_{\lambda,i}}\right), \\
        & \beta_{\lambda,i} = \frac{1}{\kappa_{\lambda,i}}\left(\frac{1-exp(-\Delta \tau_{\lambda,i})}{\Delta \tau_{\lambda,i}} - exp(-\Delta \tau_{\lambda,i})\right).
    \end{aligned}
\end{equation}
When, at certain wavelengths along the LOS, the optical thin condition is satisfied (i.e., $\Delta \tau_{\lambda,i} \ll 1$), a second-order Taylor expansion is employed to ensure numerical stability, yielding
\begin{equation}
    f(\Delta \tau_{\lambda,i}) = \frac{1-exp(-\Delta \tau_{\lambda,i})}{\Delta \tau_{\lambda,i}} \simeq 1 - \frac{\Delta \tau_{\lambda,i}}{2} + \frac{(\Delta \tau_{\lambda,i})^2}{6}, \enspace \Delta \tau_{\lambda,i} \ll 1.
\end{equation}
In this case
\begin{equation}
    \left.\begin{aligned}
        & \alpha_{\lambda,i} = \frac{1}{\kappa_{\lambda,i}}\left[1-f(\Delta \tau_{\lambda,i})\right], \\
        & \beta_{\lambda,i} = \frac{1}{\kappa_{\lambda,i}}\left[f(\Delta \tau_{\lambda,i}) - exp(-\Delta \tau_{\lambda,i})\right]
    \end{aligned}\right\}, \Delta \tau_{\lambda,i} \ll 1.
\end{equation}
Starting from the farthest point $s_0$, the RTE is integrated sequentially over each segment between adjacent LOS nodes and marched forward to the target point $s_\text{max}$, resulting in $I_{\lambda,0}$.
After obtaining the radiative intensity $I_{\lambda,0}$ at the target point along each LOS, the radiative heat flux at the target location $O$ can be evaluated by integration:
\begin{equation}
    \begin{split}
        & S(O)=\int_0^{2 \pi} \int_0^{\pi/2} S^{\varphi, \theta}(O) \sin \varphi \cos \varphi \mathrm{d} \varphi \mathrm{~d} \theta \\
        & S^{\varphi, \theta}(O) = \int_{\lambda_1}^{\lambda_2}  I_{\lambda, 0}(\varphi, \theta) \mathrm{d} \lambda
    \end{split}
\end{equation}
where $\left[\lambda_1, \lambda_2\right]$ denotes the wavelength range of interest. In the numerical implementation, the LOS-based results obtained from the Fibonacci sphere method are first interpolated onto a discretized angular space defined by uniform polar angle $\Delta \varphi$ and azimuthal angle $\Delta \theta$. A summation is then performed (assuming that $S^{\varphi, \theta}$ is constant in the integration over $\varphi$):
\begin{equation}
    \begin{aligned}
        & S(O) = \sum_{t=1}^{N_{\theta, \max }-1} \sum_{p=1}^{N_{\varphi, \max}-1} \frac{1}{4}\left(S^{\varphi_p, \theta_t}+S^{\varphi_{p+1}, \theta_t}+S^{\varphi_p, \theta_{t+1}}+S^{\varphi_{p+1}, \theta_{t+1}}\right) \\
        & \quad \times\left(\cos \varphi_{p+1}-\cos \varphi_p\right)\left(\theta_{t+1}-\theta_t\right).
    \end{aligned}
\end{equation}

\subsection{RAPRAL framework}
\begin{figure}[hbt!]
    \centering
    \includegraphics[width=1.0\textwidth]{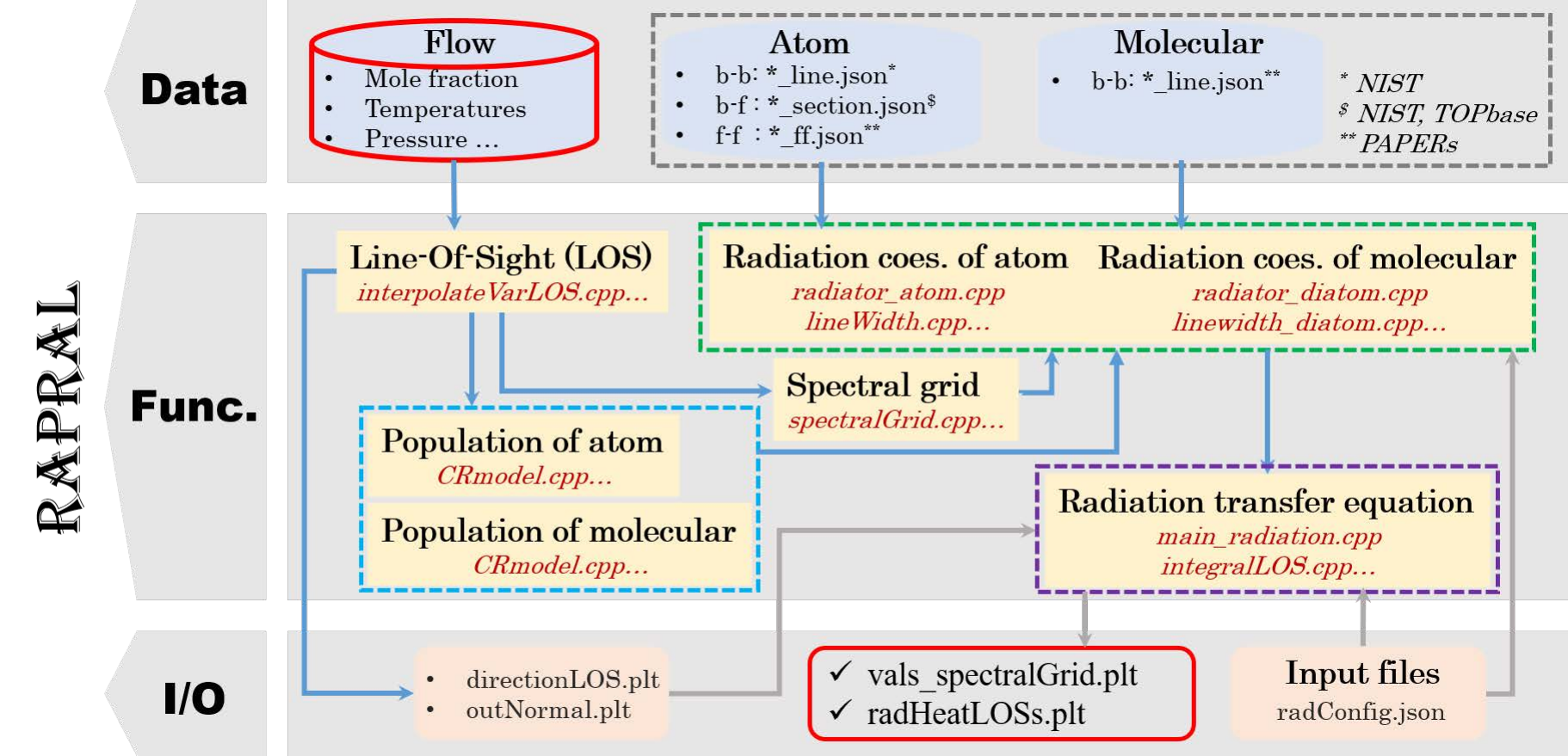}
    \caption{Framework of the RAPRAL solver.}
    \label{fig:sketchRAPRAL}
\end{figure}
The radiative solver RAPRAL adopts a layered architecture, as illustrated in Figure~\ref{fig:sketchRAPRAL}. The first layer consists of fundamental data, including the flowfield obtained from third-party solvers (currently limited to structured grids) and the species database for radiative calculations (constructed using Python), which contains energy level data and transition probabilities.
The second layer comprises the core functionalities. Specifically, LOS are generated for ray-tracing by interpolating flow properties from the flowfield; the emission and absorption coefficients at each LOS node are then evaluated; the RTE is integrated along each LOS to obtain the radiative intensity at points near the wall; and finally, integration over the solid angle yields the radiative heat flux at the wall.
The third layer corresponds to input/output (I/O) operations, including the management of process control and the reading and writing of configuration files.

\section{Results}
\label{sec:others}
In this section, two categories of test cases are considered to assess the reliability of the present solver. The first category focuses on the bulk spectral characteristics of the major atomic and molecular radiators in air, including atomic and molecular B-B transitions. The performance of RAPRAL is evaluated through comparisons with results obtained from the Spark code \cite{lopez2016spark} under identical conditions.
The second category examines the capability of solving the RTE by predicting the wall radiative heat flux for the Fire II flight experiment and comparing the results with measurements.
These test cases collectively address the key aspects of interest in the current uncoupled flow-radiation simulation framework and provide a comprehensive assessment of the solver’s capability.

\subsection{Emission and absorption coefficients}
Existing shock-tube experiments and numerical simulations indicate that the dominant radiative sources during hypersonic reentry exhibit significant dependence on the flight velocity. In the strongly compressed region behind the bow shock, when the reentry velocity is below 7 km/s, the post-shock temperature is relatively low and the dissociation of \ch{N2} and \ch{O2} remains limited. Under such conditions, radiation in the UV/Vis spectral range within the nonequilibrium region is dominated by molecular B-B transitions, primarily from species such as \ch{N2} and \ch{NO} \cite{gimelshein2019validation,tumuklu2021modeling,glenn2024radiation}. 
As the velocity increases, the degree of dissociation and ionization becomes more pronounced. The number density of \ch{N2+} produced through ionization increases and gradually becomes one of the dominant radiators (first negative band system) \cite{cruden2017measurement}. When the velocity exceeds 10 km/s, molecules are nearly fully dissociated, and a large fraction of the resulting high-energy atoms becomes ionized. In this regime, molecular radiation can be largely neglected, and atomic B-B transitions emerge as the dominant radiative source \cite{cruden2019analysis}. Furthermore, the contributions from B-F and F-F transitions, forming the continuum radiation, become increasingly significant as the velocity continues to rise.
In contrast to the strongly compressed region behind the bow shock, the flow that passes around the vehicle shoulder and enters the expansion-cooled region is characterized by a high fraction of energetic atoms and molecules, but with relatively low temperature and high flow velocity. The reduced collision frequency leads to lower rates of de-excitation and recombination, resulting in an extended nonequilibrium region dominated by atomic radiation, with secondary contributions from recombined molecular species (primarily \ch{O2} and \ch{NO}, concentrated in the VUV/UV spectral range) \cite{johnston2015features,johnston2016refinements}.
Table~\ref{tab:diatomicRadTrans} summarizes the major diatomic radiators in air and selected parameters adopted in the present code for computing their spectral characteristics.
\begin{table}[hbt!]
\caption{\label{tab:diatomicRadTrans} Diatomic systems considered in the present work.}
\centering
\begin{tabular}{p{1.5cm}cp{2.2cm}p{2.3cm}p{2.5cm}p{2.0cm}}
\hline
Diatomic species & System name & Transition designation & Included bands (0:$v_{u,max}$; 0:$v_{l,max}$) & Major features or range (nm) & $Re'$ Reference\\\hline
\ch{N2}& First-Positive& $\rm B^3\Pi_g-A^3\Sigma_u^+$& (0:19; 0:17)& 500-750 (678.7)& Liang et al.\cite{liang2021radiative}\\
\ch{N2}& Second-Positive& $\rm C^3\Pi_u-B^3\Pi_g$& (0:5; 0:19)& 282-466.6 (297.7, 315.9, 337.1, 357.7, 380.5, 405.9) & Liang et al.\cite{liang2021radiative}\\
\ch{N2+}& First-Negative& $\rm B^2\Sigma_u^+-X^2\Sigma_g^+$& (0:9; 0:21)& 300-500 (330.8, 358.2, 391.4, 427.8, 470.9)& Liang et al.\cite{liang2021radiative}\\
\ch{NO}& $\rm \gamma$& $\rm A^2\Sigma^+-X^2\Pi_r$& (0:8; 0:23)& 200-300 (237, 247.9, 259.6, 272.2, 286)& Liang et al.\cite{liang2021radiative}\\
\ch{NO}& $\rm \beta$& $\rm B^2\Pi_r-X^2\Pi_r$& (0:22; 0:23)& 200-300& Chauveau et al.\cite{chauveau2002contributions}\\
\ch{NO}& $\rm \delta$& $\rm C^2\Pi_r-X^2\Pi_r$& (0:9; 0:23)& 200-300& Liang et al.\cite{liang2021radiative}\\
\ch{NO}& $\rm \varepsilon$& $\rm D^2\Sigma^+-X^2\Pi_r$& (0:6; 0:23)& 200-300& Liang et al.\cite{liang2021radiative}\\
\ch{O2}& Schumann-Runge& $\rm B^3\Sigma_u^--X^3\Sigma_g^-$& (0:9; 0:22)& 175-455& Liang et al.\cite{liang2021radiative}\\
\hline
\end{tabular}
\end{table}

The population distributions of internal energy states for the radiating species are assumed to satisfy local equilibrium in this subsection. Specifically, the electronic states of atoms and molecules, as well as the vibrational and rotational states of molecules, are assumed to follow Boltzmann distributions characterized by $T_{ee}$, $T_v$, and $T_r$, respectively. 
Based on these assumptions, the emission and absorption coefficients of the partial radiators listed in Table~\ref{tab:diatomicRadTrans} are computed and compared with the results obtained from Spark \cite{lopez2016spark}.

\subsubsection{Atoms}

\begin{figure}[hbt!]
\centering
\begin{subfigure}{0.48\textwidth}
    \includegraphics[width=\linewidth]{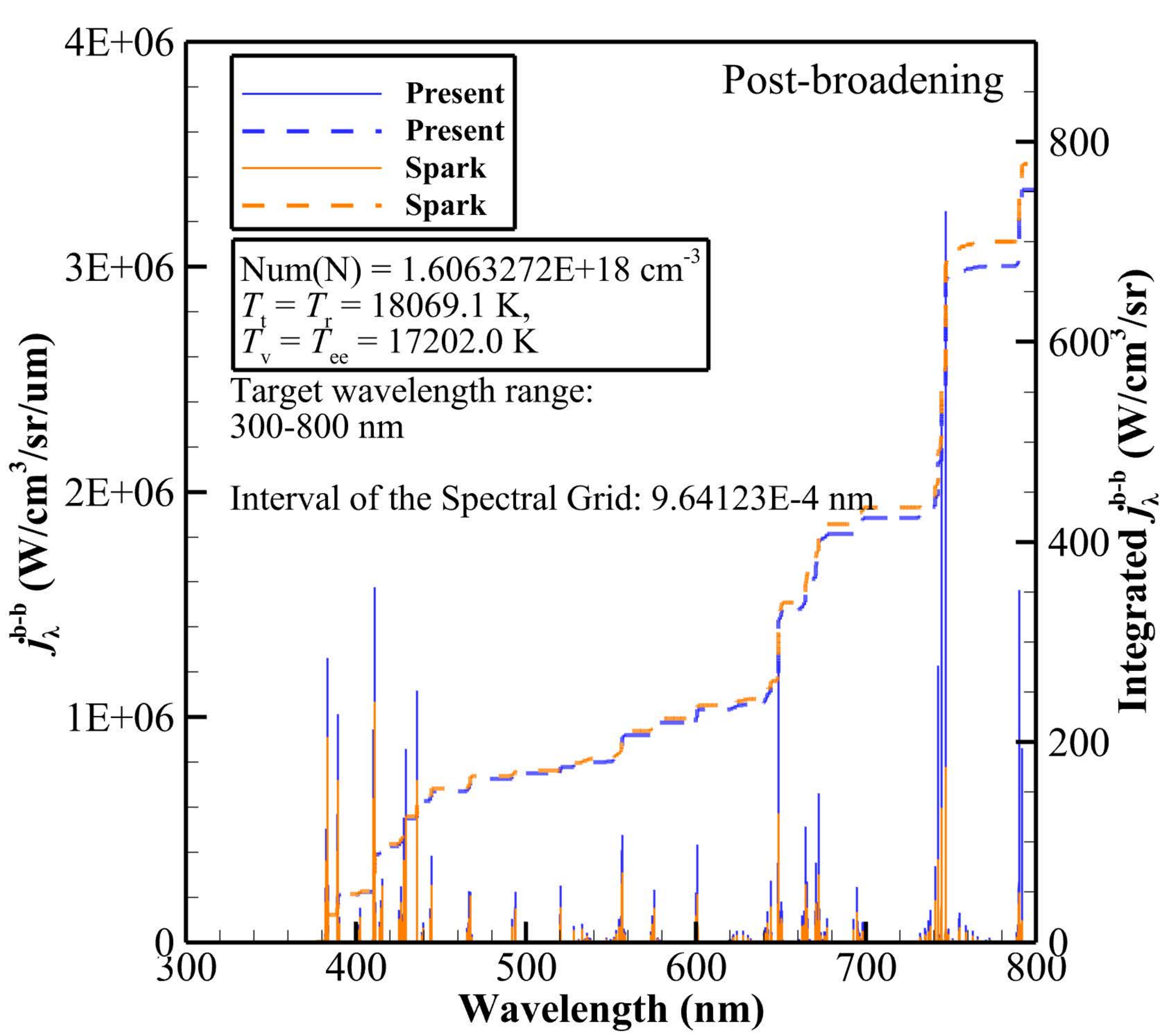}
    \caption{Emission coefficients}
\end{subfigure}
\begin{subfigure}{0.48\textwidth}
    \includegraphics[width=\linewidth]{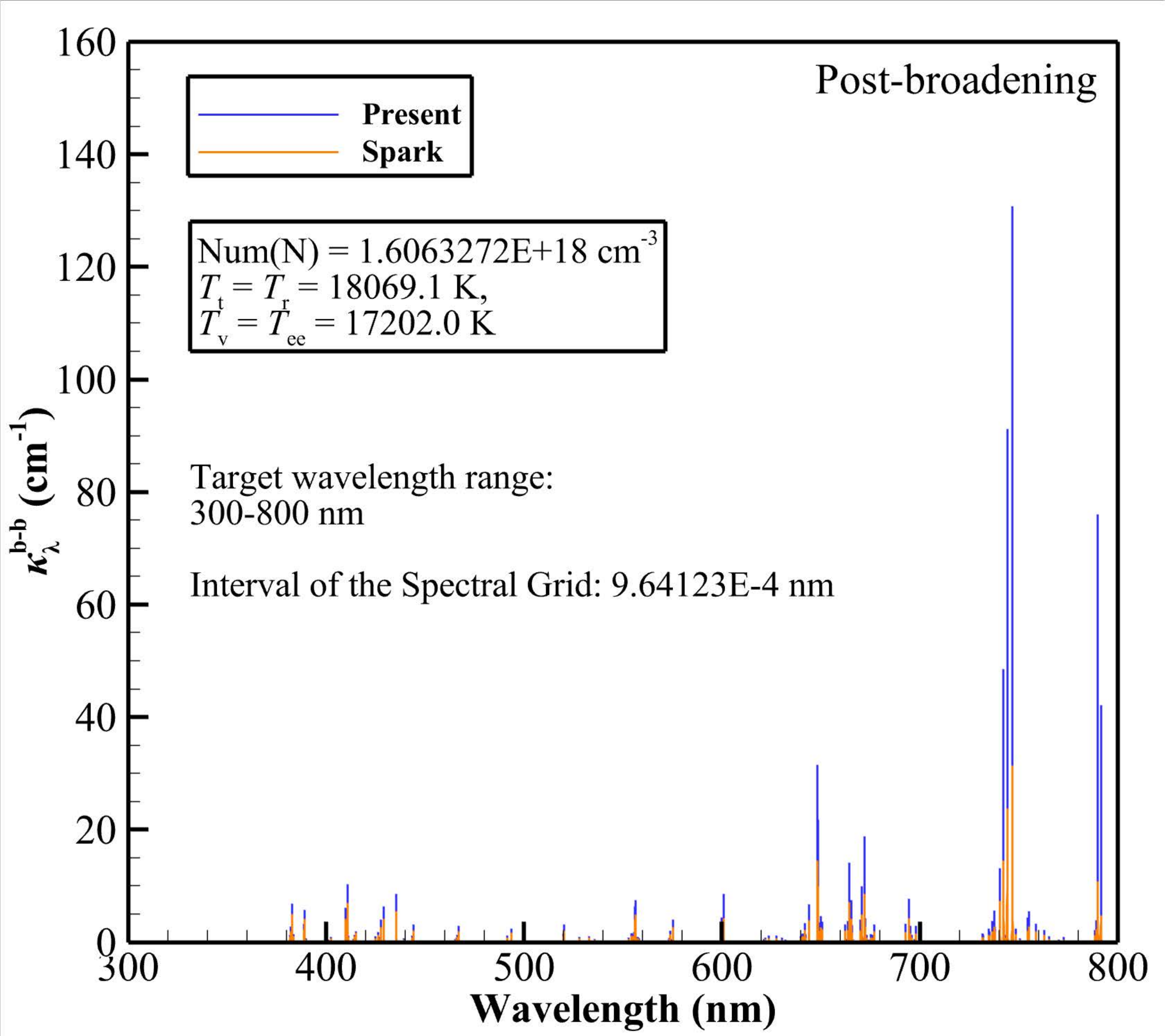}
    \caption{Absorption coefficients}
\end{subfigure}
\caption{Comparison of the B-B emission spectra of atomic \ch{N} predicted by Spark and RAPRAL.}
\label{fig:b-b_N_spectrum}
\end{figure}

Figures~\ref{fig:b-b_N_spectrum} present the emission and absorption coefficients of atomic N in the UV/Vis spectral range under the specified conditions, corresponding to B-B transitions. The dashed lines denote the integrated emission coefficients.
Comparison with the Spark results under identical conditions shows that the broadened spectral features predicted by the present code are higher than those of Spark at certain wavelengths. This discrepancy is attributed to the use of updated line-shape models, such as those proposed by Johnston \cite{johnston2006nonequilibrium}, which are based on detailed calculations of Stark broadening half-widths for \ch{N} and \ch{O} atoms.
Additionally, a finer spectral grid is employed in the present calculations, as indicated by the spectral grid interval shown in the figure. This increased resolution leads to sharper emission and absorption peaks, which contributes to the slightly lower integrated emission coefficient compared to that predicted by Spark over certain wavelength ranges.

\subsubsection{Molecules}
\begin{figure}[hbt!]
\centering
\begin{subfigure}{0.48\textwidth}
    \includegraphics[width=\linewidth]{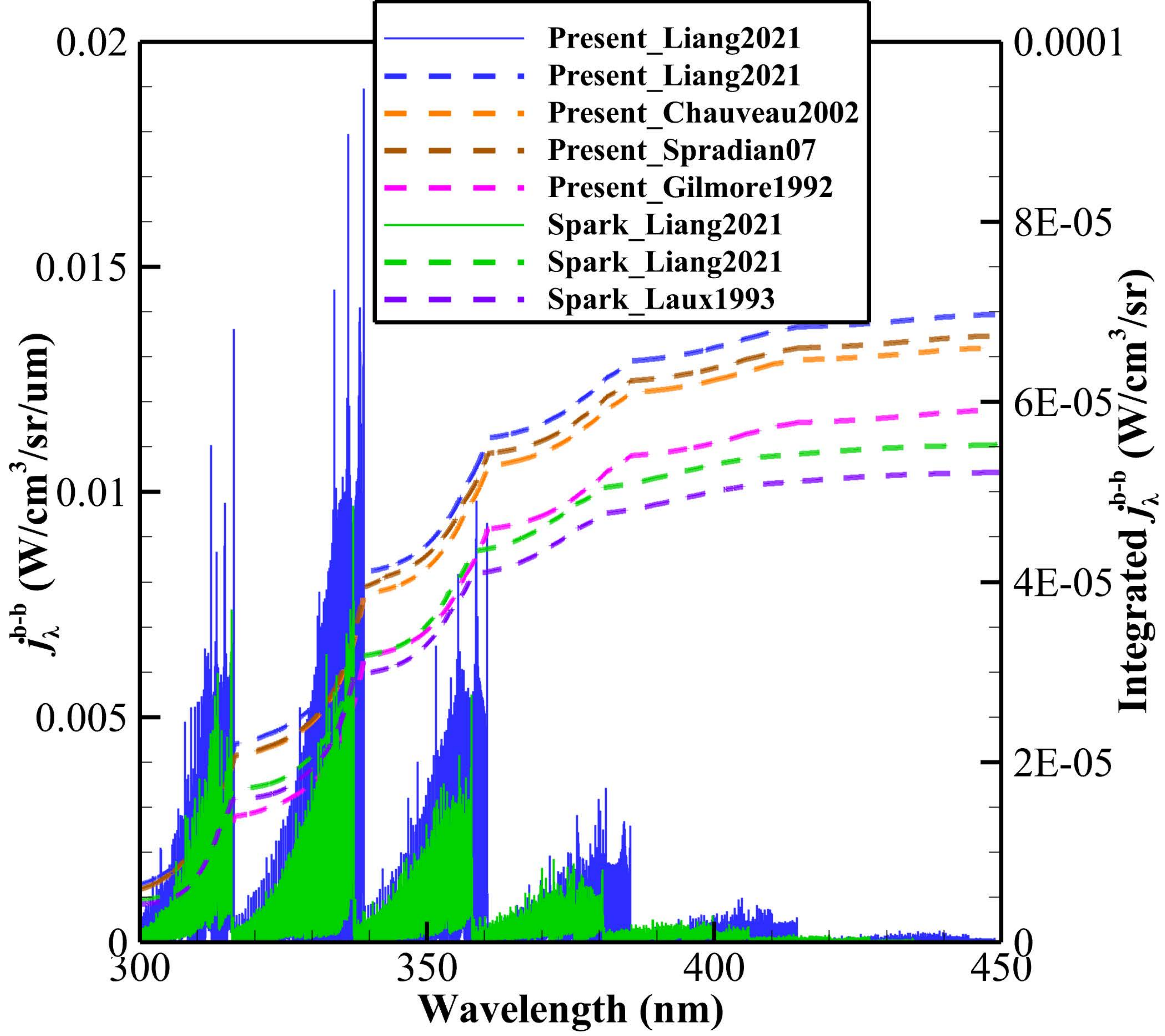}
    \caption{Emission coefficients}
\end{subfigure}
\begin{subfigure}{0.48\textwidth}
    \includegraphics[width=\linewidth]{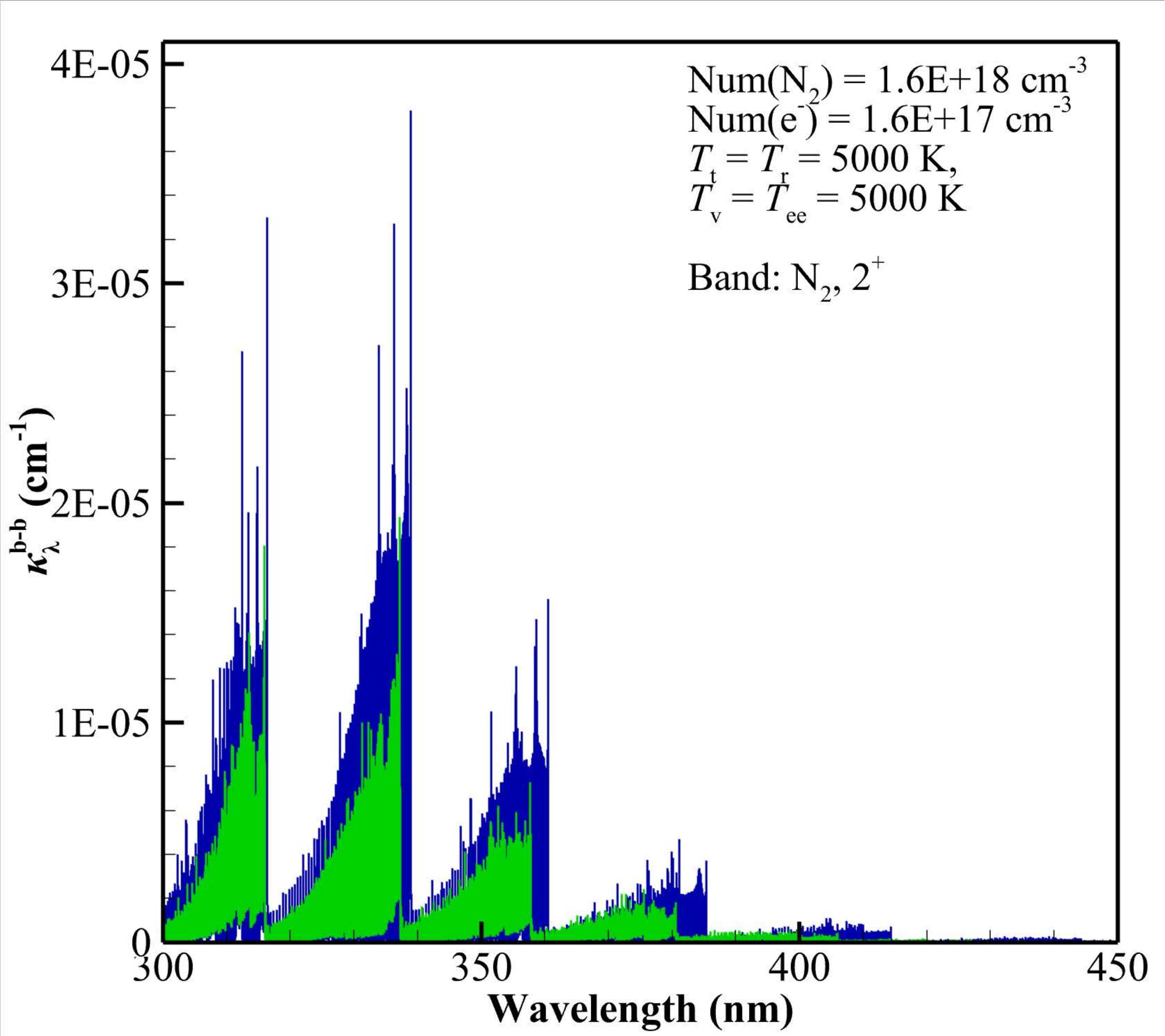}
    \caption{Absorption coefficients}
\end{subfigure}
\caption{Comparison of the B-B emission spectra of \ch{N2} Second Positive band predicted by Spark and RAPRAL.}
\label{fig:b-b_N2_2p_spectrum}
\end{figure}

\begin{figure}[hbt!]
\centering
\begin{subfigure}{0.48\textwidth}
    \includegraphics[width=\linewidth]{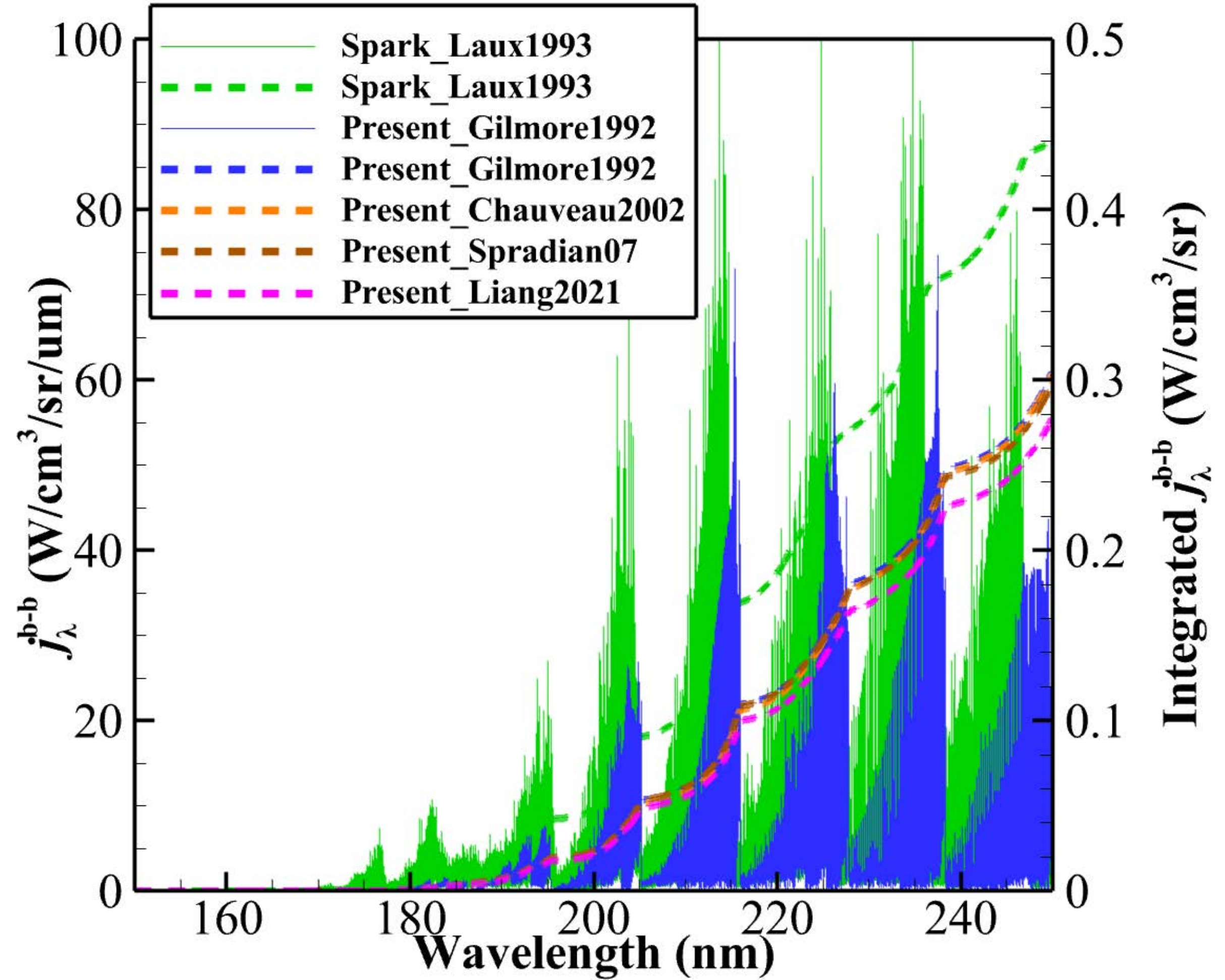}
    \caption{Emission coefficients}
\end{subfigure}
\begin{subfigure}{0.48\textwidth}
    \includegraphics[width=\linewidth]{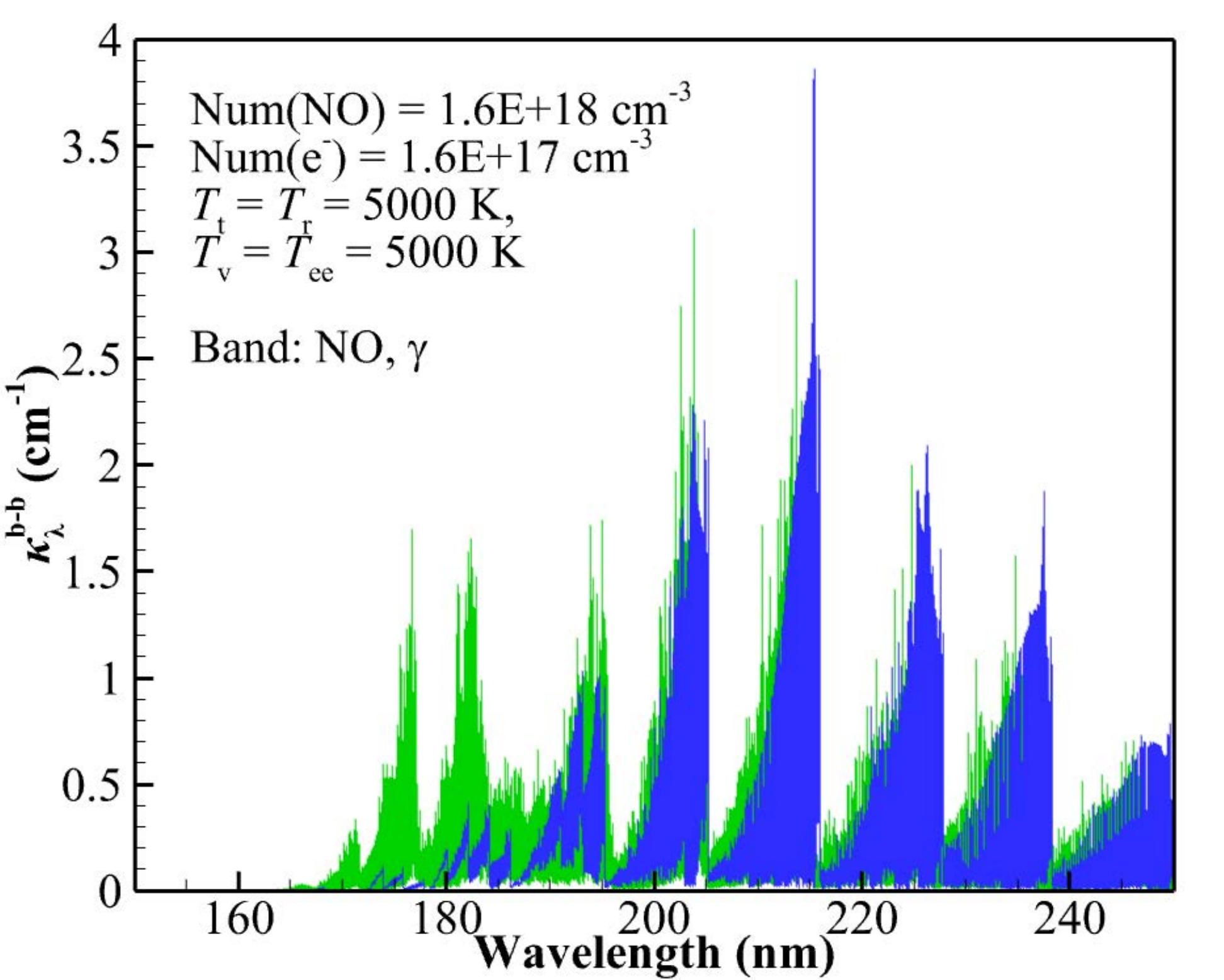}
    \caption{Absorption coefficients}
\end{subfigure}
\caption{Comparison of the B-B emission spectra of \ch{NO} $\gamma$ band predicted by Spark and RAPRAL.}
\label{fig:b-b_NO_gamma_spectrum}
\end{figure}

\begin{figure}[hbt!]
\centering
\begin{subfigure}{0.48\textwidth}
    \includegraphics[width=\linewidth]{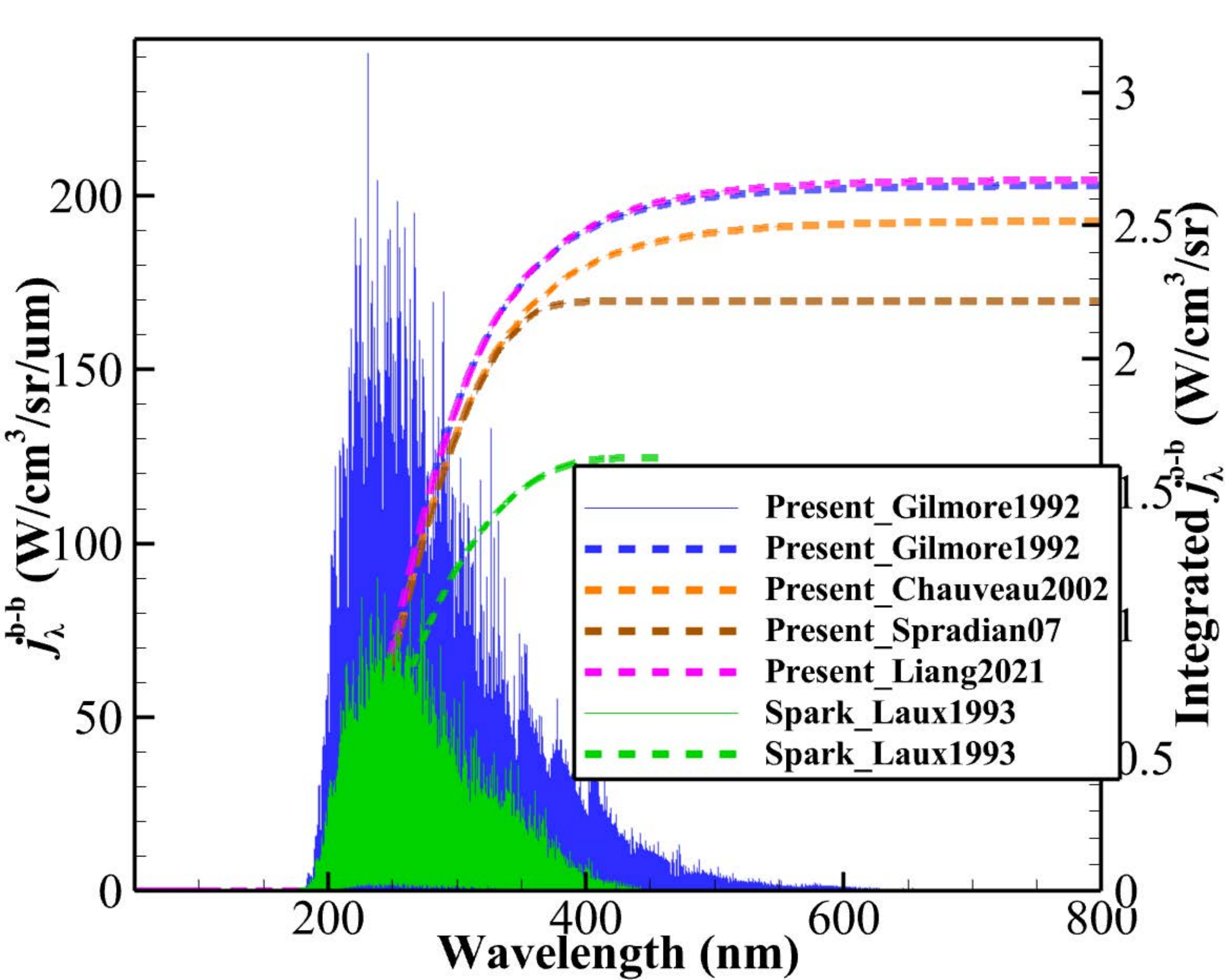}
    \caption{Emission coefficients}
\end{subfigure}
\begin{subfigure}{0.48\textwidth}
    \includegraphics[width=\linewidth]{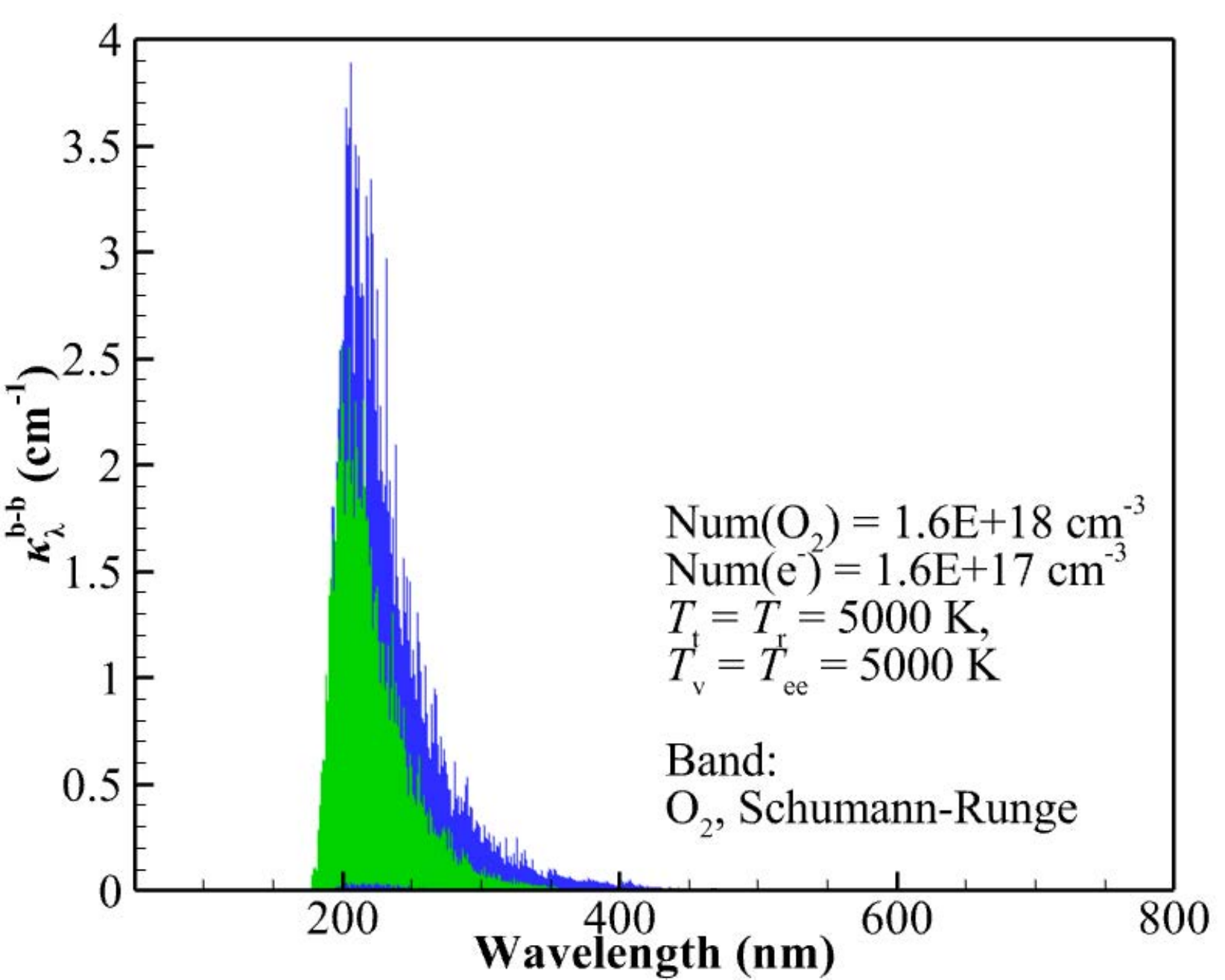}
    \caption{Absorption coefficients}
\end{subfigure}
\caption{Comparison of the B-B emission spectra of \ch{O2} Schumann-Runge band predicted by Spark and RAPRAL.}
\label{fig:b-b_O2_SR_spectrum}
\end{figure}

Figures~\ref{fig:b-b_N2_2p_spectrum} to \ref{fig:b-b_O2_SR_spectrum} present the emission and absorption coefficients of the major band systems of common radiators in air (see Table~\ref{tab:diatomicRadTrans}). 
The differences between the present code and Spark are examined using different $Re'$ data \cite{gilmore1992franck, chauveau2002contributions,laux2003optical,liang2021radiative} (including both the electronic transition moment functions [ETMF, denoted as $Re$] and the Franck-Condon factors [denoted as $q$]). 

On the one hand, discrepancies in overall intensity are observed. For example, in Figure~\ref{fig:b-b_N2_2p_spectrum}, the absorption/emission coefficients at the same wavelengths are higher in RAPRAL. A portion of these discrepancies arises from the use of different $Re'$, as evidenced by the integrated emission coefficients shown by the dashed lines in the figure.
Specifically, compared with the early work of Gilmore et al. \cite{gilmore1992franck}, where $Re$ is obtained via semi-empirical methods (e.g., inference from experimental data or the $r$-centroid approximation) and $q$ is computed through direct numerical integration, Chauveau et al. \cite{chauveau2002contributions} abandoned the $r$-centroid approximation and selected $Re$ values that best match experimental data through comparison of multiple sources (including $ab$ $initio$ calculations and extrapolated experimental data). The Franck-Condon factors $q$ were then computed using the Rydberg-Klein-Rees (RKR) method, in which potential energy curves are reconstructed from experimental spectroscopic constants. As a result, higher accuracy was achieved. However, due to rapid advancements in experimental techniques and computational capabilities, the data from Chauveau et al. \cite{chauveau2002contributions} are now somewhat outdated.
Liang et al. \cite{liang2021radiative} employed recent $ab$ $initio$ calculations as a consistent source of ETMF data and combined them with updated higher-order Dunham spectroscopic constants. The potential energy curves were reconstructed using the RKR method, and vibrational wavefunctions were obtained by numerically solving the radial Schrödinger equation, from which $q$ was evaluated. This approach provides improved consistency and physical reliability across different molecular systems. Fujita and Abe \cite{fujita1997spradian}, as implemented in Spradian07, adopted a combination of the most recent data available at the time, resulting in an accuracy level intermediate between the semi-empirical data of Gilmore et al. \cite{gilmore1992franck} and the mixed dataset of Chauveau et al. \cite{chauveau2002contributions}.
Therefore, as summarized in Table~\ref{tab:diatomicRadTrans}, the present code primarily adopts the dataset from Liang et al. \cite{liang2021radiative} for subsequent calculations, with the exception of the \ch{NO} $\beta$ band system ($\rm B^2\Pi_r \rightarrow X^2\Pi_r$). This exception is made because Liang et al. \cite{liang2021radiative} reported that the vibrational radiative lifetimes of the $\rm B^2\Pi_r$ state as a function of vibrational quantum number are consistent with those of Chauveau et al. \cite{chauveau2002contributions}; however, the resulting integrated emission coefficients differ by nearly a factor of 160 and are significantly higher than those based on Gilmore et al. \cite{gilmore1992franck}. This discrepancy suggests a possible typographical error, and therefore the present work retains the $Re'$ values derived from the data of Chauveau et al. \cite{chauveau2002contributions}.
The primary source of discrepancies in the peak intensities of the emission and absorption coefficients, however, lies in the variations in partition function evaluations (e.g., updated Dunham coefficients), which lead to deviations in the populations of vibrational and rotational excited states.

On the other hand, differences in spectral line shapes are observed within the same vibrational band ($v'\rightarrow v''$) between RAPRAL and Spark. 
For instance, in Figure~\ref{fig:b-b_NO_gamma_spectrum}, RAPRAL predicts a more compact rovibrational band, characterized by fewer weak lines and an earlier truncation toward the violet (shaded/degraded) end, whereas in Figure~\ref{fig:b-b_O2_SR_spectrum} the opposite trend is observed, with Spark yielding the more compact distribution. These differences arise primarily from the selection of the maximum rotational quantum number \cite{potter2011modelling}, which determines the number of rotational lines within a given vibrational band. In addition, the integrated emission coefficients associated with different sources of $Re'$ exhibit noticeable variations, mainly due to differences in data acquisition methods and underlying assumptions, as discussed above.

\subsection{Radiative transfer for Fire II flight test}
\begin{figure}[!htb]
    \centering
    \includegraphics[width=0.5\linewidth]{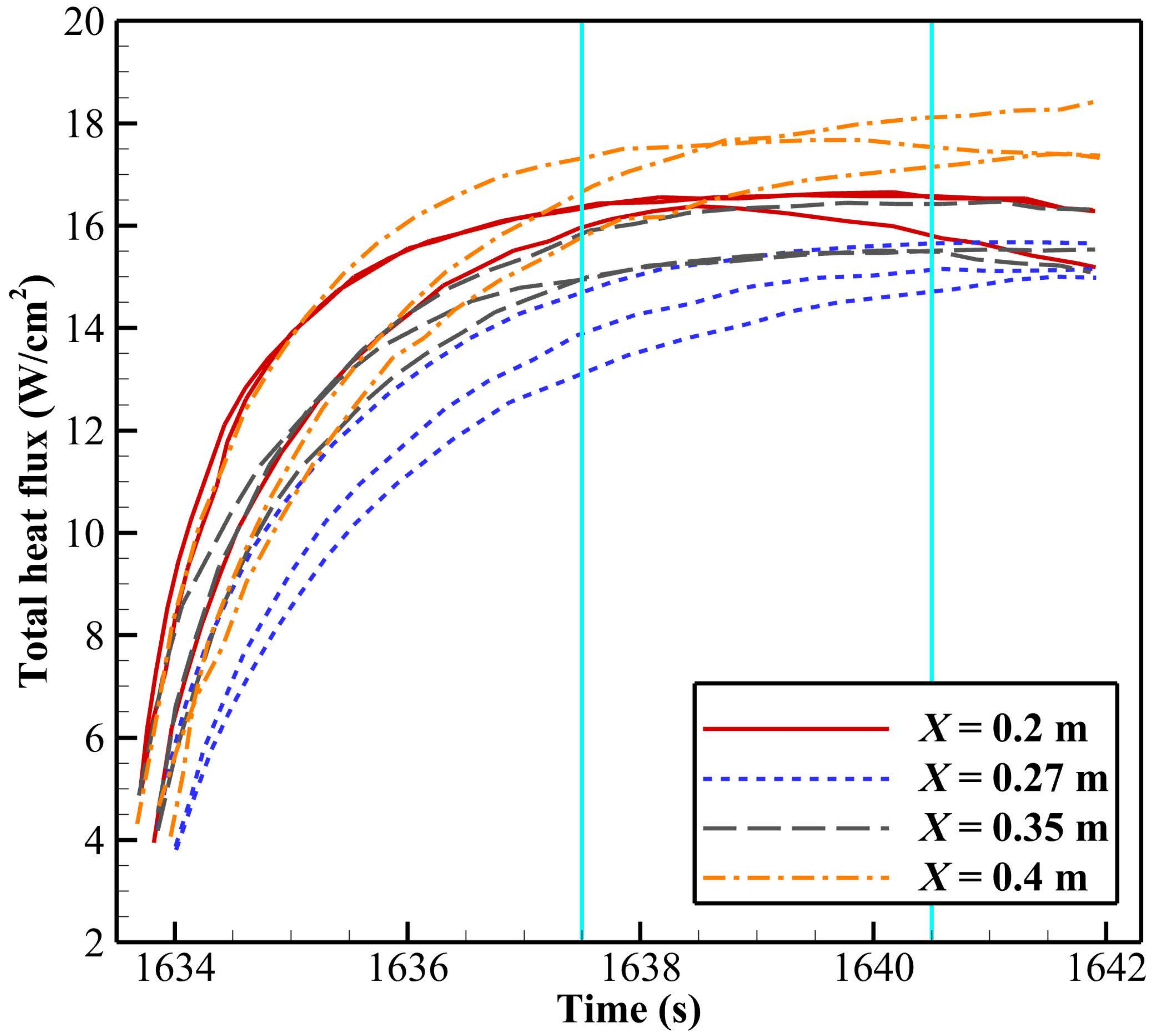}
    \caption{\enspace Calorimeter measurements at different locations on the Fire II vehicle afterbody as a function of flight time, with each axial position $X$ corresponding to three calorimeters arranged around the vehicle circumference.}
    \label{fig:totalHeatFlux_Exp_Fig11a-d_Wright2001}
\end{figure}

The FIRE II (Flight Investigation of Reentry Environment II) flight experiment \cite{minovitch1965project, lewis1966flight, cornette1966calorimeter, slocumb1966project, cauchon1967radiative} is a representative reentry aerothermal test conducted during the Apollo program in the 1960s. Its primary objective was to obtain convective and radiative heating data in high-enthalpy, strongly nonequilibrium flows under conditions approaching lunar return velocities. 
The test vehicle was a scaled Apollo command module configuration with a diameter of 0.67 m. Beryllium calorimeters and radiometers were installed along the reentry trajectory to measure the total heat flux (convective + radiative) and the shock-layer radiative intensity, respectively. Measurements were obtained at representative altitudes of 71 km, 53 km, and 37 km, corresponding to peak velocities up to 11.31 km/s and freestream densities as low as $8.57 \times 10^{-5}$ kg/m$^3$. 
The FIRE II dataset possesses several notable characteristics: 1. the high velocity and high enthalpy conditions encompass regimes with strong dissociation, significant ionization, and pronounced vibrational-electronic nonequilibrium, providing a comprehensive testset for finite-rate chemistry and multi-temperature models; 2. the flight measurements include both total heat flux and radiative contributions, offering independent validation for the convective and radiative heat transfer prediction models; 3. the relatively simple geometry and strong axisymmetric features facilitate 2D steady-state simulations and cross-validation among different computational codes. Consequently, the FIRE II dataset has been widely used for code validation in hypersonic nonequilibrium aerothermodynamics \cite{hash2007fire}.
Figure~\ref{fig:totalHeatFlux_Exp_Fig11a-d_Wright2001} presents the time histories of the total heat flux measured by calorimeters installed on the afterbody of the vehicle. In the present study, two representative time instances along the reentry trajectory are selected (indicated by light blue solid lines in the figure), with the corresponding flight conditions listed in Table~\ref{tab:conditionsFireII}. 
Because the angle of attack at both instants was less than $1^\circ$ \cite{slocumb1966project}, the axisymmetric assumption can be applied without introducing significant error.
The flowfield around the vehicle is first computed using the in-house Artist-CFD solver \cite{wang2023accuracy} (with radiative cooling neglected and a fully catalytic wall assumption). 
A two-temperature model ($T_{tr}$ and $T_{vee}$) is adopted, with the vibration-dissociation coupling model proposed by Park \cite{park1989nonequilibrium}. The chemical reaction model and turbulence closure are modeled using the Park2001 model \cite{park2001chemical} and the Spalart-Allmaras model, respectively. 
Subsequently, the RAPRAL solver is employed to evaluate and analyze the radiative heat flux at different locations on the afterbody.

\begin{table}[htb]
    \caption{\enspace Flow conditions at two selected trajectory points of the Fire II flight experiment.}
    \footnotesize
    \setlength{\tabcolsep}{4pt}
    \renewcommand{\arraystretch}{1.5}
    \centering
    \begin{tabular}{lcccc}
        \hline
        Times, s & Density, kg/m$^3$ & Velocity, km/s & Freestream temperature, K & Wall temperature, K\\
        \hline
        1637.5 & 1.47$\times10^{-4}$ & 11.25 & 228 & 1030\\
        1640.5 & 3.86$\times10^{-4}$ & 10.97 & 254 & 1560\\
        \hline
    \end{tabular}
    \label{tab:conditionsFireII}
\end{table}

\begin{figure}[!htb]
    \centering
    \includegraphics[width=0.7\linewidth]{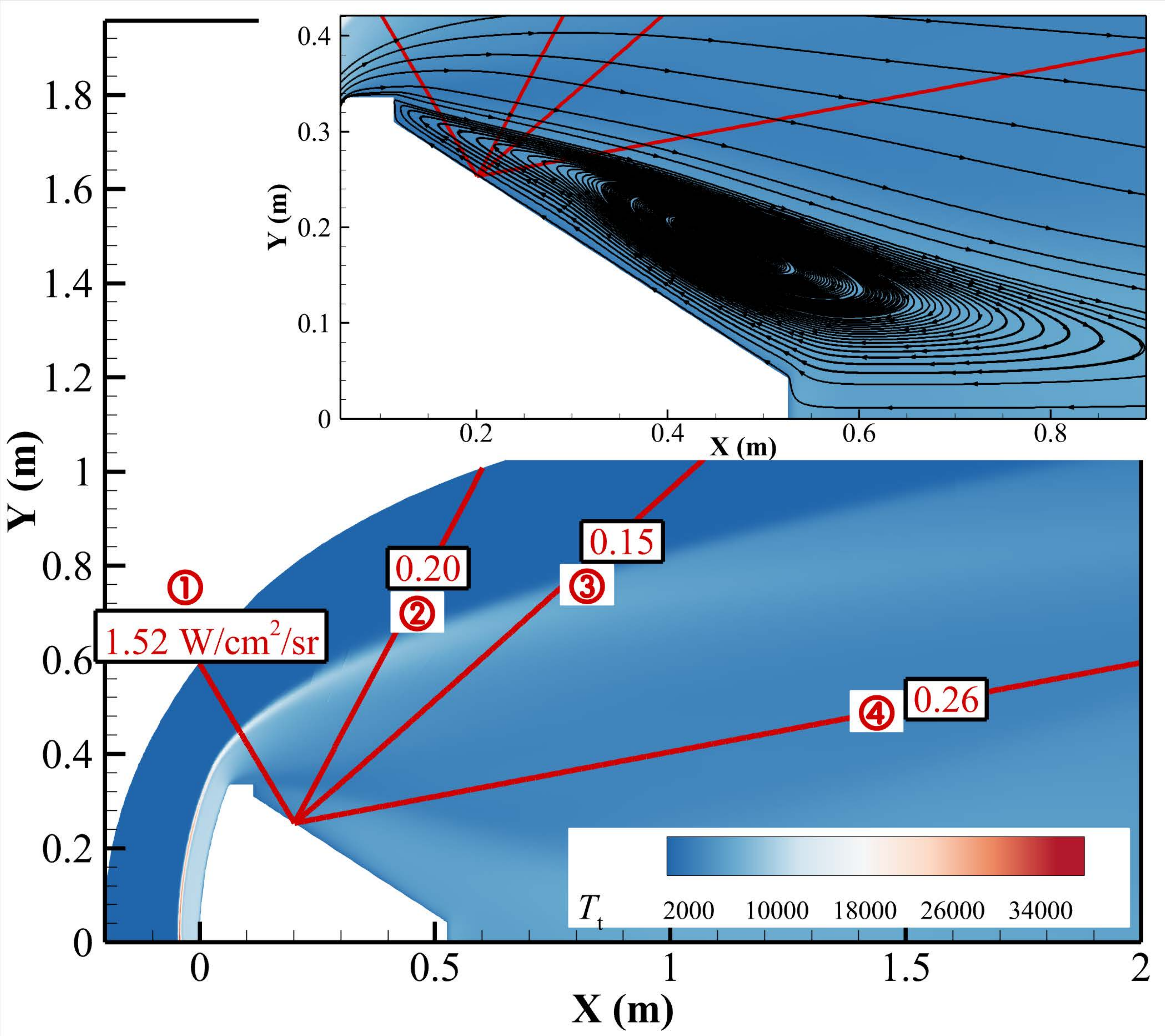}
    \caption{\enspace Flowfield temperature contour around the vehicle and streamlines on the afterbody section for $t = 1637.5~\rm s$.}
    \label{fig:flowField-radLOS_1637p5_20cm_FireII}
\end{figure}

Figure~\ref{fig:flowField-radLOS_1637p5_20cm_FireII} shows the temperature contours of the flowfield around the vehicle and the streamlines near the afterbody at $t = 1637.5~\rm s$. Four representative LOS impinging on the wall at $X=0.2$ m are also indicated, where the annotated values correspond to the radiative intensity $I(\Omega, \lambda=100-1000~\rm nm)$ obtained by integrating the RTE along each LOS.
Examination of the expansion region downstream of the shoulder reveals that, on one hand, the high-temperature region extends over a large spatial domain and persists up to the boundary of the computational domain (approximately three times the forebody diameter), where the translational temperature remains in the range of 3000-6000 K. On the other hand, a large recirculation zone develops from the shoulder step to approximately 1 m downstream. The associated vortex structures transport high-temperature gas toward the wall, significantly enhancing convective heat transfer and potentially leading to local heat flux peaks. Meanwhile, this behavior introduces strong spatial non-uniformity in thermochemical nonequilibrium. Consequently, the use of a tangent-slab approximation for integrating the RTE may result in substantial errors in this region \cite{johnston2015features,johnston2018impact}.

\begin{figure}[!htb]
    \centering
    \includegraphics[width=0.7\linewidth]{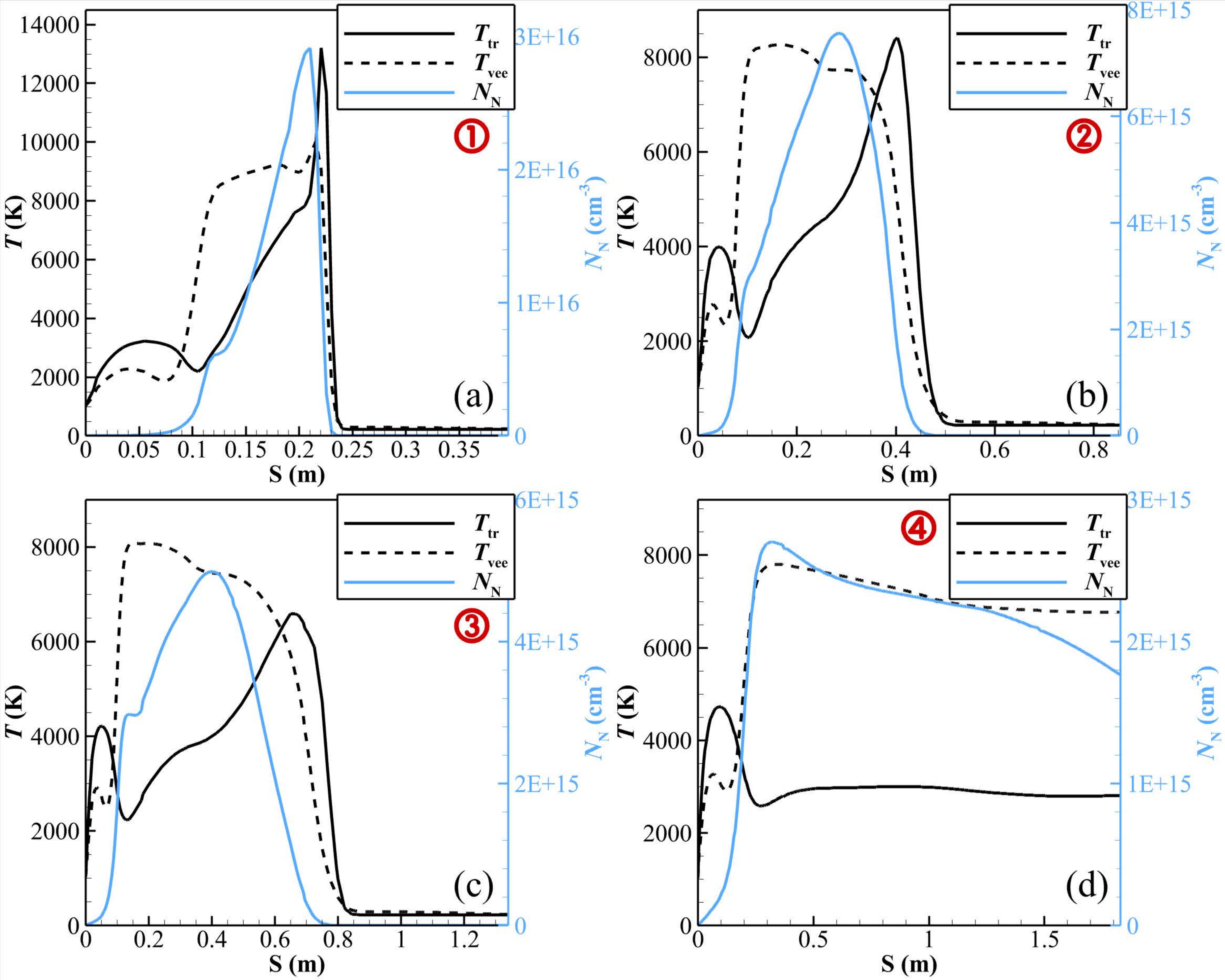}
    \caption{\enspace Temperatures and atomic N number density distributions along four LOSs indicated in Figure~\ref{fig:flowField-radLOS_1637p5_20cm_FireII} for $t = 1637.5~\rm s$.}
    \label{fig:tems-NumN-radLOS_1637p5_20cm_FireII}
\end{figure}
The distributions of temperatures and atomic N number density along the four LOSs shown in the Figure~\ref{fig:flowField-radLOS_1637p5_20cm_FireII} are presented in Figure~\ref{fig:tems-NumN-radLOS_1637p5_20cm_FireII}, where $S=0$ denotes the vehicle wall and $S=S_{max}$ represents the intersection of the LOS with the far-field boundary. The temperature profiles along all four LOSs exhibit characteristics markedly different from those along the stagnation line. Specifically, in most of the thermochemical nonequilibrium region, the vibrational-electronic-electronic temperature exceeds the translational-rotational temperature, which is a typical feature of the expansion region in hypersonic flows.
This phenomenon primarily arises from the disparity between the timescales of vibrational energy relaxation and flow expansion. After passing through the strong compression and high-frequency collisions induced by the bow shock near the forebody, a substantial portion of translational energy is transferred to internal energy modes through molecular collisions, leading to significant excitation of the internal energy. 
As the flow passes over the shoulder and enters the leeside region, rapid expansion occurs, resulting in a sharp decrease in gas density and collision frequency, and consequently a rapid drop of $T_{tr}$. However, under low-pressure conditions, the vibrational relaxation proceeds in a lower rate and may exceed the characteristic flow timescale in certain regions. As a result, the transfer of vibrational-electronic-electronic energy back to translational mode becomes inefficient and exhibits a freezing behavior, maintaining a high level of internal energy excitation. This leads to a much slower decrease in $T_{vee}$ compared to $T_{tr}$, and in many places, $T_{vee}$ even exceeds $T_{tr}$.
Furthermore, from LOS \textcircled{1} to \textcircled{4}, although the peak temperature and peak atomic N number density generally decrease, the spatial extent of the nonequilibrium region characterized by $T_{vee}>T_{tr}$ increases. According to the analysis of Johnston and Mazaheri \cite{johnston2015features}, radiation in the leeside region during Earth reentry of hypersonic vehicles is dominated by atomic emission in the VUV spectral range (e.g., N: 149.33 nm, 174.36 nm; O: 130.35 nm), contributing more than 55\% to the integrated radiative intensity. The variation of integrated radiative intensity along the four LOSs in Figure~\ref{fig:flowField-radLOS_1637p5_20cm_FireII} is then expected to be consistent with the trend of atomic N number density. It is noted that although LOS \textcircled{4} has the lowest peak N number density, its longer integration path results in a higher integrated radiative intensity compared to LOS \textcircled{2} and \textcircled{3}.

\begin{figure}[!htb]
    \centering
    \includegraphics[width=0.7\linewidth]{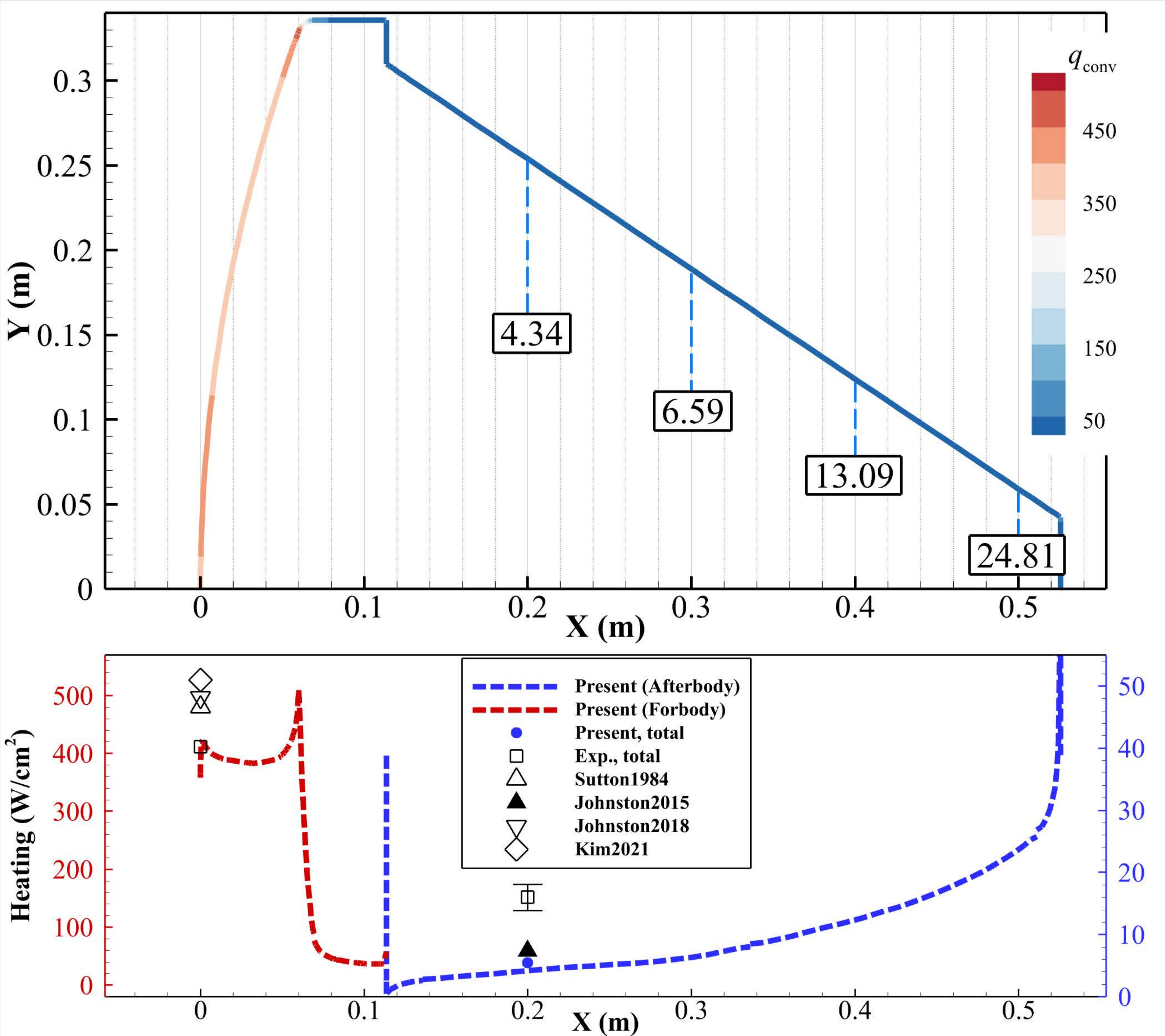}
    \caption{\enspace Spatial distribution of wall heat flux (with $X = 0$ m corresponding to the stagnation point) for $t = 1637.5~\rm s$.}
    \label{fig:conv-radHeat_1637p5_FireII}
\end{figure}

Figure~\ref{fig:conv-radHeat_1637p5_FireII} presents the distributions of convective heat flux (shown as contour and dashed lines) and total heat flux along the vehicle symmetry axis obtained from the present numerical simulation, together with experimental measurements \cite{cornette1966forebody,cauchon1967radiative,slocumb1966project} and selected results from the literature \cite{johnston2015features,sutton1984air,johnston2018impact,kim2021modification}.
First, the predicted stagnation-point convective heat flux is close to, and only slightly lower than, the experimentally measured total heat flux, which is dominated by convection in this region \cite{capra2013total}, thereby supporting the reliability of the present simulation. In addition, the predicted convective heat flux at the afterbody location $X = 0.2~\rm m$ is slightly lower than the numerical results reported in the literature \cite{johnston2015features}. However, at this location, the total heat flux (i.e., the sum of convective and radiative contributions) remains significantly lower than the experimental measurements.
The underprediction of radiative heat flux in the afterbody region can be attributed to multiple factors. On the one hand, the quasi-steady characteristics of the expansion region on the afterbody are governed by convection-dominated transport along streamlines, where the non-Boltzmann populations are primarily determined by frozen excited and ionized states originating from the high-temperature forebody region. This behavior is fundamentally different from that in the post-shock compression region, where local reaction kinetics dominate and excited-state populations evolve according to local thermodynamic conditions \cite{johnston2019features,johnston2015features}. Consequently, quasi-steady models developed based on the thermochemical nonequilibrium characteristics of the compression region may not accurately predict excited-state populations in the expansion region.
On the other hand, the present simulation neglects the effects of wall ablation products. Radiation contributions solely due to ablation species can increase the radiative heat flux by nearly 25\% \cite{erb2021features}, which further explains the discrepancy between numerical predictions and experimental observations.

\begin{figure}[!htb]
    \centering
    \includegraphics[width=0.7\linewidth]{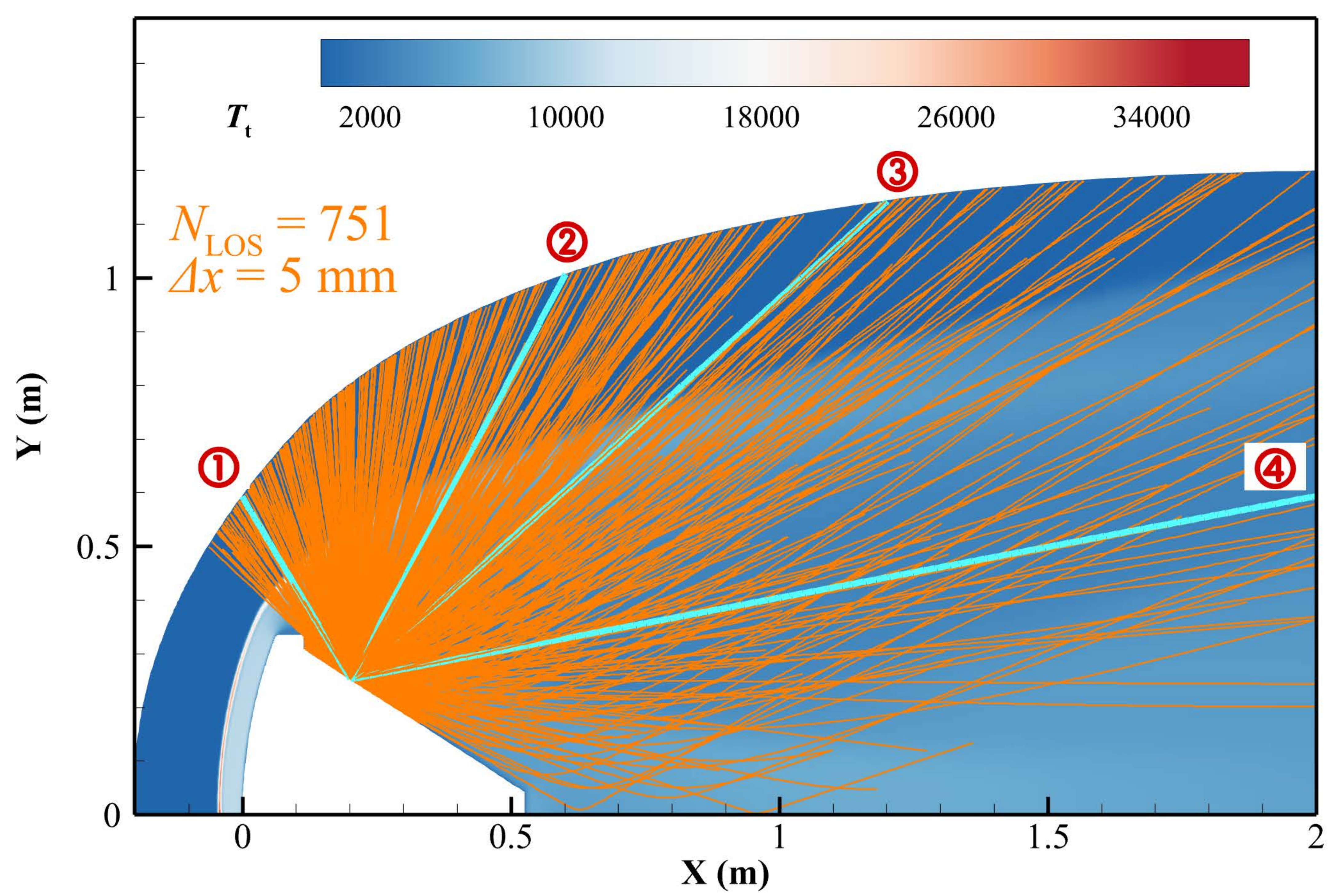}
    \caption{\enspace LOSs used for calculating wall radiative heat flux at $X = 0.2$ m on the vehicle afterbody for $t = 1637.5~\rm s$.}
    \label{fig:flowField-LOSs_1637p5s_20cm_FireII}
\end{figure}

\begin{figure}[!htb]
    \centering
    \includegraphics[width=0.7\linewidth]{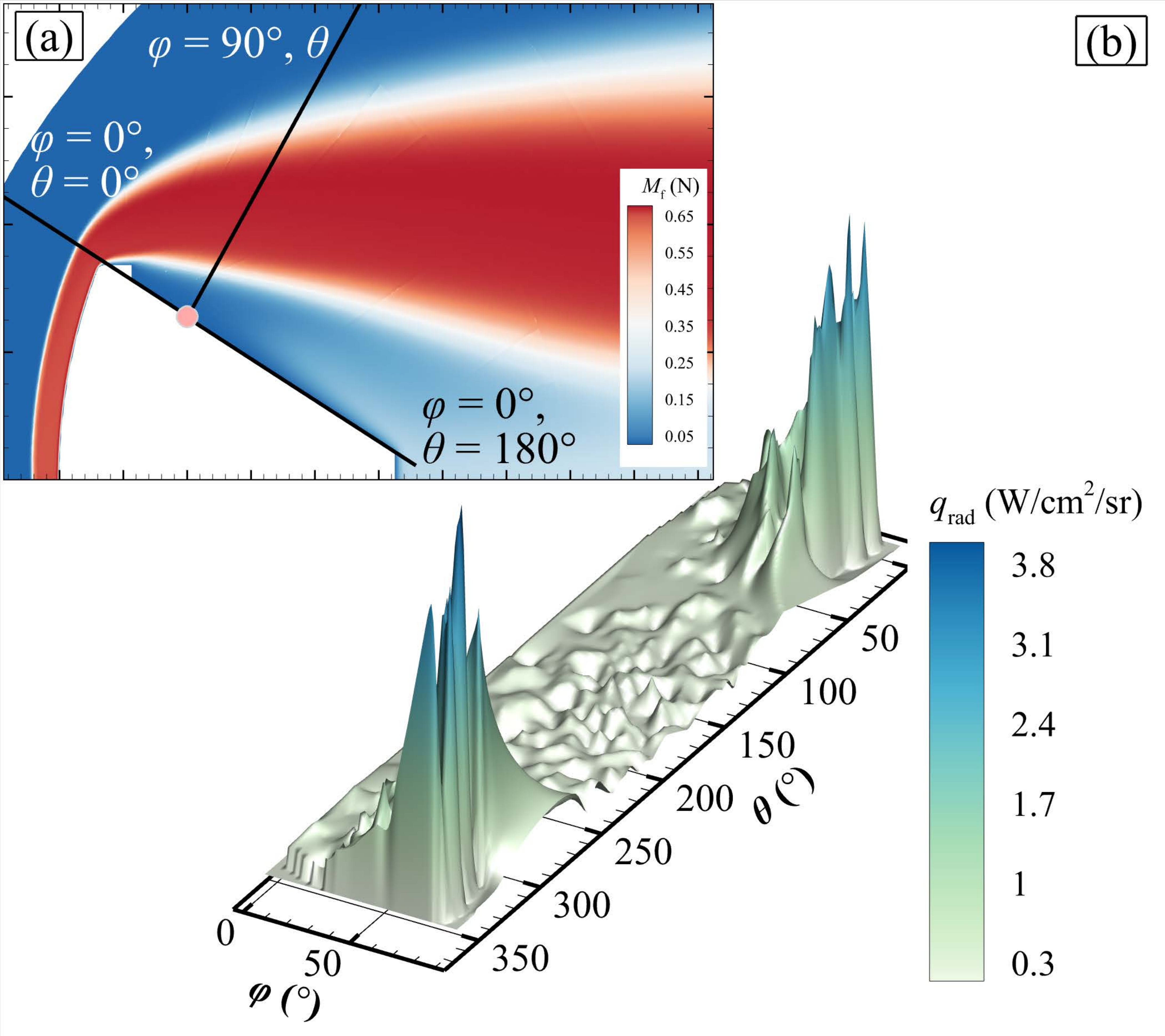}
    \caption{\enspace Radiation distribution along LOS, integrated over the wavelength range $\lambda = 100-1000$ nm, in different directions at $X = 0.2$ m on the vehicle afterbody for $t = 1637.5~\rm s$.}
    \label{fig:radHeatDistribution_LOS_1637p5_20cm}
\end{figure}
Figure~\ref{fig:flowField-LOSs_1637p5s_20cm_FireII} illustrates the distribution of the LOSs employed in the present ray-tracing procedure within the flowfield, which effectively cover the regions in the leeside where the contribution to wall radiative heat flux is most significant. Figure~\ref{fig:radHeatDistribution_LOS_1637p5_20cm} shows the spatial distribution of the integrated radiative intensity $I(\Omega, \lambda=100-1000~\rm nm)$ obtained by integrating the RTE along each LOS.
It is observed that, within the two solid-angle regions defined by $30^{\circ}<{\varphi}<90^{\circ}$, $0^{\circ}<{\theta}<100^{\circ}$ and $30^{\circ}<{\varphi}<90^{\circ}$, $250^{\circ}<{\theta}<360^{\circ}$ (which largely coincide with the regions of relatively high \ch{N} mass fraction $M_f(\rm N)$ shown in the figure) the contribution to the radiative heat flux is significantly higher than that from other spatial domains. In addition, it is found that the continuum radiation arising from B-F and F-F transitions contributes more than 50\% of the integrated radiative intensity along certain LOSs, indicating that the degree of ionization in these regions is also relatively high.

\begin{figure}[!htb]
    \centering
    \includegraphics[width=0.7\linewidth]{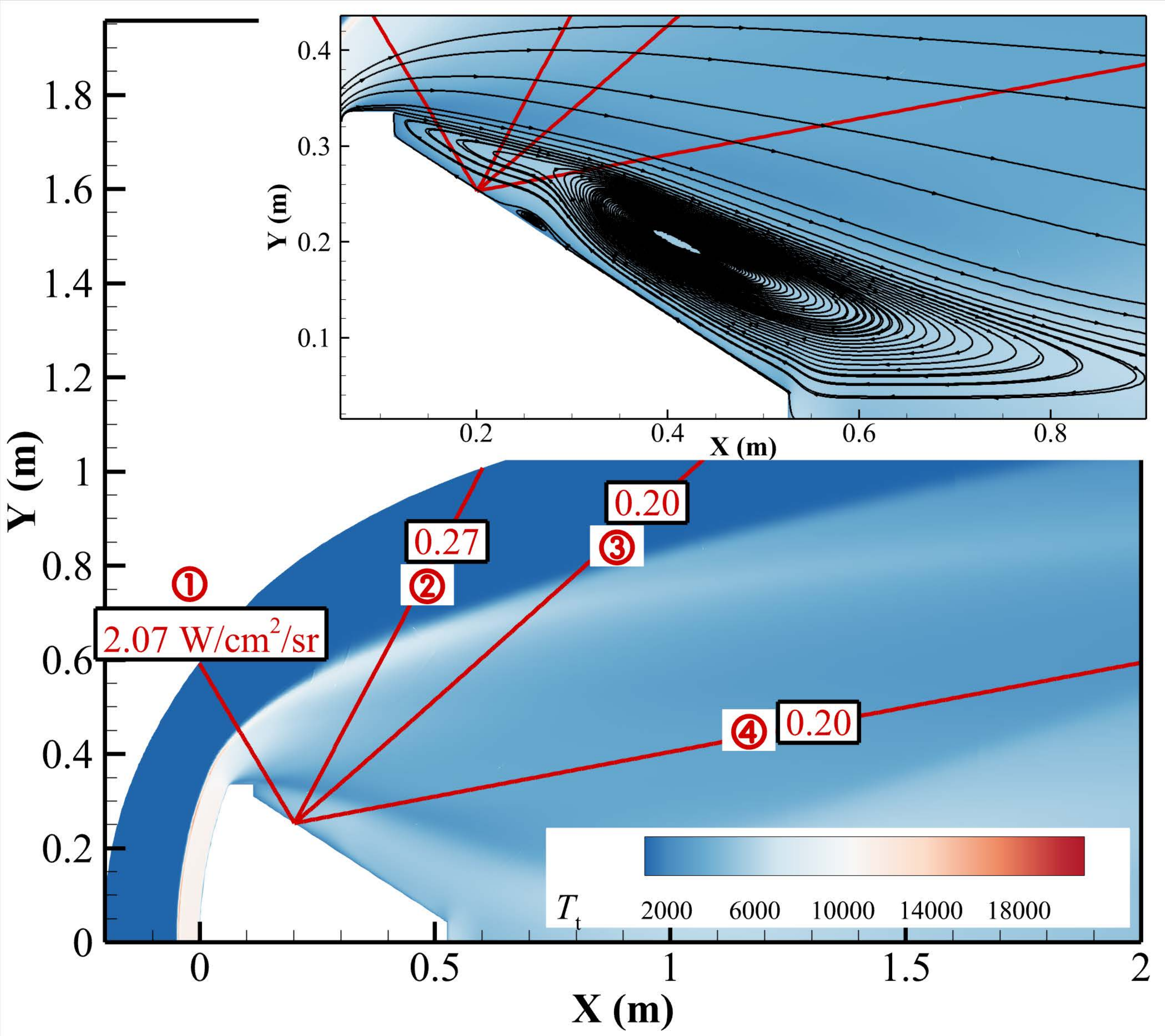}
    \caption{\enspace Flowfield temperature contour around the vehicle and streamlines on the afterbody section for $t = 1640.5~\rm s$.}
    \label{fig:flowField-radLOS_1640p5_FireII}
\end{figure}
\begin{figure}[!htb]
    \centering
    \includegraphics[width=0.7\linewidth]{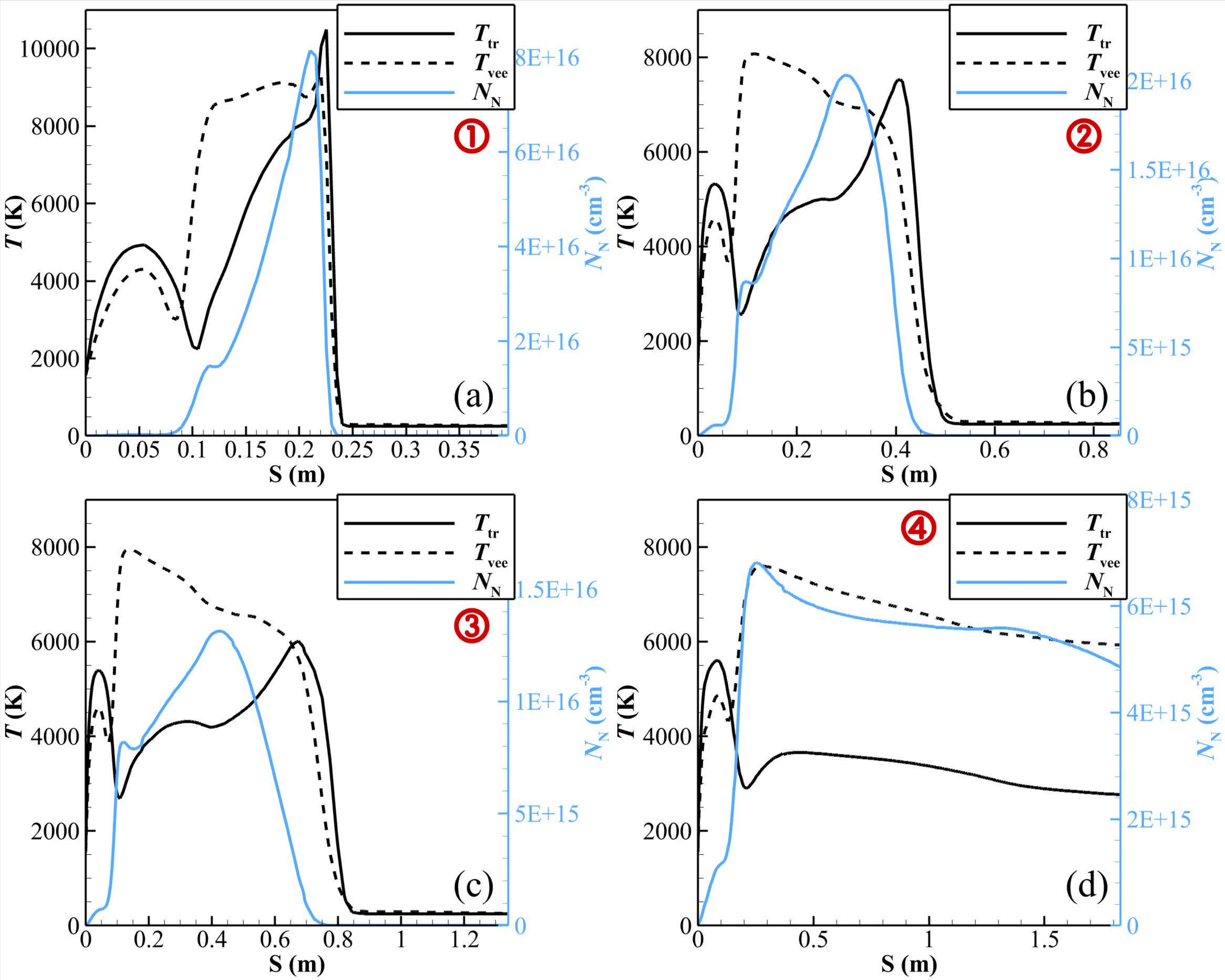}
    \caption{\enspace Temperature and atomic N number density distributions along four LOSs indicated in Figure~\ref{fig:flowField-radLOS_1640p5_FireII} for $t = 1640.5~\rm s$.}
    \label{fig:tems-NumN-radLOS_1640p5_FireII}
\end{figure}
For the case of $t = 1640.5~\rm s$, Figure~\ref{fig:flowField-radLOS_1640p5_FireII} shows the flowfield temperature contour and local streamlines near the afterbody, with four LOSs incident on the wall at $X=0.2$ m indicated. The numbers for each LOS denote the integrated radiative intensity $I(\Omega, \lambda=100-1000 \rm nm)$ obtained by solving the RTE along the LOS.
Compared with the previous time instant, the peak flow temperatures are significantly reduced, and secondary vortices have formed within the separation region near the afterbody. These vortices enhance local momentum and energy transport, increasing the convective heat flux at the wall and potentially producing local heat flux peaks near the reattachment region of the secondary vortices (around $X=0.2$ m). Nevertheless, the common feature remains that the flow maintains relatively high temperatures at the far-field boundaries.

Figure~\ref{fig:tems-NumN-radLOS_1640p5_FireII} shows the temperature and atomic \ch{N} number density distributions along the four LOSs. Similar to the previous time instant, the majority of the thermal nonequilibrium regions along all four LOSs exhibit $T_{vee}$ that are higher than $T_{tr}$. From LOS \textcircled{1} to \textcircled{4}, the peak temperatures and peak \ch{N} number densities decrease, and correspondingly, the wall-directed radiative heat flux integrated along each LOS also gradually diminishes (see Figure~\ref{fig:flowField-radLOS_1640p5_FireII}).
It is noteworthy, however, that at this time instant, the \ch{N} number densities along all four LOSs are higher than those at the previous instant, while $T_{vee}$ remains at similarly elevated levels. This indicates that the radiative heat contribution from atomic emission is stronger at this time.

\begin{figure}[!htb]
    \centering
    \includegraphics[width=0.7\linewidth]{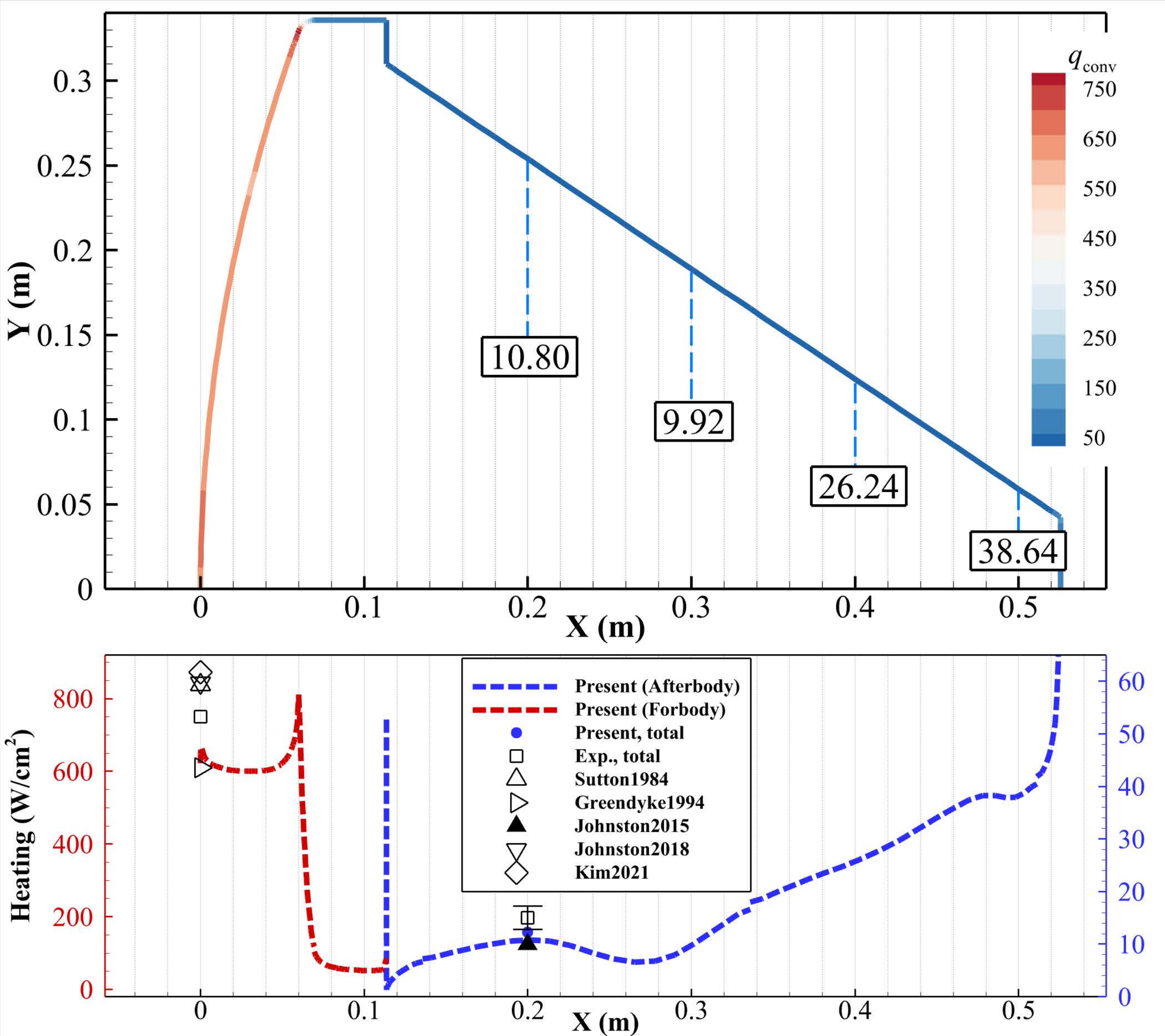}
    \caption{\enspace Spatial distribution of wall heat flux (with $X = 0$ m representing the stagnation point) for $t = 1640.5~\rm s$.}
    \label{fig:heatFluxFireII_1640p5}
\end{figure}
\begin{figure}[!htb]
    \centering
    \includegraphics[width=0.7\linewidth]{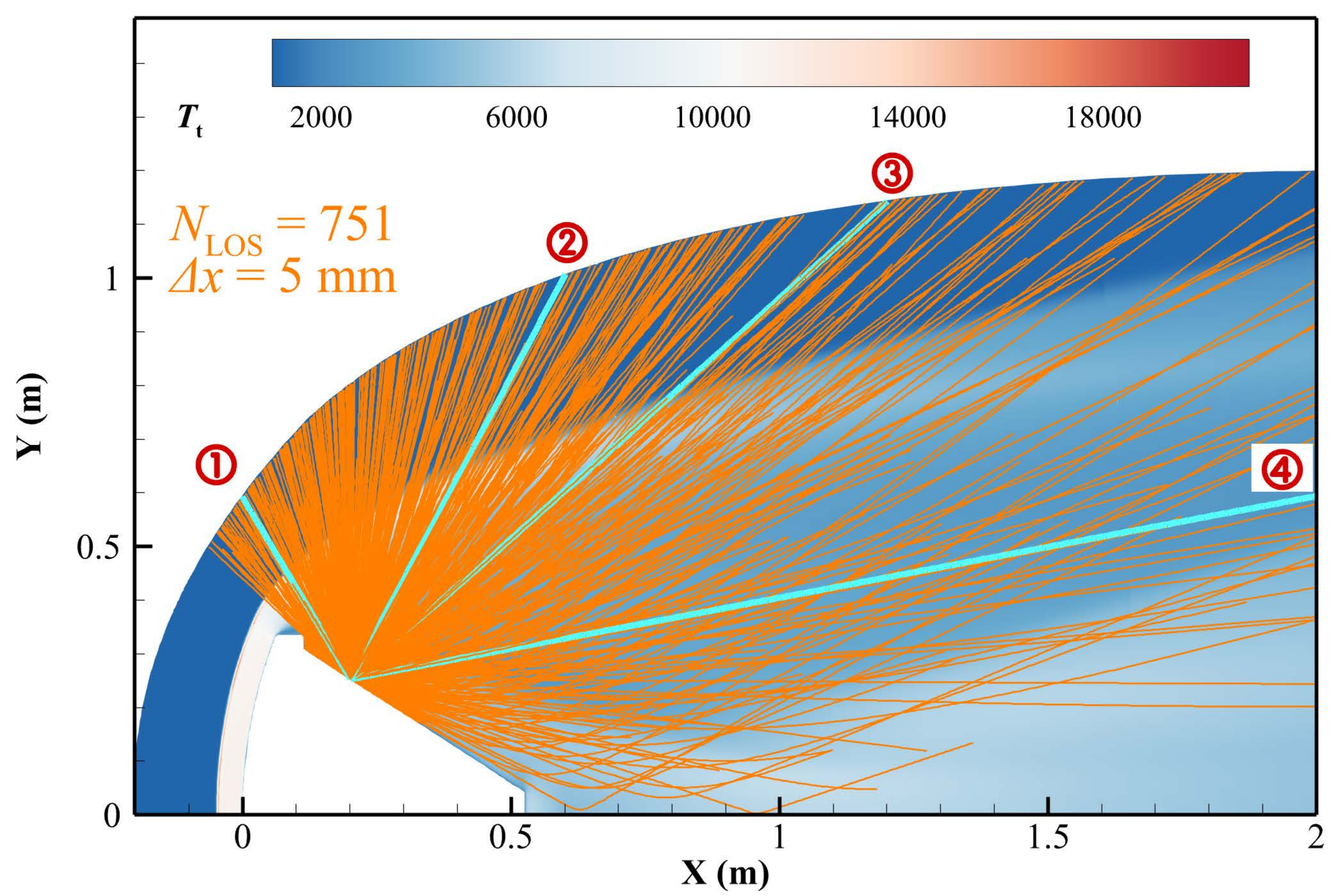}
    \caption{\enspace LOSs used for calculating wall radiative heat flux at $X = 0.2$ m on the vehicle afterbody for $t = 1640.5~\rm s$.}
    \label{fig:LOSsketch_0p2m_1640p5}
\end{figure}
Figure~\ref{fig:heatFluxFireII_1640p5} presents the distributions of convective heat flux (shown as contour and dashed lines) and total heat flux along the vehicle symmetry axis. First, the predicted convective heat flux at the afterbody is generally consistent with literature results \cite{johnston2015features} but is much lower than those on the vehicle forebody (note that the peak forebody convective heat occurs at the vehicle shoulder rather than at the stagnation point). However, the convective heat flux predictions at the stagnation point reported in different studies \cite{sutton1984air,johnston2018impact,kim2021modification,greendyke1994convective} show significant scatter; the present simulation aligns closely with the results of Greendyke \cite{greendyke1994convective} but is substantially lower than the experimental measurement. After adding the radiative heat flux computed using the ray-tracing method, the total heat flux at $X = 0.2~\rm m$ obtained from the current simulation remains slightly below the experimental value.

\begin{figure}[!htb]
    \centering
    \includegraphics[width=0.7\linewidth]{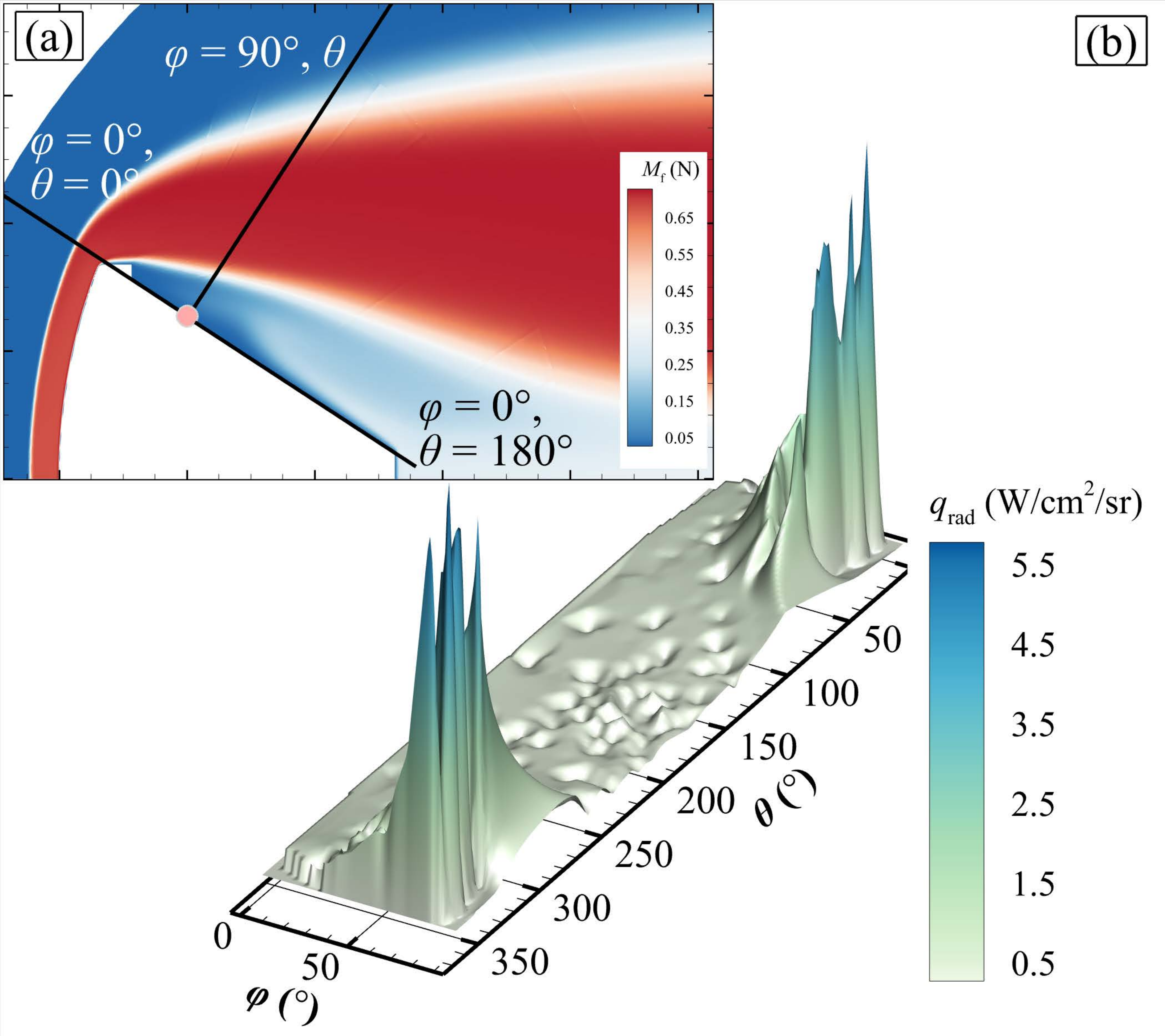}
    \caption{\enspace Radiation distribution along LOS, integrated over the wavelength range $\lambda = 100-1000$ nm, in different directions at $X = 0.2$ m on the vehicle afterbody for $t = 1640.5~\rm s$.}
    \label{fig:radFluxLOS_0p2m_1640p5}
\end{figure}
The distribution of the LOSs used in the current ray-tracing calculation is shown in Figure~\ref{fig:LOSsketch_0p2m_1640p5}, which essentially covers the regions in the wake where the contribution to wall radiative heat flux is significant. Combining with Figure~\ref{fig:radFluxLOS_0p2m_1640p5}, the spatial distribution of the integrated radiative intensity $I(\Omega, \lambda=100-1000~\rm nm)$ along each LOS can be seen more clearly. Similarly, in the directions $30^{\circ}<{\varphi}<90^{\circ}$, $0^{\circ}<{\theta}<100^{\circ}$ and $30^{\circ}<{\varphi}<90^{\circ}$, $250^{\circ}<{\theta}<360^{\circ}$, the contributions to the radiative heat flux are significantly higher than in other spatial regions.

\section{Conclusion}
This work presents the development and validation of a radiation solver for high-temperature gas flows, with detailed modeling of atomic and molecular radiative processes.
Comparison with the open-source Spark code shows that, for B-B transitions of representative atom and diatomic molecules, the emission and absorption coefficients exhibit noticeable differences in both peak intensities and rovibrational band shapes. These discrepancies are primarily attributed to the finer spectral grid and updated line-broadening parameters employed in RAPRAL, as well as differences in the treatment of the maximum rotational quantum number.
For radiative transfer, a ray-tracing-based solver of the RTE has been implemented and assessed against flight data from the Fire II experiment. 
The predicted radiative heat flux in the afterbody region agrees well with measurements at $t = 1640.5~\mathrm{s}$, but is significantly underestimated at $t = 1637.5~\mathrm{s}$. This discrepancy may be attributed to the neglect of ablation products and the breakdown of the QSS assumption in the afterbody region. Further detailed analysis is required to quantify the impact of these factors.

Future efforts will focus on extending the capability of RAPRAL to include molecular continuum radiation, particularly B-F and F-F transitions, as well as incorporating additional species relevant to planetary entry, such as polyatomic molecules (e.g., \ch{CO2} and \ch{CH4}). In addition, the integration of the newest radiative datasets will be pursued.
These developments are expected to further improve the accuracy and applicability of the code for radiative heating predictions in multi-species, high-enthalpy, nonequilibrium flows.

\section*{Acknowledgment}
This work was supported by the Strategic Priority Research Program of the Chinese Academy of Sciences (Grant Nos. XDB0620201 and XDB0620401).

\bibliographystyle{unsrt}  
\bibliography{references}  

\end{document}